\documentclass[journal]{IEEEtran}
\usepackage[dvips]{graphicx,color}
\usepackage{latexsym}
\usepackage{amssymb}
\usepackage{amsmath, bm}
\usepackage{array}

\AtBeginDvi{}

\begin{document}

\arraycolsep 0.5mm



%
%
\newcommand{\qed}{\hfill$\square$}
\newcommand{\suchthat}{\mbox{~s.t.~}}
\newcommand{\markov}{\leftrightarrow}
%
%
\newenvironment{pRoof}{%
 \noindent{\em Proof.\ }}{%
 \hspace*{\fill}\qed \\
 \vspace{2ex}}


\newcommand{\ket}[1]{| #1 \rangle}
\newcommand{\bra}[1]{\langle #1 |}
\newcommand{\bol}[1]{\mathbf{#1}}
\newcommand{\rom}[1]{\mathrm{#1}}
\newcommand{\san}[1]{\mathsf{#1}}
\newcommand{\mymid}{:~}
\newcommand{\argmax}{\mathop{\rm argmax}\limits}
\newcommand{\argmin}{\mathop{\rm argmin}\limits}
%
%
%
%
\newcommand{\bc}{\begin{center}}  %
\newcommand{\ec}{\end{center}}
\newcommand{\befi}{\begin{figure}[h]}  %
\newcommand{\enfi}{\end{figure}}
\newcommand{\bsb}{\begin{shadebox}\begin{center}}   %
\newcommand{\esb}{\end{center}\end{shadebox}}
\newcommand{\bs}{\begin{screen}}     %
\newcommand{\es}{\end{screen}}
\newcommand{\bib}{\begin{itembox}}   %
\newcommand{\eib}{\end{itembox}}
\newcommand{\bit}{\begin{itemize}}   %
\newcommand{\eit}{\end{itemize}}
\newcommand{\defeq}{:=}
\newcommand{\Qed}{\hbox{\rule[-2pt]{3pt}{6pt}}}
\newcommand{\beq}{\begin{equation}}
\newcommand{\eeq}{\end{equation}}
\newcommand{\beqa}{\begin{eqnarray}}
\newcommand{\eeqa}{\end{eqnarray}}
\newcommand{\beqno}{\begin{eqnarray*}}
\newcommand{\eeqno}{\end{eqnarray*}}
\newcommand{\ba}{\begin{array}}
\newcommand{\ea}{\end{array}}
\newcommand{\vc}[1]{\mbox{\boldmath $#1$}}
\newcommand{\lvc}[1]{\mbox{\scriptsize \boldmath $#1$}}
\newcommand{\svc}[1]{\mbox{\scriptsize\boldmath $#1$}}

\newcommand{\wh}{\widehat}
\newcommand{\wt}{\widetilde}
\newcommand{\ts}{\textstyle}
\newcommand{\ds}{\displaystyle}
\newcommand{\scs}{\scriptstyle}
\newcommand{\vep}{\varepsilon}
\newcommand{\rhp}{\rightharpoonup}
\newcommand{\cl}{\circ\!\!\!\!\!-}
\newcommand{\bcs}{\dot{\,}.\dot{\,}}
\newcommand{\eqv}{\Leftrightarrow}
\newcommand{\leqv}{\Longleftrightarrow}

\newcommand{\prev}{}
\newtheorem{co}{Corollary} 
\newtheorem{lm}{Lemma} 
\newtheorem{Ex}{Example} 
\newtheorem{Th}{Theorem}
\newtheorem{df}{Definition} 
\newtheorem{pr}{Property} 
\newtheorem{pro}{Proposition} 
\newtheorem{rem}{Remark}

\newcommand{\lcv}{convex } 

\newcommand{\hugel}{{\arraycolsep 0mm
                    \left\{\ba{l}{\,}\\{\,}\ea\right.\!\!}}
\newcommand{\Hugel}{{\arraycolsep 0mm
                    \left\{\ba{l}{\,}\\{\,}\\{\,}\ea\right.\!\!}}
\newcommand{\HUgel}{{\arraycolsep 0mm
                    \left\{\ba{l}{\,}\\{\,}\\{\,}\vspace{-1mm}
                    \\{\,}\ea\right.\!\!}}
\newcommand{\huger}{{\arraycolsep 0mm
                    \left.\ba{l}{\,}\\{\,}\ea\!\!\right\}}}
\newcommand{\Huger}{{\arraycolsep 0mm
                    \left.\ba{l}{\,}\\{\,}\\{\,}\ea\!\!\right\}}}
\newcommand{\HUger}{{\arraycolsep 0mm
                    \left.\ba{l}{\,}\\{\,}\\{\,}\vspace{-1mm}
                    \\{\,}\ea\!\!\right\}}}

\newcommand{\hugebl}{{\arraycolsep 0mm
                    \left[\ba{l}{\,}\\{\,}\ea\right.\!\!}}
\newcommand{\Hugebl}{{\arraycolsep 0mm
                    \left[\ba{l}{\,}\\{\,}\\{\,}\ea\right.\!\!}}
\newcommand{\HUgebl}{{\arraycolsep 0mm
                    \left[\ba{l}{\,}\\{\,}\\{\,}\vspace{-1mm}
                    \\{\,}\ea\right.\!\!}}
\newcommand{\hugebr}{{\arraycolsep 0mm
                    \left.\ba{l}{\,}\\{\,}\ea\!\!\right]}}
\newcommand{\Hugebr}{{\arraycolsep 0mm
                    \left.\ba{l}{\,}\\{\,}\\{\,}\ea\!\!\right]}}
\newcommand{\HUgebr}{{\arraycolsep 0mm
                    \left.\ba{l}{\,}\\{\,}\\{\,}\vspace{-1mm}
                    \\{\,}\ea\!\!\right]}}

\newcommand{\hugecl}{{\arraycolsep 0mm
                    \left(\ba{l}{\,}\\{\,}\ea\right.\!\!}}
\newcommand{\Hugecl}{{\arraycolsep 0mm
                    \left(\ba{l}{\,}\\{\,}\\{\,}\ea\right.\!\!}}
\newcommand{\hugecr}{{\arraycolsep 0mm
                    \left.\ba{l}{\,}\\{\,}\ea\!\!\right)}}
\newcommand{\Hugecr}{{\arraycolsep 0mm
                    \left.\ba{l}{\,}\\{\,}\\{\,}\ea\!\!\right)}}

\newcommand{\hugepl}{{\arraycolsep 0mm
                    \left|\ba{l}{\,}\\{\,}\ea\right.\!\!}}
\newcommand{\Hugepl}{{\arraycolsep 0mm
                    \left|\ba{l}{\,}\\{\,}\\{\,}\ea\right.\!\!}}
\newcommand{\hugepr}{{\arraycolsep 0mm
                    \left.\ba{l}{\,}\\{\,}\ea\!\!\right|}}
\newcommand{\Hugepr}{{\arraycolsep 0mm
                    \left.\ba{l}{\,}\\{\,}\\{\,}\ea\!\!\right|}}

\newcommand{\MEq}[1]{\stackrel{
{\rm (#1)}}{=}}

\newcommand{\MLeq}[1]{\stackrel{
{\rm (#1)}}{\leq}}

\newcommand{\ML}[1]{\stackrel{
{\rm (#1)}}{<}}

\newcommand{\MGeq}[1]{\stackrel{
{\rm (#1)}}{\geq}}

\newcommand{\MG}[1]{\stackrel{
{\rm (#1)}}{>}}

\newcommand{\MPreq}[1]{\stackrel{
{\rm (#1)}}{\preceq}}

\newcommand{\MSueq}[1]{\stackrel{
{\rm (#1)}}{\succeq}}

\newcommand{\Ch}{{\Gamma}}
\newcommand{\Rw}{{W}}

\newcommand{\Cd}{{\cal R}_{\rm d}(\Ch)}
\newcommand{\CdB}{{\cal R}_{\rm d}^{\prime}(\Ch)}
\newcommand{\CdBB}{{\cal R}_{\rm d}^{\prime\prime}(\Ch)}

\newcommand{\Cdi}{{\cal R}_{\rm d}^{\rm (in)}(\Ch)}
\newcommand{\Cdo}{{\cal R}_{\rm d}^{\rm (out)}(\Ch)}

\newcommand{\tCdi}{\tilde{\cal R}_{\rm d}^{\rm (in)}(\Ch)}
\newcommand{\tCdo}{\tilde{\cal R}_{\rm d}^{\rm (out)}(\Ch)}
\newcommand{\hCdo}{  \hat{\cal R}_{\rm d}^{\rm (out)}(\Ch)}

\newcommand{\Cs}{{\cal R}_{\rm s}(\Ch)}
\newcommand{\CsB}{{\cal R}_{\rm s}^{\prime}(\Ch)}
\newcommand{\CsBB}{{\cal R}_{\rm s}^{\prime\prime}(\Ch)}

\newcommand{\Csi}{{\cal R}_{\rm s}^{\rm (in)}(\Ch)}
\newcommand{\Cso}{{\cal R}_{\rm s}^{\rm (out)}(\Ch)}
\newcommand{\tCsi}{\tilde{\cal R}_{\rm s}^{\rm (in)}(\Ch)}
\newcommand{\tCso}{\tilde{\cal R}_{\rm s}^{\rm (out)}(\Ch)}
\newcommand{\cCsi}{\check{\cal R}_{\rm s}^{\rm (in)}(\Ch)}
\newcommand{\Cds}{{\cal C}_{\rm ds}(\Ch)}
\newcommand{\Cdsi}{{\cal C}_{\rm ds}^{\rm (in)}(\Ch)}
\newcommand{\Cdso}{{\cal C}_{\rm ds}^{\rm (out)}(\Ch)}
\newcommand{\tCdsi}{\tilde{\cal C}_{\rm ds}^{\rm (in)}(\Ch)}
\newcommand{\tCdso}{\tilde{\cal C}_{\rm ds}^{\rm (out)}(\Ch)}
\newcommand{\hCdso}{\hat{\cal C}_{\rm ds}^{\rm (out)}(\Ch)}
\newcommand{\Css}{{\cal C}_{\rm ss}(\Ch)}
\newcommand{\Cssi}{{\cal C}_{\rm ss}^{\rm (in)}(\Ch)}
\newcommand{\Csso}{{\cal C}_{\rm ss}^{\rm (out)}(\Ch)}
\newcommand{\tCssi}{\tilde{\cal C}_{\rm ss}^{\rm (in)}(\Ch)}
\newcommand{\tCsso}{\tilde{\cal C}_{\rm ss}^{\rm (out)}(\Ch)}
\newcommand{\Cde}{{\cal R}_{\rm d1e}(\Ch)}
\newcommand{\Cdei}{{\cal R}_{\rm d1e}^{\rm (in)}(\Ch)}
\newcommand{\Cdeo}{{\cal R}_{\rm d1e}^{\rm (out)}(\Ch)}
\newcommand{\tCdei}{\tilde{\cal R}_{\rm d1e}^{\rm (in)}(\Ch)}
\newcommand{\tCdeo}{\tilde{\cal R}_{\rm d1e}^{\rm (out)}(\Ch)}
\newcommand{\hCdeo}{  \hat{\cal R}_{\rm d1e}^{\rm (out)}(\Ch)} 
\newcommand{\Cse}{{\cal R}_{\rm s1e}(\Ch)}
\newcommand{\Csei}{{\cal R}_{\rm s1e}^{\rm (in)}(\Ch)}
\newcommand{\Cseo}{{\cal R}_{\rm s1e}^{\rm (out)}(\Ch)}
\newcommand{\tCsei}{\tilde{\cal R}_{\rm s1e}^{\rm (in)}(\Ch)}
\newcommand{\tCseo}{\tilde{\cal R}_{\rm s1e}^{\rm (out)}(\Ch)}

\newcommand{\Capa}{C}

\newcommand{\ZeTa}{\zeta(S;Y,Z|U)}
\newcommand{\ZeTaI}{\zeta(S_i;Y_i,Z_i|U_i)}

\newcommand{\SP}{\mbox{{\scriptsize sp}}}
\newcommand{\mSP}{\mbox{{\scriptsize sp}}}
\newcommand{\CEreg}{\irBr{rate} }
\newcommand{\CEregB}{rate\MarkOh{-equivocation }}

\newcommand{\Cls}{class NL}
\newcommand{\vSpa}{\vspace{0.3mm}}
\newcommand{\Prmt}{\zeta}
\newcommand{\pj}{\omega_n}

\newfont{\bg}{cmr10 scaled \magstep4}
\newcommand{\bigzerol}{\smash{\hbox{\bg 0}}}
\newcommand{\bigzerou}{\smash{\lower1.7ex\hbox{\bg 0}}}
\newcommand{\nbn}{\frac{1}{n}}
\newcommand{\ra}{\rightarrow}
\newcommand{\la}{\leftarrow}
\newcommand{\ldo}{\ldots}
\newcommand{\ep}{\epsilon }
\newcommand{\typi}{A_{\epsilon }^{n}}
\newcommand{\bx}{\hspace*{\fill}$\Box$}
\newcommand{\pa}{\vert}
\newcommand{\ignore}[1]{}
%
%

%
%

%
%
%

\newenvironment{jenumerate}
	{\begin{enumerate}\itemsep=-0.25em \parindent=1zw}{\end{enumerate}}
\newenvironment{jdescription}
	{\begin{description}\itemsep=-0.25em \parindent=1zw}{\end{description}}
\newenvironment{jitemize}
	{\begin{itemize}\itemsep=-0.25em \parindent=1zw}{\end{itemize}}
\renewcommand{\labelitemii}{$\cdot$}

\newcommand{\iro}[2]{{\color[named]{#1}#2\normalcolor}}
 \newcommand{\irr}[1]{{\color[named]{Black}#1\normalcolor}}

\newcommand{\irrTypo}[1]{{\color[named]{Black}#1\normalcolor}}
\newcommand{\irrAE}[1]{{\color[named]{Black}#1\normalcolor}}
\newcommand{\irbRevA}[1]{{\color[named]{Black}#1\normalcolor}}
\newcommand{\irOlgRevB}[1]{{\color[named]{Black}#1\normalcolor}}
\newcommand{\irPkAu}[1]{{\color[named]{Black}#1\normalcolor}}



\newcommand{\irrAEb}[1]{{\color[named]{Black}#1\normalcolor}}

%
%


\newcommand{\irOlgRevBb}[1]{{\color[named]{Black}#1\normalcolor}}
\newcommand{\irbRevAb}[1]{{\color[named]{Black}#1\normalcolor}}
\newcommand{\irPkAub}[1]{{\color[named]{Black}#1\normalcolor}}

\newcommand{\irPkAubc}[1]{{\color[named]{Black}#1\normalcolor}}
%

\newcommand{\irrTypoRevA}[1]{{\color[named]{Black}#1\normalcolor}}
\newcommand{\irrComRevA}[1]{{\color[named]{Black}#1\normalcolor}}
\newcommand{\irOlgTypoRevB}[1]{{\color[named]{Black}#1\normalcolor}}
\newcommand{\irOlgComRevB}[1]{{\color[named]{Black}#1\normalcolor}}
\newcommand{\irPkOh}[1]{{\color[named]{Black}#1\normalcolor}}
\newcommand{\irOOh}[1]{{\color[named]{Black}#1\normalcolor}}

\newcommand{\clrTypoRevA}[1]{{\color[named]{Black}#1\normalcolor}}
\newcommand{\clrTypoRevB}[1]{{\color[named]{Black}#1\normalcolor}}
\newcommand{\clrMComRevB}[1]{{\color[named]{Black}#1\normalcolor}}
\newcommand{\clrTypoOh}[1]{{\color[named]{Black}#1\normalcolor}}
\newcommand{\clrOh}[1]{{\color[named]{Black}#1\normalcolor}}
\newcommand{\clrMjRvOh}[1]{{\color[named]{Black}#1\normalcolor}}

\newcommand{\irOAuc}[1]{{\color[named]{Black}#1\normalcolor}}

\newcommand{\irg}[1]{{\color[named]{Green}#1\normalcolor}}

\newcommand{\irb}[1]{{\color[named]{Blue}#1\normalcolor}}

\newcommand{\irBl}[1]{{\color[named]{Black}#1\normalcolor}}
\newcommand{\irWh}[1]{{\color[named]{White}#1\normalcolor}}

\newcommand{\irY}[1]{{\color[named]{Yellow}#1\normalcolor}}
\newcommand{\irO}[1]{{\color[named]{Orange}#1\normalcolor}}
\newcommand{\irBr}[1]{{\color[named]{Black}#1\normalcolor}}

\newcommand{\IrBr}[1]{{\color[named]{Purple}#1\normalcolor}}
\newcommand{\irBw}[1]{{\color[named]{Brown}#1\normalcolor}}
\newcommand{\irPk}[1]{{\color[named]{Magenta}#1\normalcolor}}
\newcommand{\irCb}[1]{{\color[named]{CadetBlue}#1\normalcolor}}

\newcommand{\irMho}[1]{{\color[named]{Mahogany}#1\normalcolor}}
\newcommand{\irOlg}[1]{{\color[named]{Black}#1\normalcolor}}

\newcommand{\irBg}[1]{{\color[named]{BlueGreen}#1\normalcolor}}
\newcommand{\irCy}[1]{{\color[named]{Cyan}#1\normalcolor}}
\newcommand{\irRyp }[1]{{\color[named]{RoyalPurple}#1\normalcolor}}

\newcommand{\irAqm}[1]{{\color[named]{Aquamarine}#1\normalcolor}}
\newcommand{\irRyb}[1]{{\color[named]{RoyalBule}#1\normalcolor}}
\newcommand{\irNvb}[1]{{\color[named]{NavyBlue}#1\normalcolor}}
\newcommand{\irSkb}[1]{{\color[named]{SkyBlue}#1\normalcolor}}
\newcommand{\irTeb}[1]{{\color[named]{TeaBlue}#1\normalcolor}}
\newcommand{\irSep}[1]{{\color[named]{Sepia}#1\normalcolor}}
\newcommand{\irReo}[1]{{\color[named]{RedOrange}#1\normalcolor}}
\newcommand{\irRur}[1]{{\color[named]{RubineRed}#1\normalcolor}}
\newcommand{\irSa }[1]{{\color[named]{Salmon}#1\normalcolor}}
\newcommand{\irAp}[1]{{\color[named]{Apricot}#1\normalcolor}}


%
\newenvironment{indention}[1]{\par
\addtolength{\leftskip}{#1}\begingroup}{\endgroup\par}
%
\newcommand{\namelistlabel}[1]{\mbox{#1}\hfill} 
\newenvironment{namelist}[1]{%
\begin{list}{}
{\let\makelabel\namelistlabel
\settowidth{\labelwidth}{#1}
\setlength{\leftmargin}{1.1\labelwidth}}
}{%
\end{list}}
%
%
\newcommand{\bfig}{\begin{figure}[t]}
\newcommand{\efig}{\end{figure}}
\setcounter{page}{1}

\newtheorem{theorem}{Theorem}
\newcommand{\Ep}{\mbox{\rm e}}

\newcommand{\Exp}{\exp
}
\newcommand{\idenc}{{\varphi}_n}
\newcommand{\resenc}{
{\varphi}_n}
\newcommand{\ID}{\mbox{\scriptsize ID}}
\newcommand{\TR}{\mbox{\scriptsize TR}}
\newcommand{\Av}{\mbox{\sf E}}

\newcommand{\Vl}{|}
\newcommand{\Ag}{(R,P_{X^n}|W^n)}
\newcommand{\Agv}[1]{({#1},P_{X^n}|W^n)}
\newcommand{\Avw}[1]{({#1}|W^n)}

\newcommand{\Jd}{X^nY^n}
\newcommand{\IdR}{r_n}

\newcommand{\Index}{{n,i}}

\newcommand{\cid}{C_{\mbox{\scriptsize ID}}}
\newcommand{\cida}{C_{\mbox{{\scriptsize ID,a}}}}
\newcommand{\rmOH}{\empty
}

\newcommand{\NoizeVar}{\sigma^2}

\newcommand{\GN}{
\frac{{\rm e}^{-\frac{(y-x)^2}{2\NoizeVar}}}
{\sqrt{2\pi {\NoizeVar}}}}

\arraycolsep 0.5mm
\date{}

\newcommand{\eXchange}{\empty}

\newcommand{\BiBGauArXiv}{
\bibitem{SingleStConvGauArXiv17}
Y. Oohama, ``The reliability function for 
the additive white Gaussian noise channel at 
rates above the capacity,'' 
available at https://arxiv.org/abs/1701.06357.
}

\date{}
\title{
Exponent Function for Stationary Memoryless Channels with 
Input Cost at Rates above the Capacity 
}

\author{Yasutada Oohama
\thanks{Dept. of Communication Engineering 
  and Informatics, University of Electro-Communications,
  1-5-1 Chofugaoka Chofu-shi, Tokyo 182-8585, Japan.   
}
}

\markboth{
}
{
}
\maketitle
\newcommand{\irPkAubb}[1]{{\color[named]{Black}#1\normalcolor}}
\begin{abstract}
We consider the stationary memoryless channels with input cost. 
We prove that for transmission rates above the capacity 
the correct probability of decoding tends to zero exponentially 
as the block length $n$ of codes tends to infinity. 
In the case where both of channel input and output sets 
are finite, we determine the optimal exponent function on 
the above exponential decay of the correct probability. 
To derive this result we use a technique called 
the recursive method, 
which the author has recently introduced to study 
the strong converse theorems for several communication 
systems including multiterminal 
systems. The recursive method is based on the information spectrum approach and 
utilizes a certain recursive structure on the information spectrum 
quantities. In this paper to obtain {\it the optimal} exponent 
function we propose a new method called {\it the improved} 
recursive method. 
We further give an example of discrete memoryless channel 
for which the exponent function obtained by the previous 
recursive method can by no means be the optimal exponent 
function. 
\end{abstract}


\section{Introduction}
\noindent

A certain class of noisy channels has a property 
that the error probability 
of decoding goes to one as the block 
length $n$ of transmitted codes 
tends to infinity at rates above 
the channel capacity. This property is 
called the strong converse property. 
In the case of DMCs without cost Arimoto \cite{arimoto:73} proved that the 
error probability of decoding goes to one exponentially and derived a 
lower bound of the exponent function. Subsequently, Dueck and K\"orner
\cite{dueck:79} determined the optimal exponent function for the 
error probability of decoding to go to one. They derived the result by
using a combinatorial method \clrTypoRevA{based} 
on the type of sequences \cite{csi2011information}.
The equality of the lower bound of Arimoto \cite{arimoto:73} to 
that of the optimal bound of Dueck and K\"orner \cite{dueck:79} 
was proved by the author \cite{oohama2015two}. A simple 
derivation of the exponent function in the problem set up 
of quantum channel coding was given by
\clrTypoRevB{Hayashi and Nagaoka \cite{hayashi:03}}.
In the derivation they used the information spectrum method
introduced by Han \cite{Han98InfSpec} and a min-max expression 
of the channel capacity. 

In this paper we determine the optimal exponent function on the correct 
probability of decoding at rates above capacity for DMCs with input cost. 
The optimal exponent function can be obtained by a method quite parallel 
with the method Dueck and K\"orner \cite{dueck:79} used to 
obtain the optimal exponent function in the case without input cost. 
Instead of using their method, we use two other methods based on the 
information spectrum method.
{One is a method, which is regarded as a minor extension of 
the method of 
\clrTypoRevB{Hayashi and Nagaoka 
\cite{hayashi:03}} to the case where we have 
the constraint on the input cost. 
We hereafter call this method the NH method.
}
The other is a method called the recursive method. This method includes 
a certain recursive algorithm for a single letterization of exponent 
functions. The recursive method is a general powerful tool to prove 
strong converse theorems for several coding problems in information 
theory. In fact, this method plays important roles in deriving exponential 
strong converse exponent for 
communication systems treated in 
\cite{Oohama15a},
\cite{Oohama15b},
\cite{OhIsita16abc},
\cite{oohama:18}, and 
\cite{oohama:19}.

In this paper \irrAE{we propose a new method} 
which includes an {\it improved} recursive algorithm. 
We \irPkAu{hereafter} \irrTypo{call} this method the 
improved recursive method. 
\irrAE{``Improvement" in the improved recursive method means 
that it is an improvement from the recursive method proposed by 
the author in his previous works 
\cite{Oohama15a},
\cite{Oohama15b},
\cite{OhIsita16abc},
\cite{oohama:18}, and 
\cite{oohama:19}.}

{In the following argument, to distinguish the improved recursive 
method from the previous recursive method, we simply call the latter 
the recursive method. Our main results on the lower bound of 
the optimal exponent function are stated as two theorems. One is 
a result obtained by the NH method and the other is a result 
obtained by the improved recursive method. The former 
is stated in Theorem \ref{Th:main0} and the latter is 
stated in Theorem \ref{Th:main}. By some algebra we can show that those 
lower bounds are equal to the optimal exponent function 
\irrTypoRevA{obtained} 
by the method of Dueck and K\"orner \cite{dueck:79}.}

We also derive a lower bound of the optimal exponent function 
using the recursive method. This result is stated in 
Proposition \ref{pro:oldrecursive}. From this proposition we can 
see that the exponent function derived by the recursive 
method serves as an exponent function of the strong converse 
theorem for DMC with input cost. We further 
give an example of DMC for which this exponent function 
can not be the optimal exponent function.

In this paper, to compare the improved recursive method 
with the NH method and demonstrate that the former is 
quite different from the latter, we separately prove 
Theorems \ref{Th:main0} and \ref{Th:main}. 
A discussion on the difference between the above two methods 
is provided in Section \ref{sec:Secaa}. Summary of 
the discussion is as follows.
\begin{itemize}
\item[1.] Multi-letter expression on the lower bound of the optimal 
exponent function is expressed in the form of a max-min 
optimization on the 
moment generating function with respect to the information spectral 
quantity.

\item[2.] In the single letterization of the multi-letter 
lower bound, the NH method first relaxes the max-min 
expression of the lower bound to the 
min-max expression and then evaluate the single letterized lower bound. 
Therefore, the resulting lower bound has a form of min-max optimization.   
On the other hand, the improved recursive method preserves 
the max-min structure in the single characterization of the 
multil-letter lower bound. Therefore, the resulting single 
letterized lower bound has a form of max-min optimization.

\item[3.] Since the min-max optimization does not exceed the max-min 
optimization, the NH method is by no means better 
than the improved recursive method. When the 
minimax \eXchange theorem \cite{csiszar:95} holds 
both methods \irrTypoRevA{yield} the same result. 
\end{itemize}
The above difference between the NH method and 
the improved recursive method has a close connection 
with the fact that the former 
does \irrTypoRevA{not} seem to be applied to 
the proof of the converse coding theorems of 
the multi-terminal communication systems but 
the latter can be applied. 

As we mentioned previously, there have been three different methods
to derive the optimal \irPkAu{exponent} function for the correct 
probability of decoding. Those three methods are the 
method by Arimoto \cite{arimoto:73}, 
the method by Dueck and K\"orner \cite{dueck:79}, and 
\irPkAu{the NH method} \clrTypoRevB{\cite{hayashi:03}}.
Hence the improved recursive method proposed in this paper 
can be regarded as the fourth new method to derive the result.


The NH method \clrTypoRevB{\cite{hayashi:03}}
and the improved recursive method are based on the information 
spectrum method. Those two methods have 
a common advantage that they also work for the derivation of 
the exponent function for general memoryless 
channels \irrTypo{(GMCs)}, where the channel 
input and outputs are real lines. 
On the other hand, the method of type used by Dueck and K\"orner 
\cite{dueck:79} only works for DMCs where channel input and 
output sets are finite. Using the advantage mentioned above, we derive a 
new lower bound of the optimal exponent function for GMCs. This lower 
bound is thought to be useful for deriving explicit lower bounds of the 
optimal exponent functions for several examples of GMCs.

\newcommand{\Memorandom}{
aaa
}

\section{Exponent Functions on the Correct Probability of Decoding}

\subsection{Capacity Results for 
the Discrete Memoryless Channels with Input Cost
}

\noindent

We consider a stationary discrete memoryless channel \irrTypo{(SDMC)} with 
the input set ${\cal X}$ and the output set ${\cal Y}$. 
We assume that ${\cal X}$ and ${\cal Y}$ are finite sets. 
A case where ${\cal X}$ and ${\cal Y}$ are real lines will be 
treated in Section \ref{sec:GeneralCase}.
The SDMC is specified with the following stochastic matrix:
\beq
{W} \defeq \{ {W}(y|x)\}_{
(x,y) 
\in    {\cal X}
\times {\cal Y} 
}.
\eeq
Let $X^n$ be a random variable taking values in ${\cal X}^n$. 
We write an element of ${\cal X}^n$ as 
$x^n=x_{1}x_{2}$$\cdots x_{n}.$ 
Suppose that $X^n$ has a probability distribution on ${\cal X}^n$ 
denoted by 
$p_{X^n}=$ 
$\left\{p_{X^n}(x^n) 
\right\}_{{x^n} \in {\cal X}^n}$.
Similar notations are adopted for other random variables. 
Let $Y^n \in {\cal Y}^n$ be a random variable obtained 
as the channel output by connecting $X^n$ to the 
input of channel. We write a conditional 
distribution of $Y^n$ on given $X^n$ as 
$$
W^n=
\left\{W^n(y^n|x^n)\right\}_{(x^n,y^n) 
\in {\cal X}^n \times {\cal Y}^n}.
$$
Since the channel is memoryless, we have 
\beq
W^n({y}^n|x^n)=\prod_{t=1}^nW (y_t|x_t).
\label{eqn:sde0}
\eeq
Let $K_n$ be uniformly distributed random variables 
taking values in message sets ${\cal K}_n $. 
The random variable $K_n$ is a message sent to the receiver.
A sender transforms $K_n$ into a transmitted sequence 
$X^n$ using an encoder function and sends it to the receiver. 
In this paper we assume that the encoder function $\varphi^{(n)}$ 
is a deterministic encoder. In this case, $\varphi^{(n)}$ 
is a one-to-one mapping from ${\cal K}_n$ into ${\cal X}^n$. 
The joint probability mass function on 
$
{\cal X}^n$ 
$\times {\cal Y}^n$ 
is given by
$$
\Pr\{(X^n,Y^n)=(x^n,y^n)\}
=\frac{1}{\pa{\cal K}_n\pa }
\prod_{t=1}^n W\left(y_t\left|x_t(k)\right.\right),
$$
where $x_t(k)=[\varphi^{(n)}(k)]_t$, $t=1,2,\cdots,n$ 
are the $t$-th components of $x^n=x^n(k)$ $=\varphi^{(n)}(k)$
and $\pa {\cal K}_n \pa$ is a cardinality 
of the set ${\cal K}_n$. 
The decoding function 
at the receiver is denoted by ${\psi}^{(n)}$. 
This function is formally defined by
$
{\psi}^{(n)}: {\cal Y}^{n} \to {\cal K}_n.
$
Let $c: {\cal X} \to [0,\infty)$ be a cost function.
The average cost on output of $\varphi^{(n)}$ 
must not exceed $\Gamma$. This condition is given by
\irrAEb{$\varphi^{(n)}({\cal K}_n) \subseteq {\cal S}_{\Gamma}^{(n)}$}, where 
\beqno
{\cal S}_{\Gamma}^{(n)} 
&\defeq & \biggl\{x^n\in {\cal X}^n: 
\frac{1}{n}\sum_{t=1}^n c(x_t) \leq \Gamma 
\biggr\}.
\eeqno
\irrComRevA{Set $\Gamma_0 \defeq 
\min_{x\in {\cal X}}c(x)$. Since  
$$
\frac{1}{n}\sum_{t=1}^n c(x_t)\geq \Gamma_0,  
$$
we have ${\cal S}_{\Gamma}^{(n)}=\emptyset$ if 
$\Gamma<\Gamma_0$. Hence we assume $\Gamma\geq\Gamma_0$ 
throughout this paper.}
The average error \clrTypoRevA{probability} of decoding at the 
receiver is defined by 
\beqno
{\rm P}_{\rm e}^{(n)}
&=&{\rm P}_{\rm e}^{(n)}(\varphi^{(n)},\psi^{(n)}|W)
\defeq \Pr\{ \psi^{(n)}(Y^n)\neq K_n \}
\\
&=&1-\Pr\{\psi^{(n)}(Y^n)= K_n \}.
\eeqno
Set 
${\rm P}^{(n)}_{\rm c}(\varphi^{(n)},\psi^{(n)}|W)
\defeq 1-{\rm P}^{(n)}_{\rm e}(\varphi^{(n)},\psi^{(n)}|W).$
The quantity ${\rm P}^{(n)}_{\rm c}
={\rm P}^{(n)}_{\rm c}(\varphi^{(n)},\psi^{(n)}|W)$ 
is called the average correct probability of decoding.
For $k\in {\cal K}_n$, set
$
{\cal D}(k)\defeq \{ y^n: \psi^{(n)}(y^n)=k \}.
$
The families of sets 
$\{ {\cal D}(k) \}_{k\in {\cal K}_n}$ 
is called the decoding regions. Using the decoding region, 
${\rm P}_{\rm c}^{(n)}$ can be written as
\beqno
& &{\rm P}_{\rm c}^{(n)}
=\frac{1}{|{\cal K}_n|} 
\sum_{k\in {\cal K}_n }
W^n\left( {\cal D}(k) \left| \varphi^{(n)}(k) \right.\right).
\eeqno
\newcommand{\OmitAAA}{
${\rm P}_{\rm e}^{(n)}$ can be written as
\beqno
& &
{\rm P}_{\rm e}^{(n)}
\\
&=&1-\frac{1}{|{\cal K}_n|}\sum_{k\in {\cal K}_n }
\Pr\{Y^n \in {\cal D}(k)|X^n=\varphi^{(n)}(k))\}
\\
&=&1-\frac{1}{|{\cal K}_n|}\sum_{k\in {\cal K}_n }
\sum_{\scs y^n \in {\cal D}(k)}
W^n\left(y^n \left| \varphi^{(n)}(k) \right.\right)
\\
&=&1-\frac{1}{|{\cal K}_n|}\sum_{k\in {\cal K}_n }
W^n\left( {\cal D}(k) \left| \varphi^{(n)}(k) \right.\right).
\eeqno
Set
\beqno
& &{\rm P}^{(n)}_{\rm c}={\rm P}^{(n)}_{\rm c}(\varphi^{(n)},\psi^{(n)}|W)
\defeq 
1-{\rm P}^{(n)}_{\rm e}(\varphi^{(n)}, \psi^{(n)} | W ).
\eeqno
}
\irrComRevA{
In this paper we assume that 
all logarithms are taken to the base natural.}
For given $\varepsilon$ $\in (0,1)$, 
$R$ is $\varepsilon$-{\it achievable} under $\Gamma$
if for any $\delta>0$, there exist a positive integer 
$n_0=n_0(\varepsilon,\delta)$ and a sequence of pairs 
$
\{(\varphi^{(n)},\psi^{(n)}): 
\irrAE{\varphi^{(n)}({\cal K}_n) \subseteq {\cal S}_{\Gamma}^{(n)}} 
\}_{n=1}^{\infty}
$ 
such that for any $n\geq n_0(\varepsilon,\delta)$, 
\beqa 
{\rm P}_{{\rm e}}^{(n)}
(\varphi^{(n)},\psi^{(n)} | W )
&\leq &\varepsilon, 
\quad
\nbn \log \pa {\cal K}_n \pa  \geq  R-\delta.
\eeqa
The supremum of all $\varepsilon$-achievable 
$R$ under $\Gamma$ is denoted by 
${C}_{\rm DMC}(\varepsilon, \Gamma | W)$.
We set
$$
{C}_{\rm DMC}(\Gamma|W)
\defeq \inf_{\varepsilon\in (0,1)}
C_{\rm DMC}(\varepsilon,\Gamma|W),
$$
which is called the channel capacity. 
\newcommand{\Zsa}{
The maximum error probability of decoding is 
defined by as follows:
\beqno
{\rm P}_{{\rm e,\irOlg{m}}}^{(n)}&=&
   {\rm P}_{{\rm e,\irOlg{m}}}^{(n)}(\varphi^{(n)},\psi^{(n)}|W)
\\
& \defeq & \max_{k \in {\cal K}_n}
\Pr\{\psi^{(n)}(Y^n)\neq k|K_n=k\}.
\eeqno
Based on this quantity, we define the \irOlg{maximum capacity} 
as follows. For a given $\varepsilon \in (0,1)$, 
$\irb{R}$ is $\varepsilon$-{\it achievable} under 
$\Gamma$, if for any $\delta>0$, there exist a positive integer 
$n_0=n_0(\varepsilon,\delta)$ and a sequence of pairs 
$
\{(\varphi^{(n)},\psi^{(n)}): 
\irrTypo{\varphi^{(n)}({\empty K}_n) \subseteq {\cal S}_{\Gamma}^{(n)}} 
\}_{n=1}^{\infty}$ 
such that for any $n\geq n_0(\varepsilon,\delta)$, 
\beqa 
{\rm P}_{{\rm e},\irOlg{\rm m}}^{(n)}
(\varphi^{(n)},\psi^{(n)} | W)
&\leq &\varepsilon, 
\quad
\nbn \log \pa {\cal K}_n \pa 
\geq \irb{R}-\delta.
\eeqa
The supremum of all $\varepsilon$-achievable 
rates under $\Gamma$ is denoted by $C_{\irOlg{\rm m},{\rm DMC}}
(\varepsilon,\Gamma|\irBr{W})$. 
We set 
$$
{C}_{\irOlg{\rm m},\rm DMC}(\Gamma | \irBr{W})
=\inf_{\varepsilon\in (0,1)} 
C_{\irOlg{\rm m},\rm DMC}(\varepsilon,\Gamma |\irBr{W})
$$
which is called the \irOlg{maximum} capacity of the DMC.
}
Set
\beq
C(\Gamma | W)=\max_{ \scs p_X \in {\cal P}({\cal X}):
          \atop{\scs
           {\rm E}_{p_X} c(X) \leq \Gamma 
          }
     }I(p_X,W), 
\eeq
where ${\cal P}({\cal X})$ is \irrTypo{the} set of probability 
\irrTypo{distributions} on 
${\cal X}$ and $I(p_X,W)$ stands for \irrAEb{the} mutual information 
between $X$ and $Y$ when input distribution of $X$ is $p_X$.
The following is a well known result.
\begin{Th}
\label{th:ddirect} 
{\rm For any DMC $W$, we have
$$
{C}_{\rm DMC}(\Gamma|W)={C}(\Gamma| W).
$$
}
\end{Th}

Han \cite{Han98InfSpec} established the strong converse theorem for 
DMCs with input cost. His result is as follows. 

\begin{Th}[Han \cite{Han98InfSpec}] 
If $R>C(\Gamma | W)$, then for any 
$\{(\varphi^{(n)},\psi^{(n)}): 
\irrAE{
\varphi^{(n)}({\cal K}_n) \subseteq {\cal S}_{\Gamma}^{(n)}
} \}_{n=1}^{\infty}
$ 
satisfying 
$$
\liminf_{n\to \infty}\irOlgRevBb{\frac{1}{n}}
\irrTypo{  \log \pa {\cal K}_n \pa } \geq R,
$$ 
we have 
$$
\lim_{n\to\infty}{\rm P}_{\rm e}^{(n)}
(\varphi^{(n)}, \psi^{(n)}|W)=1.
$$ 
\end{Th}

The following corollary immediately follows from this theorem. 
\begin{co}
For each fixed $\varepsilon$ $ \in (0,1)$ and any DMC 
$W$, we have 
$$
{C}_{\rm DMC}(\varepsilon,\Gamma |W)={C}(\Gamma| W).
$$
\end{co}

\irrAE{This corollary stated as Theorem 6.11 in Csisz\'ar 
and K\"orner \cite{csi2011information}.}

\subsection{Problem Set-up and Previous Results}
\noindent

To examine an asymptotic behavior of 
${\rm P}^{(n)}_{\rm c}(\varphi^{(n)},$ $\psi^{(n)})$ 
for large $n$ at $R>C(\Gamma|W)$, we define 
the following quantities:
\beqno
& & 
G^{(n)}(R,\Gamma|W)
\\
&&\defeq
\min_{\scs 
(\varphi^{(n)},\psi^{(n)}):
    \atop{\scs 
         \varphi^{(n)}(\irOlgRevBb{{\cal K}_n}) 
           \irOlgRevBb{ \subseteq }{\cal S}_{\Gamma}^{(n)},
          \atop{\scs
               (1/n)\log M_n \geq R 
               }    
         }
    }
\hspace*{-2mm}
\left(-\frac{1}{n}\right)
\log {\rm P}_{\rm c}^{(n)}(\varphi^{(n)},\psi^{(n)}|W),
\\
& &
{G}^{*}(R,\Gamma|W) \defeq \lim_{n \to \infty} G^{(n)}(R,\Gamma|W).
\eeqno
On the above exponent functions, we have the following property. 
\begin{pr}\label{pr:pr0}{$\quad$
\begin{itemize}
\item[a)] By definition we have that for each fixed $n\geq 1$, 
$G^{(n)}(R,\Gamma|W)$ is a monotone increasing function of $R\geq0$ 
and satisfies $G^{(n)}(R,\Gamma|W) \leq R$. 

\item[b)] The sequence $\{G^{(n)}(R,\Gamma|W)$ $\}_{n\geq 1}$ of 
exponent functions satisfies the following subadditivity property:
\begin{align}
 & G^{(n+m)}(R,\Gamma|{W}) 
\nonumber\\
&\leq \frac{nG^{(n)}(R,\Gamma|W)+mG^{(m)}(R,\Gamma|W)}{n+m},
\label{eqn:Zxcxx} 
\end{align}
from which we have that ${G}^{*}(R,\Gamma|W)$ exists and is equal to 
$\inf_{n\geq 1}G^{(n)}(R,\Gamma|W)$. 

\item[c)]
For fixed $R>0$, the function ${G}^{*}(R,\Gamma|W)$ is a monotone 
decreasing function 
of $\Gamma$. 
For fixed $\Gamma>\Gamma_0=\min_{x\in{\cal X}}c(x) $, the function ${G}^{*}(R,\Gamma|W)$ a monotone 
increasing function of $R$ and satisfies 
\beq
G^{*}(R,\Gamma|W)\leq R.
\label{eqn:Zxc}
\eeq

\item[d)] The function ${G}^{*}(R,\Gamma|W)$ is a convex function of 
$(R,\Gamma)$. 
\end{itemize}
}
\end{pr}

Proof of Property \ref{pr:pr0} is given in Appendix \ref{sub:ApdaAAAA}. 
\newcommand{\ApdaAAAA}{
\subsection{
General Properties on ${G}^{*}(R,\Gamma|W)$
} 
\label{sub:ApdaAAAA}

In this appendix we prove Property \ref{pr:pr0} describing 
general properties on ${G}^{*}(R,\Gamma|W)$. 

{\it Proof of Property \ref{pr:pr0}:} 
By definition it is obvious that for fixed $\Gamma>0$, 
${G}^{(n)}(R,\Gamma |W)$ 
is a monotone increasing function of $R>0$ and that 
for fixed $R>0$, ${G}^{(n)}(R,\Gamma |W)$ 
is a monotone increasing function of $\Gamma>0$. 
We prove the part b). By time sharing we have that 
\begin{align}
 & G^{(n+m)}\left(\left.
\frac{n R + m R^{\prime}}{n+m},
\frac{n \Gamma + m \Gamma^{\prime}}{n+m}
\right|W\right) 
\nonumber\\
&\leq \frac{nG^{(n)}(R,\Gamma|W)
  +mG^{(m)}(R^{\prime},\Gamma^{\prime}|W)}{n+m}.
\label{eqn:Sdd} 
\end{align}
The part b) follows by letting $R=R^{\prime}$ and $\Gamma=\Gamma^{\prime}$ in (\ref{eqn:Sdd}). 
We next prove the part c). 
By definition it is obvious that for fixed $\Gamma>0$, ${G}^{*}(R,\Gamma |W)$ 
is a monotone decreasing function of $R>0$ 
and that for fixed $R>0$, ${G}^{*}(R,\Gamma |W)$ 
is a monotone increasing function of $\Gamma>0$. 
It is obvious that the worst pair of $(\varphi^{(n)},\psi^{(n)})$ is that for $M_n=
\lfloor {\rm e}^{nR}\rfloor$, the decoder $\psi^{(n)}$ always outputs 
a constant message $m_0 \in {\cal M}_n$.
In this case we have 
\begin{align*}
&\lim_{n\to \infty }\left(-\frac{1}{n}\right)
\log {\rm P}_{\rm c}^{(n)}(\varphi^{(n)},\psi^{(n)}{}|W)
\\
&=\lim_{n\to \infty }\left(-\frac{1}{n}\right)\log {M_n}=R.
\end{align*}
Hence  we have $(\ref{eqn:Zxc})$ in the part c).
We finally prove the part d). 
Let $ \lfloor a  \rfloor $ be an integer part of $a$.   
Fix any $\alpha\in [0,1]$. Let $\bar{\alpha}=1-\alpha$. 
We choose $(n,m)$ so that 
$$
n=k_\alpha\defeq \lfloor k\alpha \rfloor,\: 
m=k_{\bar{\alpha}} \defeq  \lfloor k \bar{\alpha}\rfloor.
$$ 
For this choice of $n$ and $m$, we have
\beq
\left.
\ba{l}
\ds \left(1-\frac{1}{k}\right)\alpha \leq 
\frac{n}{n+m}\leq \frac{k}{k-1}\alpha
\vspace*{1mm}\\
\ds \left(1-\frac{1}{k}\right)\bar{\alpha} \leq 
\frac{m}{n+m}\leq \frac{k}{k-1}\bar{\alpha}\\
\ea
\right\}
\label{eqn:Sddaa} 
\eeq
Fix small positive $\tau$ arbitrary. Then, for any 
$$
k> \max\{ (\alpha R+\bar{\alpha}R^{\prime})/\tau,
          (\alpha \Gamma+\bar{\alpha}\Gamma^{\prime})/\tau\},
$$ 
we have the following chain of inequalities:
\begin{align}
 & G^{(k_\alpha+k_{\bar{\alpha}})}\left(\left.
\alpha R+\bar{\alpha}R^{\prime}-\tau, 
\alpha \Gamma +\bar{\alpha}\Gamma^{\prime}-\tau 
\right|W\right) 
\nonumber\\
&\MLeq{a}
G^{(k_\alpha+k_{\bar{\alpha}})}\left(
\left(1-\frac{1}{k}\right)
\left(\alpha R+\bar{\alpha}R^{\prime}\right), \right.
\nonumber\\
& \qquad \qquad \:\:\:\:\:\quad
\left. \left. 
\left(1-\frac{1}{k}\right)
\left(\alpha \Gamma+\bar{\alpha}\Gamma^{\prime}\right)
\right| W\right) 
\nonumber\\
&\MLeq{b}G^{(n+m)}\left(\left.
\frac{n R + m R^{\prime}}{n+m},
\frac{n \Gamma + m \Gamma^{\prime}}{n+m}
\right|W\right) 
\nonumber\\
&\MLeq{c} \frac{nG^{(n)}(R,\Gamma|W)
  +mG^{(m)}(R^{\prime},\Gamma^{\prime}|W)}{n+m}
\nonumber\\
&\MLeq{d}\left(\frac{k}{k-1}\right)\left[
\alpha G^{(k_\alpha)}(R,\Gamma|W)\right.
\notag\\
& \qquad\qquad \qquad\left. +\bar{\alpha}G^{(k_{\bar{\alpha}})}(R^{\prime},\Gamma^{\prime}|W)\right].
\quad \:\label{eqn:Sddzzz}
\end{align}
Step (a) follows from the part a) and 
$$
k> \max\{ (\alpha R+\bar{\alpha}R^{\prime})/\tau,
          (\alpha \Gamma+\bar{\alpha}\Gamma^{\prime})/\tau\}.
$$ 
Step (b) follows from the part a).
Step (c) follows from (\ref{eqn:Sdd}).
Step (d) follows from (\ref{eqn:Sddaa}).
Letting $k\to\infty$ in (\ref{eqn:Sddzzz}), we have
\begin{align}
 & G^{*}\left( \alpha R+\bar{\alpha} R^{\prime}-\tau, 
\alpha \Gamma+\bar{\alpha} \Gamma^{\prime}-\tau|W \right)
\nonumber\\
&\leq \alpha G^*(R,\Gamma|W) + \bar{\alpha} G^*(R^{\prime},\Gamma^{\prime}|W),
\label{eqn:adf}
\end{align}
where $\tau$ can be taken arbitrary small. 
We choose $R^{\prime}$, $\Gamma^{\prime}$, 
and $\alpha$, as 
\beq
\left.
\ba{l}
R^{\prime}=R+2\sqrt{\tau},\quad \Gamma^{\prime}=\Gamma +2\sqrt{\tau},
\\
\alpha =1-\sqrt{\tau}.
\ea
\right\}
\label{eqn:SddZ}
\eeq
For the above choice of 
$R^{\prime}$, $\Gamma^{\prime}$, and $\alpha$, we have 
\beq
\alpha R+\bar{\alpha} R^{\prime}=R+2\tau, \quad
\alpha \Gamma +\bar{\alpha} \Gamma^{\prime}=\Gamma +2\tau.
\label{eqn:SddZz}
\eeq
Then we have the following chain of inequalities:
\begin{align}
 &   G^{*}\left(R+\tau, \Gamma+\tau|W \right)
\nonumber\\
&\MEq{a} G^{*}\left(\alpha R+\bar{\alpha} R^{\prime}-\tau,
              \alpha \Gamma +\bar{\alpha} \Gamma^{\prime}-\tau
|W \right)
\nonumber\\
&\MLeq{b} \alpha G^*(R,\Gamma|W)
+\bar{\alpha} G^*(R^{\prime},\Gamma^{\prime}|W)
\nonumber\\
&\MLeq{c} \alpha G^*(R,\Gamma|W)+\bar{\alpha}R^{\prime}
\nonumber\\
&\MEq{d}(1-\sqrt{\tau})G^*(R,\Gamma|W)+\sqrt{\tau}
R+2\tau
\nonumber\\
&\leq G^*(R,\Gamma|W)+\sqrt{\tau}R+2\tau.
\label{eqn:adfZxxS}
\end{align}
Step (a) follows from (\ref{eqn:SddZz}). 
Step (b) follows from (\ref{eqn:adf}). 
Step (c) follows from (\ref{eqn:Zxc}). 
Step (d) follows from (\ref{eqn:SddZ}). 
For any positive $\tau$, we have the following chain of inequalities:
\begin{align}
&   G^{*}\left(\alpha R+\bar{\alpha} 
R^{\prime},\alpha \Gamma+\bar{\alpha} \Gamma^{\prime}|W \right)
\nonumber\\
&=G^{*}\left( \alpha R+\bar{\alpha} 
R^{\prime}-\tau+\tau,\alpha \Gamma+\bar{\alpha} \Gamma^{\prime}-\tau+\tau |W \right)
\nonumber\\
&\MLeq{a}  G^*(\alpha R+\bar{\alpha} R^{\prime}-\tau,\alpha \Gamma+\bar{\alpha}\Gamma^{\prime}-\tau |W)
\nonumber\\
&      \qquad +\sqrt{\tau}(\alpha R+\bar{\alpha} R^{\prime}-\tau)+2\tau
\nonumber\\
&\MLeq{b} \alpha G^*(R,\Gamma|W) + \bar{\alpha} G^*(R^{\prime},\Gamma^{\prime}|W)
\nonumber\\
&   \qquad +\sqrt{\tau}(\alpha R+\bar{\alpha} R^{\prime})
+\tau(2-\sqrt{\tau}).
\label{eqn:adfZxxxS}
\end{align}
Step (a) follows from (\ref{eqn:adfZxxS}). 
Step (b) follows from (\ref{eqn:adf}). 
Since $\tau>0$ can be taken arbitrary small in (\ref{eqn:adfZxxxS}), we have  
\begin{align*}
&  G^{*}\left( \alpha R+\bar{\alpha} R^{\prime},
\alpha \Gamma+\bar{\alpha} \Gamma^{\prime}|W \right)
\\
&\leq     \alpha G^*(R,\Gamma|W) 
   + \bar{\alpha}G^*(R^{\prime},\Gamma^{\prime}|W),
\end{align*}
which implies the convexity of $G^*(R,\Gamma|W)$ 
on $(R,\Gamma)$. 
\hfill\IEEEQED
}

Set $\Gamma_{\max}\defeq \max_{x\in {\cal X}}c(x)$.
When $\Gamma \geq \Gamma_{\max}$, we have no cost constraint. 
In this case ${G}^{*}(R,\Gamma|W)$ does not depend on $\Gamma$ and 
becomes equal to the optimal exponent function in the case without 
input cost. We write this exponent function as $G^{*}(R|W)$.  
\newcommand{\OmitSuggestedByAE}{
A formal definition of $G^{*}(R|W)$ is as follows.
\beqno
& & 
G^{(n)}(R|W)
\\
&&\defeq
\min_{\scs 
(\varphi^{(n)},\psi^{(n)}):
    \atop{\scs 
          \atop{\scs
               (1/n)\log M_n \geq R 
               }    
         }
    }
\hspace*{-2mm}
\left(-\frac{1}{n}\right)
\log {\rm P}_{\rm c}^{(n)}(\varphi^{(n)},\psi^{(n)}|W),
\\
& &
{G}^{*}(R|W) \defeq \lim_{n \to \infty} G^{(n)}(R|W).
\eeqno
}
\newcommand{\prmt}{\lambda}

Arimoto \cite{arimoto:73} derived a 
lower bound of $G^*(R|W)$, which 
we denote by $G_{\rm AR}(R|W)$. 
To describe \irrAEb{his result} 
we define some functions. For $\irOOh{\rho}\in [0,1)$ 
and $\mu\geq 0$, define 
\begin{align}
& J^{(\mu,\irOOh{\rho})}(q_X|W)
\notag\\
& 
\defeq \log 
\sum_{y \in {\cal Y} }
\left[\sum_{x \in {\cal X}} 
q_X(x)\left\{W(y|x){\rm e}^{-\mu \irOOh{\rho}c(x)}
\right\}
^{\frac{1}{\clrTypoOh{\overline{\irOOh{\rho}}}}} 
\right]^{\clrTypoOh{\overline{\irOOh{\rho}}}}.
\notag
\end{align}
\clrTypoOh{Here we set $\overline{\rho}\defeq 1-\rho$.}
When $\irOOh{\rho}=1$, we define
\begin{align*}
 J^{(\mu,1)}(q_X|W)\defeq& \lim_{\irOOh{\rho} \uparrow 1 }
 J^{(\mu,\irOOh{\rho})}(q_X|W)
\\
=&\log \sum_{y\in {\cal Y}}
       \max_{\scs x\in {\cal X}:
       \atop{\scs \irrComRevA{q_X(x)>0}}}
W(y|x){\rm e}^{-\irrTypo{\mu}c(x)}. 
\end{align*}
Furthermore, set
\begin{align*}
& {\empty}{G}_{\rm AR}(R\Vl W)
\defeq 
\max_{\irOOh{\rho}\in [0,1]}
\min_{q_X\in {\cal P}({\cal X})}
[\lambda R-{J}^{(0,\irOOh{\rho})}(q_X|W)].
\end{align*}
The function 
$J^{(\mu,\irOOh{\rho})}(q_X|W)$ for 
$\irOOh{\rho}\in [0,1]$ and $\mu \geq 0$ is related to 
$G^\ast(R,\Gamma|W)$. This function will 
appear in Sections 
\ref{sec:MainResult} and 
\ref{sec:ThreeExp}.
Arimoto \cite{arimoto:73} obtained the following result. 
\begin{Th}[Arimoto \cite{arimoto:73}] 
For any $R \geq 0$, we have
$$
G^*(R|W) \geq {G}_{\rm AR}(R \Vl W).
$$
\end{Th}

Subsequently, Dueck and K\"orner \cite{dueck:79} 
determined $G^*(R|W)$. To state their result, set
\begin{align*}
G_{\rm DK}(R|W) 
&
\defeq  
\min_{\scs q_{XY} \in \mathcal{P}(\mathcal{X}\times \mathcal{Y}):
}
\big\{ 
[R-I(q_X,q_{Y|X})]^+ \notag\\
&\qquad\qquad\qquad\qquad + D(q_{Y|X} || W | q_X)
\big\}, 
\end{align*}
where 
$\mathcal{P}
(\mathcal{X}\times \mathcal{Y} )$ is
the set of joint probability distributions
on $\mathcal{X}\times \mathcal{Y}$, 
$[t]^+ = \max\{0,t\}$, and 
\begin{align*}
I(q_X, q_{Y|X}) 
& =
{\rm E}_q \left[
\log \frac{q_{Y|X}(Y|X)}{q_{Y}(Y)}
\right], \\
D(q_{Y|X}||W| q_X)
& = 
\mathrm{E}_q \left[\log \frac{q_{Y|X}(Y|X)}{W(Y|X)}
\right].
\end{align*}
Dueck and K\"orner obtained the following result:
\begin{Th}[Dueck and K\"orner \cite{dueck:79}] 
\label{Th:DK0}
For any $R \geq 0$, 
\begin{align*}
      & G^*(R|W) = G_{\rm DK}(R|W).
\notag
\end{align*}
\end{Th}

The author \cite{oohama2015two} gave a \irrAE{rigorous}  proof of 
$G_{\irrAE{\rm AR}}$ $(R|W)=G_{\rm DK}(R|W)$. 
The simplest derivation of a lower bound of 
$G^*(R|W)$ was given 
by 
\clrTypoRevB{Hayashi and Nagaoka \cite{hayashi:03}}.
To state their result, set
\begin{align*}
& \irr{\Omega^{(\mu,\lambda)}(q_X,Q|W)}
\\
& \defeq \log\left[\sum_{(x,y)\in 
{\cal X} \times {\cal Y}}q_{X}(x)
W(y|x)\frac{W^\lambda(y|x)
{\rm e}^{-\mu\lambda c(x)}}{Q^\lambda(y)}
\right],
\\
& \tilde{\Omega}^{(\mu,\lambda)}(W)\defeq 
\min_{Q \in{\cal P}({\cal Y})}\max_{q_X \in{\cal P}({\cal X})}
\irr{\Omega^{(\mu,\lambda)}(q_X,Q|W)},
\\
& \tilde{G}_{\rmOH}(R|W)
\defeq \sup_{\lambda \geq 0} 
\frac{\lambda R - \tilde{\Omega}^{(0,\lambda)}(W)}
{1+\lambda}.
\end{align*}
\clrTypoRevB{Hayashi and 
Nagaoka \cite{hayashi:03}} obtained the following result. 
\begin{Th}
[
\clrTypoRevB{Hayashi 
and Nagaoka \cite{hayashi:03}}]
\label{Th:Nag01}

For any $R \geq 0$, we have 
\beq
G^*(R|W) \geq \tilde{G}(R \Vl W).
\label{eqn:Nag01lowerbound}
\eeq
\end{Th}


\section{Main Results}
\label{sec:MainResult}
\noindent

In this section we state our main result. Define
\begin{align*}
G_{\rm DK}(R,\Gamma|W) 
&
\defeq  
\min_{\scs q_{XY} \in \mathcal{ P}(\mathcal{X}\times \mathcal{Y}):
     \atop{ \scs 
         {\rm E}_{q_X}[c(X)]\leq \Gamma
     }
 }
\big\{ 
[R-I(q_X,q_{Y|X})]^+ \notag\\
&\qquad\qquad\qquad\qquad + D(q_{Y|X} || W | q_X)
\big\}. 
\end{align*}

Using the standard method developed by Csisz\'ar and K\"orner 
\cite{csi2011information},
we can prove the following theorem. 
\begin{Th} \label{Th:DK}
For any $R\geq 0$ and any $\Gamma \geq \Gamma_0$, 
we have 
\begin{align*}
 & G^*(R,\Gamma |W)\leq G_{\rm DK}(R,\Gamma|W).
\notag
\end{align*}
\end{Th}


Proof of this theorem is given in Appendix \ref{sub:ApdaDirectDK}. Let 
$\Gamma_{\rm max} \defeq \max_{x \in {\cal X} }c(x).$The case $\Gamma 
\geq \Gamma_{\rm max}$ corresponds to the case without cost. In this 
case Dueck and K\"orner \cite{dueck:79} 
show that 
$$
G^*(R,\Gamma |W)=G_{\rm DK}(R,\Gamma|W). 
$$
They derived the bound $G^*(R,\Gamma |W)\leq G_{\rm DK}(R,\Gamma|W)$ 
by using a combinatorial method based on the type of sequences. 
Our method to prove Theorem \ref{Th:DK} is different 
from their method since we do not use a particular 
structure of types. 

\newcommand{\ApdaDirectDK}{
\subsection{Proof of Theorem \ref{Th:DK}} 
\label{sub:ApdaDirectDK}
In this appendix we prove Theorem \ref{Th:DK}. We first describe 
some definitions necessary for the proof. For $x^n\in {\cal X}^n$, 
set
\begin{align*}
& p_{x^n}(x) \defeq \frac{|\{t:x_t=x \}|}{n}, x \in {\cal X},  
\end{align*}
The probability distribution 
$
p_{x^n} \defeq \{p_{x^n}(x)\}_{x \in{\cal X}}
$
on ${\cal X}$ is called the type of sequences in ${\cal X}^n$. 
Let ${\cal P}_n({\cal X})$ be a set of all 
types of sequences in ${\cal X}^n$.
Let ${\cal P}({\cal Y}|{\cal X})$ be a set of all 
conditional distributions $q_{Y|X}$ on ${\cal Y}$ for 
given $X\in {\cal X}$. We fix $\delta \in [0,1/2)$. 
We consider any pair 
$ (q_{X},q_{Y|X})\in {\cal P}_n({\cal X}) 
  \times {\cal P}({\cal Y}|{\cal X}) 
$
satisfying ${\rm E}_{q_{X}}c(X) \leq \Gamma$. 
For such pair of $(q_{X},q_{Y|X})$, we can construct an $n$-length 
block code $(\phi^{(n)}, \psi^{(n)})$ with message set ${\cal K}_n$ 
satisfying:
\begin{itemize}
\item[a)] ${\rm P}_{\rm c}^{(n)}(\phi^{(n)}, \psi^{(n)}|q_{Y|X})\geq 1-\delta$.
\item[b)] all codewords $\phi^{(n)}(k), k\in {\cal K}_n$ have 
the identical type $q_X$. 
\item[c)] $\frac1n \log |\mathcal{ K }_n| 
\geq \min \{ R, I(q_X, q_{Y|X}) -\delta \}$.
\end{itemize}
By the condition b), we have  
$
c(\phi^{(n)}(k))={\rm E}_{q_X}c(X)\leq \Gamma.   
$
Hence the $n$-length block code $(\phi^{(n)}, \psi^{(n)})$ 
satisfies the cost constraint. 
Furthermore, by this condition we can 
obtain the following result. 
\begin{lm}\label{lm:Aszz}
For every $k\in {\cal K}_n$, we have 
\begin{align}
 & \sum_{y^n \in \mathcal{Y}^n} 
q_{Y|X}^n(y^n|\phi^{(n)}(k)) \log 
\frac{q_{Y|X}^n(y^n|\phi^{(n)}(k))}{W^n(y^n|\phi^{(n)}(k))}
\nonumber\\
&=n D(q_{Y|X} || W | q_X) 
\label{eq.zz6}.
\end{align}
\end{lm}

{\it Proof:} For each $k \in {\cal K}_n$, we set 
$$
\phi^{(n)}(k)=x^n(k)=x_1(k)x_2(k)\cdots x_n(k). 
$$
For each $k\in {\cal K}_n$, we have 
the following chain of equalities: 
\begin{align}
& \sum_{y^n \in \mathcal{Y}^n} 
q_{Y|X}^n(y^n|\phi^{(n)}(k)) \log 
\frac{q_{Y|X}^n(y^n|\phi^{(n)}(k))}{W^n(y^n|\phi^{(n)}(k))}
\nonumber\\
&\MEq{a}
\sum_{t=1}^n \sum_{y_t \in {\cal Y} } 
q_{Y|X}(y_t|x_t(k))\log \frac{q_{Y|X}(y_t|x_t(k))}{W(y_t|x_t(k))}
\nonumber\\
&=\sum_{ a \in {\cal X} } \sum_{y \in {\cal Y} } 
{|\{ t :x_t(k)=a \}|}
q_{Y|X}(y|a)\log \frac{ q_{Y|X}(y|a)} {W(y|a)}
\nonumber\\
&=n\sum_{ a \in {\cal X} } \sum_{y \in {\cal Y} } 
p_{x^n(k)}(a)
q_{Y|X}(y|a)\log \frac{ q_{Y|X}(y|a)} {W(y|a)}
\nonumber\\
&\MEq{b}n\sum_{a \in{\cal X} } \sum_{y \in {\cal Y} } 
 q_X(a) q_{Y|X}(y|a)\log \frac{ q_{Y|X}(y|a)}{W(y|a)}
\nonumber\\
&=n D(q_{Y|X} || W | q_X). 
\nonumber
\end{align}
Step (a) follows from the memoryless property of the noisy channel.
Step (b) follows from that $p_{x^n(k)}=q_X\in {\cal P}_n({\cal X})$.
\hfill\IEEEQED

For $k\in {\cal K}_n$, we set

\begin{align*}
 \alpha_n(k)&\defeq W^n( {\mathcal{D}(k)}|\phi^{(n)}(k))=
\sum_{y^n \in \mathcal{D}(k)} W^n(y^n|\phi^{(n)}(k)),
\\
\beta_n(k)&\defeq q_{Y|X}^n( {\mathcal{D}(k)}|\phi^{(n)}(k))=
\sum_{y^n \in \mathcal{D}(k)} q_{Y|X}^n(y^n|\phi^{(n)}(k)),
\\
\overline{\alpha_n(k)}&\defeq
1-{\alpha_n(k)}=q_{Y|X}^n(\overline{\mathcal{D}(k)}|\phi^{(n)}(k)),
\\
\overline{\beta_n(k)}&\defeq 
1-{\beta_n(k)}=q_{Y|X}^n(\overline{\mathcal{D}(k)}|\phi^{(n)}(k)).
\end{align*}
Furthermore, set 
\begin{align*}
&{\alpha_n}\defeq\sum_{k\in {\cal K}_n}
 \frac{1}{|{\cal K}_n|} \alpha_n(k)
={\rm P}_{\rm c}^{(n)}(\phi^{(n)}, \psi^{(n)}|W),
\\
&{\beta_n}\defeq\sum_{k\in {\cal K}_n}
 \frac{1}{|{\cal K}_n|} \beta_n(k)
={\rm P}_{\rm c}^{(n)}(\phi^{(n)}, \psi^{(n)}|q_{Y|X}).
\end{align*}
The quantity 
${\rm P}_{\rm c}^{(n)}(\phi^{(n)}, \psi^{(n)}|W)$
has a lower bound given by the following Lemma. 
\begin{lm}\label{lm:DirectLm} For any $\delta\in [0,1/2)$, we have
\begin{align}
& {\rm P}_{\rm c} ^{(n)}( \phi^{(n)}, \psi^{(n)}|W)
=\frac1{|{\mathcal{K}}_n|}
\sum_{k \in \mathcal{K}_n} W^n(\mathcal{D}(k)|\phi^{(n)}(k)) 
\notag\\
&\geq \exp \{-n [(1-\delta)^{-1}D(q_{Y|X}||W|q_X) + \eta_n(\delta)]\}.
\label{eqn:Saab} 
\end{align}
Here we set
$
\eta_n(\delta) \defeq \frac 1 n (1-\delta)^{-1}h(1 -\delta) 
$
and $h(\cdot)$ stands for a binary entropy function.
\end{lm}

{\it Proof:} 
We have the following chain of inequalities:
\begin{align}
& nD(q_{Y|X} || W | q_X) 
\MEq{a}\frac{1} {|\mathcal{K}_n|} \sum_{k\in \mathcal{K}_n} 
\sum_{y^n \in \mathcal{Y}^n} 
q_{Y|X}^n(y^n|\phi^{(n)}(k)) 
\nonumber\\
&\qquad \times \log 
\frac{q_{Y|X}^n(y^n|\phi^{(n)}(k))}{W^n(y^n|\phi^{(n)}(k))}
\nonumber\\
&\MGeq{b} 
\frac{1} {|\mathcal{K}_n|} \sum_{k\in \mathcal{K}_n} \left[
\beta_n (k) \log \frac{\beta_n (k)}{\alpha_n(k)}
+\overline{\beta_n (k)} \log \frac{\overline{\beta_n (k)}}
{\overline{\alpha_n(k)}}\right]
\nonumber\\
&=  \sum_{k\in \mathcal{K}_n} 
\left[\frac{\beta_n (k)}{|\mathcal{K}_n|}
\log \frac
{\frac{\beta_n (k)}{|\mathcal{K}_n|}}
{\frac{\alpha_n(k)}{|\mathcal{K}_n|}}
+\frac{\overline{\beta_n (k)}}{|\mathcal{K}_n|} 
\log \frac
{\frac{\overline{\beta_n (k)}}{|\mathcal{K}_n|}}
{\frac{\overline{\alpha_n(k)}}{|\mathcal{K}_n|}}
\right]
\nonumber\\
&\MGeq{c}
\beta_n \log \frac{\beta_n}{\alpha_n}
+\overline{\beta_n} 
\log \frac{\overline{\beta_n}}{\overline{\alpha_n}}
\geq -h(\beta_n) -\beta_n\log \alpha_n
\notag \\
&\MGeq{d} -h(1-\delta)-(1-\delta)\log \alpha_n. 
\label{eqn:Zsddv}  
\end{align}
Step (a) follows from Lemma \ref{lm:Aszz}. Steps (b) and (c) 
follow from the log-sum inequality. Step (d) follows from that 
$
\beta_n={\rm P}_{\rm c} ^{(n)}(\phi^{(n)}, \psi^{(n)}|q_{Y|X})\geq 1-\delta
$
and $\delta \in  (0,1/2].$  From (\ref{eqn:Zsddv}), we obtain   
\begin{align}
& \alpha_n= {\rm P}_{\rm c} ^{(n)}( \phi^{(n)}, \psi^{(n)}|W)
\notag\\
&\geq \exp\left(
- \frac{nD(q_{Y|X}|| W| q_X) + h(1-\delta)}
{1-\delta} \right)
\notag\\
&=\exp \{-n [(1-\delta)^{-1}D(q_{Y|X}||W|q_X) + \eta_n(\delta)]\},
\notag
\end{align}
completing the proof. \hfill\IEEEQED 

{\it Proof of Theorem \ref{Th:DK}:} 
We first consider the case where 
$R \leq I(q_X, q_{Y|X}) - \delta$. 
In this case we choose $\varphi^{(n)}=\phi^{(n)}$. Then we have 
\begin{align}
&  {\rm P}_{\rm c} ^{(n)}( \varphi^{(n)}, \psi^{(n)}|W)
  ={\rm P}_{\rm c} ^{(n)}( \phi^{(n)}, \psi^{(n)}|W)
\notag\\
&\MEq{a}\exp \{ -n[R+\delta-I(q_X,q_{Y|X})]^+
\notag\\
&\quad -n [(1-\delta)^{-1}D(q_{Y|X} || W | q_X) + \eta_n(\delta)]\}
\notag\\
&\MGeq{b}\exp \{-n[R-I(q_X,q_{Y|X})]^+
\notag\\
&\quad -n [(1-\delta)^{-1}D(q_{Y|X} || W | q_X) + \delta+ \eta_n(\delta)]\}.
\label{eqn:SaabZ}
\end{align}
Step (a) follows from the condition 
$R+\delta-I(q_X,q_{Y|X})\leq 0$. 
Step (b) follows from that
$$
[R+\delta-I(q_X,q_{Y|X})]^+ \leq [R-I(q_X,q_{Y|X})]^+ +\delta.
$$
We next consider the case where $R>I(q_X,q_{Y|X})-\delta$. 
Consider the new message set $\widehat{ \mathcal{K}}_n$ satisfying 
$|\widehat{\mathcal{K}}_n |={\rm e}^{\lfloor nR \rfloor}$. 
For new message set $\widehat{\cal K}_n$, we define 
$\varphi^{(n)}(k)$ such that 
$\varphi^{(n)}(k) = \phi^{(n)}(k)$ if $k\in \mathcal{K}_n$.
For $k \in \widehat{\mathcal{K}}_n-\mathcal{K}_n$, 
we define $\varphi^{(n)}(k)$
arbitrary sequence of ${\cal X}^n$ having the type $q_X$. 
We use the same decoder $\psi^{(n)}$ 
as that of the message set $\mathcal{K}_n$.
Then we have the following: 
\begin{align}
&{\rm P}_{\rm c} ^{(n)}(\varphi^{(n)}, \psi^{(n)}|W)
=\frac{1}{ | \widehat{\mathcal{K} }_n |}
\left[\sum_{k\in \mathcal{K}_n } 
W^n(\mathcal{D}(k)|\varphi^{(n)}(k)) 
\right. 
\notag\\
&\qquad \left. + 
\sum_{k\in \widehat{\mathcal{K}}_n - \mathcal{K}_n} 
W^n(\mathcal{D}(k)|\varphi^{(n)}(k))\right] 
\notag\\
& \geq 
\frac 1 {|\widehat{\mathcal{K}}_n|}
\sum_{k\in \mathcal{K}_n} W^n(\mathcal{D}(k)|\varphi^{(n)}(k)) 
\notag\\
& \MGeq{a}
\frac {|\mathcal{K}_n|}{{\rm e}^{nR}} 
\exp\{ -n [(1-\delta)^{-1}D(q_{Y|X}||W|q_{X}) +\eta_n({\delta})]\}
\notag\\
&\MGeq{b} \exp \left[
-n \left\{R-(I(q_X, q_{Y|X})- \delta)
\right.\right. 
\notag\\
&\left.\left.\quad\qquad+ (1-\delta)^{-1}D(q_{Y|X}||W |q_X) 
+ \eta_n(\delta)\right\} \right]
\notag\\
&\MGeq{c} \exp \left[-n \left\{
[R-I(q_X, q_{Y|X})]^+ \right.\right.
\notag\\
&\left.\left.\quad \qquad +(1-\delta)^{-1}D(q_{Y|X}||W|q_X)
+\delta+ \eta_n(\delta)\right\}\right].
\label{eqn:eq29}
\end{align}
Step (a) follows from (\ref{eqn:Saab}) in Lemma \ref{lm:DirectLm}. 
Step (b) follows from 
$|\mathcal{K}_n| \geq {\rm e}^{n [(I(q_X, q_{Y|X})-\delta]}.$
Step (c) follows from $[a]\leq [a]^+$. 
Combining (\ref{eqn:SaabZ}) and (\ref{eqn:eq29}), we have
\begin{align}
 & {\rm P}_{\rm c} ^{(n)}( \varphi^{(n)}, \psi^{(n)}|W)
\nonumber\\
&\geq  \exp \left[-n \left\{
[R-I(q_X, q_{Y|X})]^+ \right.\right.
\nonumber\\
& \left.\left.\quad +(1-\delta)^{-1}D(q_{Y|X}||W|q_X)
+\delta+ \eta_n(\delta)\right\}\right]
\label{eq.30} 
\end{align}
for any $q_X \in {\cal P}_n({\cal X})$ with 
${\rm E}_{q_X}c(X) \leq \Gamma$ and 
$q_{Y|X} \in {\cal P}($ ${\cal Y}|{\cal X})$.
Hence from (\ref{eq.30}), we have
\begin{align}
&- \frac1 n \log {\rm P}_{\rm c} ^{(n)}( \varphi^{(n)}, \psi^{(n)}|W)
\notag\\
& \leq 
\min_{\scs q_X \in \mathcal{P}_n({\cal X}), 
  \atop{\scs 
        {\rm E}_{q_X}c(X) \leq \Gamma, 
        \atop{\scs 
         q_{Y|X} \in \mathcal{P}({\cal Y}|{\cal X})
        }
  }
}
\{ [R -I(q_X, q_{Y|X}) ]^+ 
\notag\\
&\quad + (1-\delta)^{-1}D(q_{Y|X} || W | q_X) + \delta + \eta_n(\delta) \} 
\notag\\
& \leq  
(1-\delta)^{-1}\min_{\scs q_X \in \mathcal{P}_n({\cal X}), 
  \atop{\scs 
        {\rm E}_{q_X}c(X) \leq \Gamma, 
        \atop{\scs 
        q_{Y|X} \in {\cal P}({\cal Y}|{\cal X}) 
        }
  }
}
\{[R -I(q_X, q_{Y|X}) ]^+ 
\notag\\
&\quad + D(q_{Y|X} || W | q_X)\}+\delta +\eta_n(\delta)
\notag\\
& \leq (1-\delta)^{-1}G_{\rm DK}(R,\Gamma|W)+\delta 
 + \eta_n(\delta) + \varepsilon_n.
\label{eqn:Bound}
\end{align}
The quantity $\{\varepsilon_n\}_{n \geq 1}$ appearing in the last 
inequality is an error bound coming from an approximation of 
the marginal distribution $q_X^*$ of $q^*$ 
achieving $G_{\rm DK}(R,\Gamma|W)$ 
by some suitable type $q_X\in {\cal P}_n({\cal X }).$ 
Since $q_{X}\in {\cal P}_n(\cal X)$ can be made arbitrary 
close to $q^*_X$ by letting $n$ sufficiently large, 
we can choose $\varepsilon_n $ so that
$\varepsilon_n \to 0$ as $n \to \infty$. We further note that
$\eta_n(\delta)\to 0$ as $n\to\infty$. Hence by letting 
$n\to \infty$ in (\ref{eqn:Bound}), we obtain 
$$
G^*(R,\Gamma|W) \leq (1-\delta)^{-1}G_{\rm DK}(R,\Gamma|W)+\delta.
$$
Since $\delta$ can be made arbitrary small, 
we conclude that
$G^*(R,$ $\Gamma|W) \leq G_{\rm DK}(R,\Gamma|W).$
\hfill \IEEEQED

\newcommand{\DelGauDK}{

We fix $\delta \in [0,1/2)$. We consider a Gaussian channel 
with input $X$ and output $Y$, having the form 
$$
Y=\alpha X+ S, X\perp S, 
S\sim {\cal N}(0,\xi).
$$
We consider a Gaussian random pair  $(X,Y)$ obtained by letting $X$ be 
Gaussian random variable with $X\sim {\cal N}(0,\theta)$. We assume that 
$\theta \leq \Gamma$. Let $q_{XY}$ be a probability density function of 
$(X,Y)$. For the Gaussian channel specified by $q_{Y|X}$, we can 
construct an $n$-length block code $(\phi^{(n)}, \psi^{(n)})$ with 
message set ${\cal K}_n$ satisfying:
\begin{itemize}
\item[a)] ${\rm P}_{\rm c}^{(n)}(\phi^{(n)}, \psi^{(n)}|
q_{Y|X})\geq 1-\delta$.
\item[b)] all codewords $\phi^{(n)}(k), k\in {\cal K}_n$ satisfy 
$||\phi^{(n)}(k)||^2 \leq n \theta$. 
\item[c)] $\frac1n \log |\mathcal{ K }_n| 
\geq \min \{ R, I(q_X,q_{Y|X}) -\delta \}$.
\end{itemize}
By the condition b), we can obtain the following result. 
\begin{lm}\label{lm:Aszzz}
For every $k\in {\cal K}_n$, we have 
\begin{align}
& \underbrace{\int \cdots \int}_{n}{\rm d}y^n
q_{Y|X}^n(y^n|\phi^{(n)}(k)) \log 
\frac{q_{Y|X}^n(y^n|\phi^{(n)}(k))}{W^n(y^n|\phi^{(n)}(k))}
\notag\\
&\leq n D(q_{Y|X}||W|q_X) 
\label{eq.zzz6}.
\end{align}
\end{lm}

{\it Proof:} By a direct computation we have 
\begin{align}
& \underbrace{\int \cdots \int}_{n} {\rm d}y^n
q_{Y|X}^n(y^n|\phi^{(n)}(k)) \log 
\frac{q_{Y|X}^n(y^n|\phi^{(n)}(k))}{W^n(y^n|\phi^{(n)}(k))}
\notag\\
&=
\frac{1}{2}(1-\alpha)^2
\frac{||\phi^{(n)}(k))||^2}{{\NoizeVar}}
+\frac{n}{2}\left[\frac{\xi}{{\NoizeVar}}-1+\log \frac{{\NoizeVar}}{\xi}\right]
\notag\\
&\MLeq{a}
\frac{n}{2}(1-\alpha)^2 \frac{\theta}{{\NoizeVar}}
+\frac{n}{2}\left[\frac{\xi}{{\NoizeVar}}-1+\log \frac{{\NoizeVar}}{\xi}\right]
\notag\\
&=nD(q_{Y|X} || W | q_X). 
\end{align}
Step (a) follows from the condition b).
\hfill\IEEEQED
}
%
}

\irPkAu{As we mentioned in the introduction}, we propose 
two different methods to derive tight lower bounds 
of $G^*(R,\Gamma |W)$. Those methods are based on the 
information spectrum method and are quite different from 
that of Dueck and K\"orner \cite{dueck:79}. 
\irPkAu{One is the NH method, which is a minor 
extension of the method of 
\clrTypoRevB{Hayashi and Nagaoka \cite{hayashi:03}} to 
the case with input cost.} 
The other \irPkAu{is called the improved recursive 
method since it is an improvement of the recursive 
method the author introduced in his previous works 
\cite{Oohama15a},
\cite{Oohama15b},
\cite{OhIsita16abc},
\cite{oohama:18}, and 
\cite{oohama:19}.
This method} includes a new recursive algorithm 
in a single letterization of exponent functions 
serving as lower bounds of $G^*(R,\Gamma|W)$.

The lower bound obtained by \irPkAu{ the NH method} is a form which 
includes that of the exponent function obtained by 
\clrTypoRevB{\cite{hayashi:03}} in the case without input cost as a special case. 
The lower bound obtained by the \irPkAu{ improved recursive method} 
has a new form different from previous exponent functions. To 
describe those lower bounds we define several quantities. We first 
define the following function: 
\begin{align*}
\irOOh{{\Omega}^{(\mu,\lambda)}(q_X|W)
\defeq}&
\irOOh{\min_{Q \in{\cal P}({\cal Y})}
\Omega^{(\mu,\lambda)}(q_X,Q|W),
}\\
{\Omega}^{(\mu,\lambda)}(W)\irOOh{\defeq}
&\irOOh{ \max_{q_X \in{\cal P}({\cal X})} 
  {\Omega}^{(\mu,\lambda)}(q_X|W)}
\\
\irOOh{=}
&\max_{q_X \in{\cal P}({\cal X})}
  \min_{Q \in{\cal P}({\cal Y})}
{\Omega^{(\mu,\lambda)}(q_X,Q|W)}.
\end{align*}
In general we have 
\beq
\tilde{\Omega}^{(\mu,\lambda)}(W)\geq {\Omega}^{(\mu,\lambda)}(W).
\label{eqn:AsssD}
\eeq
Since 
$\irOlgRevB{\exp\{ \Omega^{(\mu,\lambda)}(q_X,Q|W)\}}$ 
is a linear function of $q_X$ and is convex with 
respect to $Q$, the \irPkAu{minimax} \eXchange theorem 
stated in \cite{csiszar:95} holds between 
$\tilde{\Omega}^{(\mu,\lambda)}(W)$ and 
${\Omega}^{(\mu,\lambda)}(W)$. Then, we have 
the equality in (\ref{eqn:AsssD}). 
We set
\begin{align*}
& \tilde{G}_{\rmOH}^{(\mu,\lambda)}(R,\Gamma|W)
\defeq \frac{\lambda (R-\mu\Gamma)
-\tilde{\Omega}^{(\mu, \lambda)}(W)}
{1+\lambda},
\\
& G_{\rmOH}^{(\mu,\lambda)}(R,\Gamma|W)
\defeq \frac{\lambda (R-\mu\Gamma) -\Omega^{(\mu, \lambda)}(W)}
{1+\lambda},
\\
& \tilde{G}_{\rmOH}(R,\Gamma|W) \defeq 
\sup_{\mu,\lambda \geq 0} 
  \tilde{G}_{\rmOH}^{(\mu,\lambda)}(R,\Gamma|W),
\\
& G_{\rmOH}(R,\Gamma|W) \defeq 
\sup_{\mu,\lambda \geq 0} 
  G_{\rmOH}^{(\mu,\lambda)}(R,\Gamma |W).
\end{align*}
Since we have the equality in (\ref{eqn:AsssD}), we have for 
any $\mu,\lambda\geq 0$, 
$
{G}_{\rmOH}^{(\mu,\lambda)}(R,\Gamma|W)
=\tilde{G}_{\rmOH}^{(\mu,\lambda)}(R,\Gamma|W)\irb{,} 
$
from which we obtain 
\beq 
{G}_{\rmOH}(R,\Gamma|W)=\tilde{G}_{\rmOH}(R,\Gamma|W).
\label{eqn:AsZxi}
\eeq
{
We can show that ${G}_{\rmOH}(R,\Gamma|W)$ and 
$\tilde{G}_{\rmOH}(R,\Gamma|W)$ satisfy the following property.
\begin{pr}
\label{pr:prOne}
When $R\geq C(\Gamma|W)+\tau$ for some $\tau>0$, 
there exists positive $\lambda_0$ such that
\begin{align}
{G}_{\rmOH}(R,\Gamma|W)=\tilde{G}_{\rmOH}(R,\Gamma|W)
\geq \frac{ \irPkAubc{\lambda_0} \tau}{2(1+\lambda_0)}.
\label{eqn:AsZza}
\end{align}
Furthermore, we have that for $R>C(\Gamma|W)$,
\begin{align}
& G_{\rmOH}(R,\Gamma|W)
= \sup_{
\scs \mu \geq 0,
\atop{\scs \lambda>0}}
\frac{\lambda (R-\mu\Gamma)-\Omega^{(\mu, \lambda)}(W)}
{1+\lambda}.
\label{eqn:AsZz}
\end{align}
\end{pr}

Since $\Omega^{(\mu,0)}(W)=0$ for any $\mu \geq 0$, the 
equality (\ref{eqn:AsZz}) is obvious. Proof of (\ref{eqn:AsZza}) 
is given in \ref{sub:ApdOne}.
}
Our main results are the followings: 
\begin{Th}\label{Th:main0} 
For any $R\geq 0$ and any $\Gamma\geq \Gamma_0$, 
\beqa
G^*(R,\Gamma|W) &\geq& \tilde{G}_{\rmOH}(R,\Gamma|W). 
\label{eqn:mainIeq0}
\eeqa
\end{Th}
\begin{Th}\label{Th:main} 
For any $R\geq 0$ and any $\Gamma\geq \Gamma_0$, 
\beqa
G^*(R,\Gamma|W) &\geq& G_{\rmOH}(R,\Gamma|W). 
\label{eqn:mainIeq}
\eeqa
\end{Th}

It follows from Property \ref{pr:prOne} and Theorems \ref{Th:main0} 
and \ref{Th:main} that $G(R,\Gamma|W)=\tilde{G}(R,\Gamma|W)$ 
serves as the strong converse exponent function for the DMC 
with input cost.

Proofs of Theorems \ref{Th:main0} and \ref{Th:main} will be 
given in Section \ref{sec:Secaa}. In this section we outline 
the proofs of Theorems \ref{Th:main0} and \ref{Th:main}. For the 
derivation of the lower bound (\ref{eqn:Nag01lowerbound}) in 
Theorem \ref{Th:Nag01}, 
\clrTypoRevB{Hayashi 
and Nagaoka \cite{hayashi:03}} used a method of converting 
the problem of the channel coding into the simple statistical 
\irrTypoRevA{hypothesis} testing. 
{
Polyanskiy et al. \cite{PolyanskiyPoorVerdu10} extended 
this \irrTypoRevA{conversion} to the case where 
we have the constraint 
on the input cost. This result is stated as Theorem 27 
in \cite{PolyanskiyPoorVerdu10}.} 
In this paper we derive a useful lemma, 
which can be regarded as an extension of the lemma obtained 
by 
\clrTypoRevB{\cite{hayashi:03}} to the case 
with input cost. This lemma corresponds 
to Lemma \ref{lm:Ohzzz}, in which we upper bound 
${\rm P}_{\rm c}^{(n)}$ by some information spectrum 
quantities. Using Lemma \ref{lm:Ohzzz} and the large 
deviation 
theory, we derive a lower bound of $G^{*}(R,\Gamma|W)$. 
This result is stated in Corollary \ref{co:corOne}. 
The exponent function stated in this corollary is not a 
single letterized form yet. To derive the single letterized 
lower bound (\ref{eqn:mainIeq0}) in Theorem \ref{Th:main0}, we use 
the same method as that 
of 
\clrTypoRevB{\cite{hayashi:03}}. 
On the other hand, to derive the single letterized 
lower bound (\ref{eqn:mainIeq}) in Theorem \ref{Th:main}, 
we use the improved recursive method, which is regarded as 
the fourth method to \irrTypo{derive} a tight lower 
bound of $G^{*}(R,\Gamma|W)$. 

Here we explain that for an example of $W$, 
the {\prev} recursive method can not yield the lower 
bound (\ref{eqn:mainIeq}) in Theorem \ref{Th:main}. 
Let $Y$ be an output random variable from $W$ when $X$ with 
the probability distribution $q_X$ is connected 
to the input of $W$. Then the output distribution $q_Y$ 
is given by
$$
q_{Y}(y)=\sum_{ x \in {\cal X}}
q_X(x)W(y|x), \mbox{ for }y\in {\cal Y}. 
$$ 
We write this output distribution as $q_Y=q_XW$. We set 
\begin{align}
\hat{\Omega}^{(\mu,\lambda)}(q_X|W)&
\defeq \Omega^{(\mu,\lambda)}(q_X,q_X W|W),
\notag\\
\hat{\Omega}^{(\mu,\lambda)}(W)&
\defeq
\max_{q_X \in {\cal P}({\cal X})}
\hat{\Omega}^{(\mu,\lambda)}(q_X|W),
\notag\\
\hat{G}(R,\Gamma|W)&
\irOOh{\defeq \sup_{\scs \mu\geq 0, \atop{\scs \lambda \geq 0}}}
\frac{\ds \lambda(R-\mu \Gamma)-\hat{\Omega}^{(\mu,\lambda)}(W)}
{1+\lambda}.
\label{eqn:hatGBda}
\end{align}
We can show that 
$\clrTypoRevB{\hat{G}_{\rmOH}}(R,\Gamma|W)$ 
satisfies the
following property.
\begin{pr}\label{pr:prTwo}$\quad$
\begin{itemize}
\item[a)]
When $R\geq C(\Gamma|W)+\tau$ for some $\tau>0$,  
there exists positive $\lambda_0$ such that
\begin{align}
\hat{G}_{\rmOH}(R,\Gamma|W)
\geq \frac{\irPkAubc{\lambda_0}\tau}{2(1+\lambda_0)}.  
\label{eqn:AsZzabb}
\end{align}
Furthermore, we have that for $R>C(\Gamma|W)$,
\begin{align}
& \hat{G}_{\rmOH}(R,\Gamma|W)
= \sup_{
\scs \mu \geq 0,
\atop{\scs \lambda>0}}
\frac{\lambda (R-\mu\Gamma)-\hat{\Omega}^{(\mu, \lambda)}(W)}
{1+\lambda}.
\label{eqn:AsZzbb}
\end{align}
\item[b)] By definition we have that for any 
$\mu, \lambda \geq 0$,
\begin{align*}
&\hat{\Omega}^{(\mu,\lambda)}(W)
= \max_{q_X \in {\cal P}({\cal X})}
\Omega^{(\mu,\lambda)}(q_X,q_X W|W)
\\
&\geq \max_{q_X \in {\cal P}({\cal X})}
\min_{Q \in {\cal P}({\cal Y})}
\Omega^{(\mu,\lambda)}(q_X,Q|W)
{=\Omega^{(\mu,\lambda)}(W)},
\end{align*} 
from which we have
$\hat{G}(R,\Gamma|W)\leq {G}(R,\Gamma|W)$ 
\irrComRevA{for $R\geq 0$}. 
\item[c)] We consider the case where 
${\cal X}=\{1,2,\cdots, |{\cal X}|\}$, 
${\cal Y}=\{0\} \cup {\cal X}$. 
We fix $\theta \in (0,1)$. 
Let $W$ be a symmetric erasure channel given by 
\beq
\ba{l}
W(x|x)=\bar{\theta},\:W(0|x)=\theta,
\mbox{ for } x \in {\cal X}.
\ea
\label{eqn:Erasure}
\eeq
For $W$ given by (\ref{eqn:Erasure}), %
we have that 
\irPkOh{$\hat{G}(R,\Gamma|W)$ is strictly 
smaller than ${G}_{\rmOH}(R,\Gamma|W)$} 
for $\irBr{\Gamma > \Gamma_0}$ and $R>C(\Gamma|W)$. 
\end{itemize}
\end{pr}

Since $\Omega^{(\mu, 0)}(W)=0$ for any $\mu \geq 0$, 
the equality (\ref{eqn:AsZzbb}) is obvious. Proof of (\ref{eqn:AsZzabb}) is given in  Appendix \ref{sub:ApdOne}.  
\irPkOh{Proof of Property \ref{pr:prTwo} part c)} 
is given in \irrTypoRevA{Section} \ref{sec:ThreeExp}. 

On the quantity $\hat{G}(R,\Gamma|W)$, we have the following 
proposition. 
\begin{pro}
\label{pro:oldrecursive}
The recursive method yields that 
for any $R\geq 0$ and any $\Gamma\geq \Gamma_0$,
\beq
G^{*}(R,\Gamma|W) \geq \hat{G}(R,\Gamma|W).
\label{eqn:AzDD}
\eeq
\end{pro} 

It follows from Proposition \ref{pro:oldrecursive} and Property 
\ref{pr:prTwo} that the lower bound $\hat{G}(R,\Gamma|W)$ of 
${G}^{*}(R,\Gamma|W)$ obtained by the {\prev} recursive method 
serves as an exponent function of the strong converse theorem 
for DMC with input cost. Furthermore, it follows from 
Theorem \ref{Th:main}, Proposition \ref{pro:oldrecursive}, 
and Property \ref{pr:prTwo} that \irPkAub{when $W$ is given 
by (\ref{eqn:Erasure}),} $\hat{G}(R,\Gamma|W)$ can not match 
the optimal exponent function $G^{*}(R,\Gamma|W)$ 
for $R>C(\Gamma|W)$. 
In the next section we explain the {\prev} recursive method and 
give the proof of Proposition \ref{pro:oldrecursive}.
\newcommand{\ApdONE}{
\subsection{{Conditions for the Exponent 
Functions to be Positive}}
\label{sub:ApdOne}
{
\noindent

In this appendix we prove (\ref{eqn:AsZza})
in Property \ref{pr:prOne} and (\ref{eqn:AsZzabb})  
in Property \ref{pr:prTwo} part a). 
Since we have ${G}_{\rmOH}(R,\Gamma|W) \geq$
$\hat{G}_{\rmOH}(R,\Gamma|W)$ 
for $R \geq C(\Gamma|W)$, it suffices to 
show (\ref{eqn:AsZzabb}). 
We first give some preliminaries necessary 
for the proof. 
According to Csisiz\'ar and K\"orner \cite{csi2011information}, 
$C(\Gamma|W)$ has the following expression:
\begin{align}
C(\Gamma|W)=&\min_{\mu \geq 0}\Bigl[
\max_{q_X \in {\cal P}({\cal X})}
\{I(q_X,W)
\notag\\
& -\mu\irOlgTypoRevB{{\rm E}_{q_X}}\left[c(X)\right]\} 
+\mu \Gamma\Bigr].
\label{eqn:aaaZZZ}
\end{align}
By simple 
computation we have 
\begin{align}
& \left(  \frac{\rm d}{\irbRevAb{ {\rm d}\lambda }}
 \Omega^{(\mu,\lambda)}(q_X,Q|W)\right)_{\lambda=0}
\notag\\
&=I(q_X,W)-\mu
  \irOlgTypoRevB{{\rm E}_{q_X}}\left[c(X)\right]
  \irOlgTypoRevB{+D(q_XW||Q)}.
\label{eqn:axxZZ}
\end{align}

{\it Proof of  (\ref{eqn:AsZzabb}) 
in Property \ref{pr:prTwo} part a):} 
Let $\mu^*$ be a \irrTypoRevA{parameter} 
$\mu$ which attains the minimum of 
$$
\max_{q_X \in {\cal P}({\cal X})}
\{I(q_X,W) -\mu{\rm E}_{p_X}\left[c(X)\right]\}
+\irbRevAb{\mu}\Gamma
$$
appearing the expression of 
$C(\Gamma|W)$ given by (\ref{eqn:aaaZZZ}).
Then we have 
\begin{align}
&C(\Gamma|W)-\mu^{*}\Gamma
=\max_{q_X \in {\cal P}({\cal X})}
\{ I(q_X,W) -\mu^*{\rm E}_{p_X}\left[c(X)\right] \}
\notag\\
&\geq I(q_X,W) -\mu^*{\rm E}_{p_X}\left[c(X)\right] 
\label{eqn:aPPaZZZ} 
\end{align}
for any $q_X \in {\cal P}({\cal X})$. 
Here we consider the case where $R>C(\Gamma|W)$. In 
this case we have $R\geq C(\Gamma|W)+\tau$ for some $\tau>0$. 
Applying the equality (\ref{eqn:axxZZ}) to 
$\hat{\Omega}^{(\mu,\lambda)}(q_X|W)$
$=\Omega^{(\mu,\lambda)}($$q_X,q_XW|W)$,
we have
\begin{align}
&\lim_{\lambda \to 0} 
\frac{\hat{\Omega}^{(\mu^*,\lambda)}(q_X|W)}{\lambda}
= \left(\frac{\rm d}{ \irbRevAb{ {\rm d} \lambda } }
 \hat{\Omega}^{(\mu^*,\lambda)}(q_X|W)\right)_{\lambda=0}
\notag\\
&=I(q_X,W)-\mu^*{\rm E}_{p_X}\left[c(X)\right].
\label{eqn:axPss}
\end{align} 
From (\ref{eqn:axPss}), we can see that 
there exists $\lambda_0>0$ such that
\begin{align}
& \frac{\hat{\Omega}^{(\mu^*,\lambda_0)}(q_X|W)}{\lambda_0}
\leq I(q_X,W)-\mu^*{\rm E}_{p_X}\left[c(X)\right]
+\frac{\tau}{2}.
\label{eqn:ZssP}
\end{align}
From (\ref{eqn:aPPaZZZ}) and (\ref{eqn:ZssP}), we have
\begin{align} 
&\frac{\hat{\Omega}^{(\mu^*,\lambda_0)}(q_X|W)}{\lambda_0}
\leq 
C(\Gamma|W)-\mu^* \Gamma+\frac{\tau}{2}.
\label{eqn:pZssPx}
\end{align}
From (\ref{eqn:pZssPx}), we have 
\beq
{\hat{\Omega}^{(\mu^*,\lambda_0)}(q_X|W)}
\leq \lambda_0\left\{C(\Gamma|W)-\mu^* \Gamma+\frac{\tau}{2}\right\}.
\label{eqn:AsEEE}
\eeq
Since (\ref{eqn:AsEEE}) holds 
for any $q_X\in {\cal P}({\cal X})$, we have  
\beq
{\hat{\Omega}^{(\mu^*,\lambda_0)}(W)}
\leq \lambda_0\left\{C(\Gamma|W)-\mu^* \Gamma+\frac{\tau}{2}\right\}.
\label{eqn:AsEE}
\eeq
Then we have the following chain of inequalities:  
\begin{align*}
&\hat{G}(R, \Gamma |W)
\\
&\geq \frac{\lambda_0[ R-\mu^* \Gamma]
-\hat{\Omega}^{(\mu^*,\lambda_0)}(W)}
{1+\lambda_0}
\MGeq{a}
\frac{\lambda_0{\tau}}{2(1+\lambda_0)}.
\end{align*}
Step (a) follows from (\ref{eqn:AsEE}) and 
$R \geq C(\Gamma|W)+\tau$.
\hfill\IEEEQED
}
}

We next provide another expression of 
$G(R,\Gamma|W)$ denoted by $G_{\rm AR}(R,\Gamma |W)$, 
which has the same form as $G_{\rm AR}(R|W)$ 
when $\Gamma \geq \Gamma_{\max}$.
To describe this exponent function we define some 
functions. For $\irPkOh{\rho}\in [0,1]$ and 
$\mu\geq 0$, define   \begin{align*}
& {J}^{(\mu,\irPkOh{\rho})}(W)
 \defeq \max_{q_X \in {\cal P}({\cal X})}
  J^{(\mu,\irPkOh{\rho})}(q_X|W),
\\
& {\empty}{G}_{\rm AR}^{(\mu,\irPkOh{\rho})}(R,\Gamma \Vl W) 
\defeq  \irPkOh{\rho}(R-\mu \Gamma)
-\irPkOh{{J}^{(\mu,\irPkOh{\rho})}(W)},
\\
& {\empty}{G}_{\rm AR}(R,\Gamma \Vl W)
\defeq 
\sup_{\scs \mu \geq 0, \atop{\scs \irPkOh{\rho} \in [0,1]}}
{\empty}{G}_{\rm AR}^{(\mu, \irPkOh{\rho})}(R,\Gamma \Vl W).
\end{align*}
\irOlgComRevB{
We have the following property.
\begin{pr}\label{pr:ApdLmTwo}
$\quad$
\begin{itemize}
 \item[a)] For each fixed $\mu\geq 0$, 
 $J^{(\mu,\rho)}(W)$ is left continuous  
 at $\rho=1$.

 \item[b)] For each fixed $\mu\geq 0$, 
 $J^{(\mu,\rho)}(W)$ is a convex function of 
 $\rho \in [0,1)$.

 \item[c)] For each fixed $\rho\in [0,1]$, 
 $J^{(\mu,\rho)}(W)$ is  a convex function of 
 $\mu \geq 0$.

 \item[d)] For $R\geq 0$ and $\Gamma > \Gamma_0$,
the supremum in the definition of $G_{\rm AR}(R,\Gamma)$
can be replaced with the maximum. That is, 
\begin{align*}
&\sup_{\scs \mu \geq 0, \atop{\scs \irPkOh{\rho} \in [0,1]}}
{\empty}{G}_{\rm AR}^{(\mu,  \irPkOh{\rho})}(R,\Gamma \Vl W)
=\max_{\scs \mu \geq 0, \atop{\scs \irPkOh{\rho}\in [0,1]}}
{\empty}{G}_{\rm AR}^{(\mu,\irPkOh{\rho})}(R,\Gamma \Vl W).
\end{align*}
\end{itemize}
\end{pr}

\clrMjRvOh{
We can prove this property by standard analytical 
arguments based on the definitions of 
$J^{(\mu,\rho)}(W)$ and $G_{\rm AR}(R,\Gamma|W)$.  
Details of the proofs are 
given in Appendix \ref{sub:ApdLmTwo}.} 
}
\newcommand{\PrApdLmTwo}{
\subsection{
Proofs of Properties \ref{pr:ApdLmTwo} 
and \ref{pr:ApdLmThr}
}
\label{sub:ApdLmTwo}

In this appendix we prove Properties \ref{pr:ApdLmTwo} 
and \ref{pr:ApdLmThr}.
Before entering into the proofs we provide 
\clrMjRvOh{several} preliminaries to prove 
those properties. 

\clrMjRvOh{
Let $U$ be an arbitrary random variable taking values in 
finite set ${\cal U}$. We assume that $U$ has a 
probability distribution $q_U$. 
Let $f_0:{\cal U} \to \mathbb{R}$ 
be an arbitrary bounded real valued function. 
For each 
$i=1,2,\cdots,k$, let 
$f_i:{\cal U}\to \mathbb{R}^{+}\defeq \{a\geq0 \}$ be 
an arbitrary bounded nonnegative function. 
For $\rho \in [0,1)$ and $\mu\geq 0$, define 
$$
B_k^{(\mu,\rho)}\defeq  
\log
\sum_{i=1}^k\left(
{\rm E}
\left\{f_i(U){\rm e}^{-\mu f_0(U)}
\right\}^{\frac{1}{\overline{\rho}}}
\right)^{\overline{\rho}}.
$$
When $\rho=1$, we define
\begin{align*}
B_k^{(\mu,1)}&\defeq \lim_{\rho \uparrow 1}
B_k^{(\mu,\rho)}
  =\log \sum_{i=1}^k
        \max_{u: q_U(u)>0}f_i(u){\rm e}^{-\mu f_0(u)}.
\end{align*}
On $B_k^{(\mu,\rho)}$, we have the following lemma:
\begin{lm}\label{lm:PrBkMuRho}
$\quad$ 
\begin{itemize}
\item[a)]For each fixed $\mu \geq0$, $B_k^{(\mu,\rho)}$ 
is a monotone increasing function of $\rho \in [0,1)$.
\item[b)] For each fixed $\mu \geq0$, $B_k^{(\mu,\rho)}$ 
is a convex function of $\rho\in [0,1)$.
\item[c)] For each fixed $\rho \in [0,1]$, $B_k^{(\mu,\rho)}$ 
is a convex function of $\mu\geq 0$.
\end{itemize}
\end{lm}

 {\it Proof:} We first prove the part a). 
 For $u\in {\cal U}$, set 
 $$
\Lambda_i^{(\mu,\rho)}(u)\defeq 
\left\{f_i(U){\rm e}^{-\mu f_0(U)}
\right\}^{\frac{1}{\overline{\rho}}}.
 $$
Then we have 
$$
B_k^{(\mu,\rho)}=\log \sum_{i=1}^k 
\left({\rm E}\Lambda_i^{(\mu,\rho)}(U)\right)^{\overline{\rho}}. 
$$
Computing partial derivative of $B_k^{(\mu,\rho)}$ with respect 
 to $\rho\in [0,1)$, we have the following:
 \begin{align*}
 \frac{\partial}{\partial \rho} 
 B_k^{(\mu,\rho)}   
 =&\exp\left[-B_k^{(\mu,\rho)}\right] 
\sum_{i=1}^k \left({\rm E}\Lambda_i^{(\mu,\rho)}(U)
               \right)^{-\rho}
\notag\\
& \times 
\left({\rm E}
\Lambda_i^{(\mu,\rho)}(U)
\log \Lambda_i^{(\mu,\rho)}(U)
\right.
\notag\\
&\quad  -{\rm E}\Lambda_i^{(\mu,\rho)}(U)
         \log{\rm E}\Lambda_i^{(\mu,\rho)}(U)   
 \Bigr)  
 \MGeq{a}0.              
\end{align*}
Step (a) follows from that $z\log z$ 
is a convex function of $z>0$. 

We next prove the part b). 
Fix any $\tau \in [0,1]$ and any 
$\rho_1,\rho_2\in [0,1)$. Set $\rho=\tau \rho_1+\overline{\tau}\rho_2$. 
By the definition of $\Lambda_i^{(\mu,\rho)}(u), u\in {\cal U}$, we have
\begin{align}
 \Lambda_i^{(\mu,\rho)}(u)= 
 [\Lambda_i^{(\mu,\rho_1)}(u)
  ]^{\frac{\tau \overline{\rho_1}}{\overline{\rho}} }
  [\Lambda_i^{(\mu,\rho_2)}(u)
  ]^{\frac{\overline{\tau}\,\overline{\rho_2}}{\overline{\rho}} }.
\label{eqn:LambdaEqa}
 \end{align}
For each $\mu \geq 0$, we have the 
following chain of inequalities:
\begin{align}
& \exp\left[B_k^{(\mu,\rho)}\right]
\notag\\
&
\MEq{a}\sum_{i=1}^k 
\left({\rm E}
 [\Lambda_i^{(\mu,\rho_1)}(U)
  ]^{\frac{\tau \overline{\rho_1}}{\overline{\rho}} }
  [\Lambda_i^{(\mu,\rho_2)}(U)
  ]^{\frac{\overline{\tau}\,\overline{\rho_2}}{\overline{\rho}} }
\right)^{\overline{\rho}}
\notag\\
&\MLeq{b}
\sum_{i=1}^k 
\left({\rm E}
  \Lambda_i^{(\mu,\rho_1)}(U)
  \right)^{\tau\overline{\rho_1}}
  \left({\rm E}  
  \Lambda_i^{(\mu,\rho_2)}(U)
  \right)^{\overline{\tau}\,\overline{\rho_2}}
 \notag\\
 &\MLeq{c}
\newcommand{\LamBound}{
\left\{\sum_{i=1}^k 
  \left({\rm E}
  \Lambda_i^{(\mu,\rho_1)}(U)
  \right)^{\overline{\rho_1}}
\right\}^{\tau}  
\left\{\sum_{i=1}^k 
  \left({\rm E}  
  \Lambda_i^{(\mu,\rho_2)}(U)
  \right)^{\overline{\rho_2}}
  \right\}^{\overline{\tau}}
\notag\\  
}
\exp\left[\tau B_k^{(\mu,\rho_1)}\right]
   \exp\left[\overline{\tau} B_k^{(\mu,\rho_2)}\right].
\label{eqn:PrApdLmTwoEqAb}
\end{align}
Step (a) follows from (\ref{eqn:LambdaEqa}).
Steps (b) and (c) follow from  H\"{o}lder inequality.
The bound (\ref{eqn:PrApdLmTwoEqAb}) implies that 
we have the part b).  

We finally prove the part c). 
Fix any $\tau \in [0,1]$ and any 
$\mu_1,\mu_2\geq 0$. Set $\mu=\tau \mu_1+\overline{\tau}\mu_2$. 
By the definition of $\Lambda_i^{(\mu,\rho)}(u), u\in {\cal U}$, we have
\begin{align}
 \Lambda_i^{(\mu,\rho)}(u)= 
 [\Lambda_i^{(\mu_1,\rho)}(u)
  ]^{\tau }
  [\Lambda_i^{(\mu_2,\rho)}(u)
  ]^{\overline{\tau}}.
\label{eqn:LambdaEqb}  
 \end{align}
For each $\rho\in [0,1)$, we have the following chain of 
inequalities:
\begin{align}
& \exp\left[B_k^{(\mu,\rho)}\right]
\MEq{a}\sum_{i=1}^k 
\left({\rm E}
 [\Lambda_i^{(\mu_1,\rho)}(U)]^{\tau }
  [\Lambda_i^{(\mu_2,\rho)}(U)]^{\overline{\tau} }
\right)^{\overline{\rho}}
\notag\\
&\MLeq{b}
\sum_{i=1}^k 
\left({\rm E}
  \Lambda_i^{(\mu_1,\rho)}(U)
  \right)^{\tau\overline{\rho}}
  \left({\rm E}  
  \Lambda_i^{(\mu_2,\rho)}(U)
  \right)^{\overline{\tau}\,\overline{\rho}}
 \notag\\
 &\MLeq{c}
\newcommand{\LamBdb}{
\left\{\sum_{i=1}^k 
  \left({\rm E}
  \Lambda_i^{(\mu_1,\rho)}(U)
  \right)^{\overline{\rho}}
\right\}^{\tau}  
\left\{\sum_{i=1}^k 
  \left({\rm E}  
  \Lambda_i^{(\mu_2,\rho)}(U)
  \right)^{\overline{\rho}}
  \right\}^{\overline{\tau}}
\notag\\ 
}
\exp\left[\tau B_k^{(\mu_1,\rho)}\right]
   \exp\left[\overline{\tau} B_k^{(\mu_2,\rho)}\right].
\label{eqn:PrApdLmTwoEqAc}
\end{align}
Step (a) follows from (\ref{eqn:LambdaEqb}).
Steps (b) and (c) follow from H\"{o}lder 
inequality.
By letting $\rho \uparrow 1$ 
in (\ref{eqn:PrApdLmTwoEqAc}), 
we have the bound (\ref{eqn:PrApdLmTwoEqAc}) also for $\rho=1$.   
The bound (\ref{eqn:PrApdLmTwoEqAc}) implies that 
we have the part c).  
\hfill\IEEEQED  

}

\clrMjRvOh{
We next provide some preliminaries related 
to Properties \ref{pr:ApdLmTwo} and \ref{pr:ApdLmThr} 
parts a) and b). For $q_X\in {\cal P}({\cal X})$, we set
$$
A^{(\mu)}(q_X)\defeq 
\sum_{x\in {\cal X}}q_X(x){\rm e}^{\mu c(x)}. 
$$
Define $\tilde{q}_X
=\{\tilde{q}_X(x)\}_{x\in{\cal X}}$ 
$\in {\cal P}({\cal X})$ by 
$$
\tilde{q}_X^{(\mu)}(x) 
\defeq \frac{q_X(x)}{A^{(\mu)}(q_X)}, 
x\in {\cal X}.   
$$
Furthermore we define
\begin{align*}
&
K^{(\mu,\rho)}(\tilde{q}_X^{(\mu)}|W)
\\
& \defeq \log
\sum_{y \in {\cal Y} }
\left[\sum_{x \in {\cal X}} 
\tilde{q}_X^{(\mu)}(x)\left\{W(y|x)
{\rm e}^{-\mu c(x)}\right\}^{\frac{1}{\overline{\rho}}} 
\right]^{\overline{\rho}},
\notag \\
&
 \hat{K}^{(\mu,\rho)}(\tilde{q}_X^{(\mu)}|W)
\defeq \log
\Hugebl\sum_{x \in {\cal X} } 
\tilde{q}_X^{(\mu)}(x)
\\
&\qquad \times \sum_{y \in {\cal Y}}(q_XW)(y)
\left\{\frac{W(y|x){\rm e}^{-\mu c(x)}}{(q_XW)(y)}
\right\}^{\frac{1}{\overline{\rho}}} 
\Hugebr^{\overline{\rho}}.
\notag
\end{align*}
When $\irOOh{\rho}=1$, we define
\begin{align*} 
K^{(\mu,1)}(\tilde{q}_X^{(\mu)}|W)
 \defeq& \lim_{\irOOh{\rho} \uparrow 1 }
 K^{(\mu,\irOOh{\rho})}(\tilde{q}_X^{(\mu)}|W),
\\
 \hat{K}^{(\mu,1)}(\tilde{q}_X^{(\mu)}|W)
 \defeq& \lim_{\irOOh{\rho} \uparrow 1 }
 \hat{K}^{(\mu,\irOOh{\rho})}(\tilde{q}_X^{(\mu)}|W).
\end{align*}
We have the following relationship between 
$J^{(\mu,\rho)}(q_X$ $|W)$ and 
$K^{(\mu,\rho)}(\tilde{q}_X|W)$ 
and that between $\hat{J}^{(\mu,\rho)}(q_X|W)$ and $\hat{K}^{(\mu,\rho)}(\tilde{q}_X|W)$:
\begin{align}
 \left.
\ba{rl}
&J^{(\mu,\rho)}({q}_X^{(\mu)}|W)
\\
&= \overline{\rho}\log A^{(\mu)}(q_X)+ 
   K^{(\mu,\rho)}(\tilde{q}_X^{(\mu)}|W),
\vspace*{1mm}\\
& \hat{J}^{(\mu,\rho)}({q}_X^{(\mu)}|W)
\\
& =\overline{\rho}\log A^{(\mu)}(q_X)+ 
\hat{K}^{(\mu,\rho)}(\tilde{q}_X^{(\mu)}|W).
\ea \right\}
\label{eqn:JandhatJEq}
\end{align}
From (\ref{eqn:JandhatJEq}), we know that
\begin{align}
 \left. 
 \ba{rl}
 J^{(\mu,1)}(      {q}_X|W)
=&K^{(\mu,1)}(\tilde{q}_X^{(\mu)}|W),\:
\vspace*{1mm}\\
\hat{J}^{(\mu,1)}( {q}_X|W)
=&\hat{K}^{(\mu,1)}(\tilde{q}_X^{(\mu)}|W).   
\ea
\right\}
\label{eqn:JandhatJEqTwo}
\end{align}
We have the following lemma:
\begin{lm}\label{lm:PrJpxandhatJpx}
$\quad$
\begin{itemize}
\item[a)]
 For any $\mu \geq 0$, 
$\rho\in [0,1)$, and any 
$q_X \in{\cal P}({\cal X})$, 
\begin{align*}
 \left. 
 \ba{rl}
& J^{(\mu,\rho)}(q_X \Vl W) \leq 
\overline{\rho}\log A^{(\mu)}(q_X) 
+ J^{(\mu,1)}(q_X \Vl W), 
\vspace*{1mm}\\
& \hat{J}^{(\mu,\rho)}(q_X \Vl W) 
\leq \overline{\rho}\log A^{(\mu)}(q_X) 
+\hat{J}^{(\mu,1)}(q_X \Vl W).
\ea 
\right\}
\end{align*}
\item[b)] For each fixed $\mu\geq 0$ and 
each $q_X\in {\cal P}({\cal X})$, 
 $J^{(\mu,\rho)}($ $q_X| W)$ and 
 $\hat{J}^{(\mu,\rho)}(q_X| W)$ 
 are convex functions of $\rho \in [0,1)$.
\item[c)]
For each fixed $\rho \in [0,1]$ and 
each $q_X\in {\cal P}({\cal X})$, 
$J^{(\mu,\rho)}($$q_X| W)$ and 
$\hat{J}^{(\mu,\rho)}(q_X| W)$
are convex functions of $\mu \geq 0$.
\end{itemize}
\end{lm}

{\it Proof:} We first observe that  
$K^{(\mu,\rho)}(\tilde{q}_X^{(\mu)}|W)$ and 
$\tilde{K}^{(\mu,\rho)}(\tilde{q}_X^{(\mu)}|W)$, 
respectively, have the same forms as 
$B_k^{(\mu,\rho)}$ 
and $B_1^{(\mu,\rho)}$. 
Hence we have the following three results:
\begin{itemize}
\item[i)]For each fixed $\mu \geq 0$,
$K^{(\mu,\rho)}(\tilde{q}_X^{(\mu)}|W)$ and 
$\tilde{K}^{(\mu,\rho)}($ $\tilde{q}_X^{(\mu)}|W)$  
are monotone increasing functions of $\rho\in [0,1)$. 
\item[ii)]
For each fixed $\mu \geq 0$, 
$K^{(\mu,\rho)}(\tilde{q}_X^{(\mu)}|W)$ and 
$\tilde{K}^{(\mu,\rho)}($ $\tilde{q}_X^{(\mu)}|W)$  
are convex functions of $\rho\in [0,1)$.

\item[iii)]
For each fixed $\rho \geq [0,1]$, 
$K^{(\mu,\rho)}(\tilde{q}_X^{(\mu)}|W)$ and 
$\tilde{K}^{(\mu,\rho)}($ $\tilde{q}_X^{(\mu)}|W)$  
are convex functions of $\mu \geq 0$.
\end{itemize}
We first prove the part a).
The result i) together with 
(\ref{eqn:JandhatJEq}) and 
(\ref{eqn:JandhatJEqTwo}) yields the part a).  

We next prove the part b).
The result ii) together with 
(\ref{eqn:JandhatJEq}) 
yields the part b). 

We finally prove the part c). It can be seen from the result iii) and (\ref{eqn:JandhatJEq}) that it suffices to prove 
that $\log A^{(\mu)}(q_X)$ is a convex function of 
$\mu\geq 0$ to prove the part c). 
For $B_k^{(\mu,0)}$, we set $k=1$ 
and choose $f_1: {\cal U}\to \{a\geq 0\}$ so that 
$f_1(u)=1, u\in {\cal U}$. For the choice of 
$k=1$ and the above choice of $f_1$, we have 
that $\log A^{(\mu)}(q_X)$ has 
the same form as $B_1^{(\mu,0)}$.
Then by Lemma \ref{lm:PrBkMuRho} part c), we have that 
$\log A^{(\mu)}(q_X)$ is a convex function of 
$\mu\geq 0$. 
\hfill \IEEEQED

We finally provide a bound related to 
Properties \ref{pr:ApdLmTwo} and \ref{pr:ApdLmThr} 
part b). This bound is stated in the following 
lemma:
\begin{lm}\label{lm:UbGAR} For any $\mu \geq 0$ 
and any $\rho\in [0,1]$, 
 \begin{align}
&
 G_{\rm AR}^{(\mu,\rho)}(R,\Gamma\Vl W)
\notag\\
& \leq 
  \rho R -\mu\rho[\Gamma-\Gamma_0]
  -\overline{\rho}\log q_{X}(x_0).
\label{eqn:upperG}
\end{align} 
\end{lm}
}

{\it Proof:} 
Let $x_0$ be a symbol such that $c(x_0)=\Gamma_0$.
We choose $q_X \in {\cal P}({\cal X})$ so that 
$q_X(x_0)>0$. By definition we have that for 
$\mu \geq 0,\rho\in [0,1]$,
\begin{align}
{J}^{(\mu,\rho)}(W)
&\geq {J}^{(\mu,\rho)}(q_X|W)
\notag\\
& \geq \overline{\rho}\log q_{X}(x_0)-\mu\rho\Gamma_0.  
\label{eqn:Jlowb}
\end{align}
From (\ref{eqn:Jlowb}), we have the bound 
(\ref{eqn:upperG}) for any $\mu \geq 0$ 
and $\rho\in [0,1]$.
\hfill\IEEEQED

The bound (\ref{eqn:upperG}) is quite useful 
to prove Properties \ref{pr:ApdLmTwo} and 
\ref{pr:ApdLmThr} part b). 

{\it Proof of Property \ref{pr:ApdLmTwo}:} 
\clrMjRvOh{
We first prove the part a). Let $q_X^{(\mu,\rho)}$ 
be a probability distribution attaining 
the maximum in the definition of 
$J^{(\mu,\rho)}(W)$. Set 
$
 A^{(\mu)}\defeq 
 \max_{q_X\in {\cal P}({\cal X})}A^{(\mu)}(q_X).
$
For any $\mu \geq 0$ and $\rho\in [0,1)$, 
we have the following chain of inequalities: 
\begin{align}
& J^{(\mu,\rho)}(W)=
J^{(\mu,\rho)}(q_X^{(\mu,\rho)}|W)
\notag\\
&\MLeq{a}
\overline{\rho}\log A^{(\mu)}(q_X^{(\mu,\rho)}) 
 +J^{(\mu,1)}(q_X^{(\mu,\rho)}|W)
\notag\\
&\leq \overline{\rho}\log A^{(\mu)} 
 +{J}^{(\mu,1)}(W). 
\label{eqn:BdPartA}
\end{align}
Step (a) follows from Lemma \ref{lm:PrJpxandhatJpx} part a). 
}
From (\ref{eqn:BdPartA}), 
we have the following chain of inequalities:  
\begin{align}
&\clrMjRvOh{-\overline{\rho}\log A^{(\mu)}} 
\leq {J}^{(\mu,1)}(W)
        -{J}^{(\mu,\rho)}(W) 
\notag\\
&\qquad  =  {J}^{(\mu,1)}(q_X^{(\mu,1)}\Vl W)
        -{J}^{(\mu,\rho)}(W) 
\notag\\        
&\qquad \leq  {J}^{(\mu,1)}(q_X^{(\mu,1)}\Vl W)
        -{J}^{(\mu,\rho)}(q_X^{(\mu,1)}\Vl W). 
\label{eqn:Jub}
\end{align}
\clrMjRvOh{The first and the fourth members 
of (\ref{eqn:Jub}) tend to zero as 
$\rho$ tends to 1.} Hence we have the part a).

We next prove the parts b) and c). 
Fix any $\tau \in [0,1]$ and any 
$\rho_1,\rho_2\in [0,1)$. Set $\rho=\tau \rho_1+\overline{\tau}\rho_2$. 
Then we have the following chain of 
inequalities:
\begin{align}
& J^{(\mu,\tau\rho_1+\overline{\tau}\rho_2)}(W)
=J^{(\mu,\tau\rho_1+\overline{\tau}\rho_2)}(q_X^{(\mu,\rho)}|W)
\notag\\
& \MLeq{a} \tau J^{(\mu,\rho_1)}(q_X^{(\mu,\rho)}|W)
+\overline{\tau}J^{(\mu,\rho_2)}(q_X^{(\mu,\rho)}|W)
\notag\\
& \leq     \tau J^{(\mu,\rho_1)}(W)
+\overline{\tau}J^{(\mu,\rho_2)}(W).
\label{eqn:PrApdLmTwoEqA}
\end{align}
Step (a) follows from Lemma \ref{lm:PrJpxandhatJpx} part b). The bound (\ref{eqn:PrApdLmTwoEqA}) implies 
the part b). We can prove the part c) by using 
Lemma \ref{lm:PrJpxandhatJpx} part c).  
The proof is quite parallel with 
that of the part b). We omit the detail. 

We finally prove the part d). 
We first consider the case where $\rho=0$. In this case, we have that for any $\mu \geq 0$.
$\clrTypoOh{G_{\rm AR}^{(\mu,0)}}(R,\Gamma\Vl W)=0$.
Hence $\clrTypoOh{G_{\rm AR}^{(\mu,0)}}(R,\Gamma\Vl W)$ 
takes the maximum value zero at some $\mu\geq 0$. 
We next consider the case where $\rho \in (0,1]$. By Lemma \ref{lm:PrJpxandhatJpx} part c), we have that for each 
fixed  $\rho \in (0,1]$, 
$G_{\rm AR}^{(\mu,\rho)}(R,\Gamma\Vl W)$ is a concave function of $\mu\geq 0$.
Furthermore, in the case of $\rho \in (0,1]$, under the assumption of $\Gamma >\Gamma_0$, 
the quantity in the right member of          
 (\ref{eqn:upperG}) goes to the negative 
 infinity as $\mu$ tends to the positive 
 infinity.
 From those arguments we know that under $\Gamma > \Gamma_0$, 
 for each $\rho \in [0,1]$,
the function $\clrTypoRevB{G_{\rm AR}^{(\mu,\rho)}}(R,\Gamma\Vl W)$
 takes the maximum value for some 
 nonnegative finite value of 
 $\mu$. Hence we have that for $\Gamma >\Gamma_0$,  
\begin{align}
 & 
   \sup_{\scs \mu \geq 0,
         \atop{\scs \rho \in [0,1]
         }}
 {\empty}G_{\rm AR}^{(\mu, \rho)}(R,\Gamma \Vl W) 
 = \sup_{\rho \in [0,1]}\max_{\scs \mu \geq 0}
 {\empty}{G}_{\rm AR}^{(\mu, \rho)}(R,\Gamma \Vl W)   
 \notag\\
 & =\max_{\scs \mu \geq 0}\sup_{\rho \in [ 0,1]}
 {\empty}{G}_{\rm AR}^{(\mu, \rho)}(R,\Gamma \Vl W)
 \MEq{a} \max_{\scs \mu \geq 0,
         \atop{\scs \rho \in [0,1]
         }}
 {\empty}{G}_{\rm AR}^{(\mu, \rho)}(R,\Gamma \Vl W).  
\notag 
 \end{align}
 Step (a) follows from that for each fixed 
 $\mu\geq 0$, ${\empty}{G}_{\rm AR}^{(\mu, \rho)}
 (R,\Gamma \Vl W)$ is a continuous 
 function of $\rho\in [0,1]$. 
 Thus the part d) is proved. 
 \hfill \IEEEQED

{\it Proof of Property \ref{pr:ApdLmThr}:} 
Proofs of Property \ref{pr:ApdLmThr} parts a), b), c) 
are quite parallel with those 
of Property \ref{pr:ApdLmTwo}. We omit the details of 
the proofs. 
\newcommand{\OmitZZ}{
We first prove the part a). Let $\hat{q}_X^{(1)}$ 
be a probability distribution attaining 
the maximum in the definition of 
$\hat{J}^{(\mu,1)}(W)$. From (\ref{eqn:BdPartA}), 
we have the following chain of inequalities:  
\begin{align}
\clrMjRvOh{-\overline{\rho}\log A^{(\mu)}} 
\leq & \hat{J}^{(\mu,1)}(W)
         -\hat{J}^{(\mu,\rho)}(W) 
\notag\\
    = &  \hat{J}^{(\mu,1)}(\hat{q}_X^{(1)}\Vl W)
        -\hat{J}^{(\mu,\rho)}(W) 
\notag\\        
   \leq  & \hat{J}^{(\mu,1)}(\hat{q}_X^{(1)}\Vl W)
          -\hat{J}^{(\mu,\rho)}(\hat{q}_X^{(1)}\Vl W). 
\label{eqn:hatJub}
\end{align}
\clrMjRvOh{The first and the fourth members of (\ref{eqn:hatJub}) 
tend to zero as $\rho$ tends to 1.} 
Hence we have the part a). 
}
We only prove the part d).  
\clrMjRvOh{
For each $q_X\in {\cal P}({\cal X})$ and 
    each $\rho\in [0,1)$, we have the following chain 
    of inequalities:
\begin{align}
&\hat{J}^{(\mu,\rho)}(q_X|W)= \overline{\rho}
 \Omega^{\left(\mu,\frac{\rho}{\overline{\rho}}\right)}
 (q_X,q_XW\Vl W)
\notag\\
&\MGeq{a}\overline{\rho}\Omega^{\left(\mu,\frac{\rho}    {\overline{\rho}}\right)}(q_X\Vl W)
\MEq{b}{J}^{(\mu,\rho)}(q_X|W)
\label{eqn:LowBhatJ}.
\end{align}
Step (a) follows from Property \ref{pr:prTwo} part b). 
Step (b) follows from Lemma \ref{lm:lmSddQ}.
From (\ref{eqn:LowBhatJ}), we have that for $\rho\in [0,1)$,
\begin{align}
&\hat{J}^{(\mu,\rho)}(W)\geq {J}^{(\mu,\rho)}(W)
\label{eqn:LowBhatJb}.
\end{align}
According to Properties 
\ref{pr:ApdLmTwo} and \ref{pr:ApdLmThr}
part a) now we have proved, we have that for each $\mu\geq 0$, 
$\hat{J}^{(\mu,\rho)}(W)$ and ${J}^{(\mu,\rho)}(W)$ are 
left continuous at $\rho=1$. 
Hence we have the bound (\ref{eqn:LowBhatJb}) also for $\rho=1$,
implying that for each $\mu\geq 0$ and each $\rho\in [0,1]$, we 
have}
\begin{align}
 &\hat{G}^{(\mu,\rho)}(R,\Gamma|W)
 \leq  {G}_{\rm AR}^{(\mu,\rho)}(R,\Gamma|W)
 \notag\\
 &\leq  
  \rho R -\mu\rho[\Gamma-\Gamma_0]
  -\overline{\rho}\log q_{X}(x_0).
\label{eqn:upperGb}
\end{align}
Based on the upper bound (\ref{eqn:upperGb}) on 
$\hat{G}^{(\mu,\rho)}(R,\Gamma|W)$ for 
$\mu\geq 0,\rho \in [0,1]$, we can prove the part b) 
in a manner quite parallel with that of the proof 
of Property \ref{pr:ApdLmTwo} part b) we have for 
${G}_{\rm AR}(R,\Gamma|W)$. 
\hfill\IEEEQED
}
\irPkOh{We have the following} proposition. 
\begin{pro}\label{pro:pro1z}
For any $R\geq 0$, any $\Gamma \geq \Gamma_0$, 
and any $\mu,\lambda\geq 0$, 
\beqa
G_{\rmOH}^{(\mu,\lambda)}(R,\Gamma|W)
&=&G_{\rm AR}^{(\mu,\frac{\lambda}{1+\lambda})}(R,\Gamma|W).
\label{eqn:mainEq00a}
\eeqa
In particular,
\beqa
& &G_{\rmOH}(R,\Gamma|W)=G_{\rm AR}(R,\Gamma|W).
\label{eqn:mainEq0xx}
\eeqa
\end{pro}

Proof of this proposition is given in 
Section \ref{sec:ThreeExp}. 
\newcommand{\LemmaForProposition}{


In this section we prove \irOlgRevBb{Proposition} \ref{pro:pro1z} 
stated in Section \ref{sec:MainResult}. The following is 
a key lemma to prove this proposition. 
\begin{lm}\label{lm:lmSddQ} For any $q_X$ 
$\in {\cal P}({\cal X})$
$$
\min_{Q \in{\cal P}({\cal Y})}\Omega^{(\mu,\lambda)}(q_X,Q|W)
=(1+\lambda)J^{(\mu,\frac{\lambda}{1+\lambda})}(q_X|W).
$$
The distribution $Q\in {\cal P}({\cal Y})$ attaining $(1+\lambda)J^{(\mu,\frac{\lambda}{1+\lambda})}($ $q_X|W)$ 
is given by 
$$
Q(y)=\kappa \left[ 
\sum_{x\in {\cal X}}q_X(x) 
W^{1+\lambda}(y|x){\rm e}^{-\mu \lambda c(x)}\right]^{\frac{1}{1+\lambda}},
$$
where $\kappa$ is a constant for normalization, having the form
\begin{align}
\kappa^{-1}
&=\sum_{y \in {\cal Y} }
\left[\sum_{x \in {\cal X}}q_X(x)
W^{1+\lambda}(y|x){\rm e}^{-\mu \lambda c(x)} 
\right]^{\frac{1}{1+\lambda}}
\notag\\
&=\exp\left[{J}^{(\mu, \frac{\lambda}{1+\lambda})}(q_X|W) \right].
\label{eqn:AzzXc}
\end{align}
\irrAE{
Furthermore, we have
\beq
\Omega^{(\mu,\lambda)}(W)
=\max_{q_X \in {\cal P}({\cal X})}
(1+\lambda)J^{(\mu,\frac{\lambda}{1+\lambda})}(q_X|W).
\label{eqn:AzzXd}
\eeq
}
\end{lm}

{\it Proof:} We observe that
\begin{align}
&\Omega^{(\mu,\lambda)}(W)
=\max_{q_X \in{\cal P}({\cal X})}
\log \biggl\{ \min_{Q \in{\cal P}({\cal Y})}
\sum_{(x,y) \in {\cal X} \times {\cal Y}}1 
\nonumber\\
&\quad \times q_{X}(x)W(y|x)
\left[\frac{W(y|x){\rm e}^{-\mu c(x)}}{Q(y)} \right]^\lambda
\biggr\}
\label{eqn:aAzzb}. 
\end{align}
On the objective function of the minimization problem 
inside the logarithm function in (\ref{eqn:aAzzb}), 
we have the following chain of inequalities:
\begin{align}
&\sum_{(x,y) \in {\cal X} \times {\cal Y}}q_{X}(x)W(y|x)
\left[\frac{W(y|x){\rm e}^{-\mu c(x)}}{Q(y)} \right]^\lambda
\nonumber\\
&=\sum_{y \in {\cal Y} }\left[
\sum_{x \in {\cal X}}q_X(x)W^{1+\lambda}(y|x)
{\rm e}^{-\mu \lambda c(x)}\right]Q^{-\lambda}(y)
\nonumber\\
&\MGeq{a}
\left\{\sum_{y \in {\cal Y} }
\left[\sum_{x \in {\cal X}  }q_X(x) W^{1+\lambda}(y|x)
{\rm e}^{-\mu \lambda c(x)}
\right]^{\frac{1}{1+\lambda}}
\right\}^{1+\lambda} 
\nonumber\\
&\quad \times
\left\{ \sum_{y \in {\cal Y}}Q(y)\right\}^{-\lambda}
\nonumber\\
& 
=\exp\left\{(1+\lambda)J^{(\mu,\frac{\lambda}{1+\lambda})}(q_X|W) \right\}.
\label{eqn:AzzzV}
\end{align}   
In (a), we have used the reverse H\"older inequality
$$
\sum_i a_ib_i\geq 
\left( \sum_{i} a_i^{\frac{1}{\alpha}}\right)^{\alpha}
\left(\sum_{i} b_i^{\frac{1}{\beta}}\right)^{\beta}
$$
which holds for \irrTypoRevA{nonnegative} $a_i,b_i$ and for $\alpha+\beta=1$ 
such that either $\alpha >1$ or $\beta >1$. In our case we 
have applied the inequality to 
$$
\left.
\ba{rcl}
i &\to& y,\:(\alpha,\beta) \to (1+\lambda, -\lambda),
\vspace*{1mm}\\
a_i &\to& \ds 
\sum_{x\in {\cal X} }q_X(x) 
W^{1+\lambda}(y|x){\rm e}^{-\mu \lambda  c(x)},\:
b_i \to Q^{-\lambda}(y). 
\ea
\right\}
$$
In the reverse H\"older inequality the equality holds if and only if 
$a_i^{\frac{1}{\alpha}}=\irPkAu{\nu} b_i^{\frac{1}{\beta}}$ for 
some constant \irPkAu{$\nu$}. In (\ref{eqn:AzzzV}), the equality holds for 
$$
Q(y)=\kappa \left[ 
\sum_{x \in {\cal X} }q_X(x) 
W^{1+\lambda}(y|x)
{\rm e}^{-\mu \lambda c(x)}\right]^{\frac{1}{1+\lambda}},
$$
where $\kappa$ is a normalized constant. 
From (\ref{eqn:aAzzb}) and (\ref{eqn:AzzzV}), we have (\ref{eqn:AzzXd}).	
\hfill\IEEEQED

{\it Proof of Proposition \ref{pro:pro1z}:} 
The equality (\ref{eqn:mainEq00a}) in Proposition \ref{pro:pro1z} 
immediately follows from Lemma \ref{lm:lmSddQ}. 
Using (\ref{eqn:mainEq00a}), we prove 
${G}_{\rmOH}(R,\Gamma\Vl$$ W)=$
${G}_{\rm \irPkAu{AR}}(R,\Gamma\Vl W)$. We have the following 
chain of inequalities:
\begin{align*}
& G_{\rmOH}(R,\Gamma \Vl W)
=\irOOh{\sup_{\mu\geq 0,\lambda\geq 0}}
G_{\rmOH}^{(\mu,\lambda)}(R,\Gamma \Vl W)
\\
&\MEq{a}\sup_{\scs \mu \geq 0}
        \sup_{\scs
          \rho=\frac{\lambda}{1+\lambda}
         \scs \in [0,1)
   }
G_{\rm AR}^{(\mu,\rho)}(R,\Gamma \Vl W)
\\
&\irOOh{
 \MEq{b}}\irOOh{\sup_{\scs \mu \geq 0}
        \sup_{\scs
         \rho\in [0,1]
        }
G_{\rm AR}^{(\mu,\rho)}(R,\Gamma \Vl W) 
=G_{\rm AR}(R,\Gamma \Vl W).}
\end{align*}
Step (a) \irOOh{follows} 
from (\ref{eqn:mainEq00a}) in Proposition \ref{pro:pro1z}. 
\irOOh{Step (b) follows 
from Property \ref{pr:ApdLmTwo} 
part a).} 
\hfill\IEEEQED

}

The following proposition states that the two 
quantities $G_{\rm AR}(R,\Gamma|W)$ and 
$G_{\rm DK}(R,\Gamma|W)$ match. 
\begin{pro}\label{pro:pro1b}
For any $R\geq0 $ and any $\Gamma\geq \Gamma_0$,   
\beqa
& &
G_{\rm AR}(R,\Gamma|W)
  =G_{\rm DK}(R,\Gamma|W).
\label{eqn:mainEq0b}
\eeqa
\end{pro}

Proof of this proposition is found in \cite{Oohama17a} 
and \cite{Oohama18}. 
From \irrComRevA{Theorems \ref{Th:DK}}, \ref{Th:main} and   
Propositions \ref{pro:pro1z}, \ref{pro:pro1b}, 
we immediately obtain the following theorem. 
\begin{Th}\label{th:Finite} 
For any $R\geq 0$ and any $\Gamma \geq \Gamma_0$, 
\begin{align}
& G^*(R,\Gamma|W)= G_{\rmOH}(R,\Gamma|W) 
\nonumber\\
&=G_{\rm AR}(R,\Gamma|W)
 =G_{\rm DK}(R,\Gamma|W).\quad
\label{eqn:mainIeq000}
\end{align}
\end{Th}

\newcommand{\barprmt}{1-\lambda}
\newcommand{\muprmt}{\mu\lambda}

\section{Proof of the Results}
\label{sec:Secaa}
\noindent

In this section we prove Theorems \ref{Th:main0} and \ref{Th:main}. 
We first prove the following lemma. 
\irPkAu{
This lemma is an extension of the lemma 
by 
\irOlgComRevB{\cite{hayashi:03} (stated as Lemma 4 in this reference)} 
to the case where we have the constraint on the input cost.
} 
\begin{lm} 
\label{lm:Ohzzz}
For any $\eta>0$ and for any $(\varphi^{(n)},\psi^{(n)})$  
satisfying 
\irPkAu{ 
$\varphi^{(n)}({\cal K}^{(n)}) \subseteq {\cal S}_{\Gamma}^{(n)}$ 
and } 
$(1/n)\log |{\cal K}_n| \geq R,$
\begin{align}
& {\rm P}_{\rm c}^{(n)}
(\varphi^{(n)},\psi^{(n)}{}|W)
\leq p_{X^nY^n}\hugel
\nonumber\\
& R \leq \frac{1}{n}\log
\frac{W^n(Y^n|X^n)}{Q_{Y^n}(Y^n)}+\eta,
\left. \Gamma \geq \frac{1}{n}c(X^n)
\right\}
\nonumber\\
&+{\rm e}^{-n\eta}.\quad\: 
\label{eqn:azsad}
\end{align}
In (\ref{eqn:azsad}), we can choose any probability 
distribution $Q_{Y^n}$ on ${\cal Y}^n$. 
\end{lm}

\irPkAu{
{\it Proof: }By 
\clrTypoRevB{\cite{hayashi:03}}, 
we have that for any $\eta>0$ and for any $(\varphi^{(n)},\psi^{(n)})$  
satisfying 
$
(1/n)\log |{\cal K}_n| \geq R,
$
\begin{align}
& {\rm P}_{\rm c}^{(n)}
(\varphi^{(n)},\psi^{(n)}{}|W)
\leq p_{X^nY^n}\hugel
\nonumber\\
& R \leq \frac{1}{n}\log
\frac{W^n(Y^n|X^n)}{Q_{Y^n}(Y^n)}+\eta
\huger +{\rm e}^{-n\eta}.\quad\: 
\label{eqn:azsadPPP}
\end{align}
In (\ref{eqn:azsadPPP}), we can choose any probability 
distribution $Q_{Y^n}$ on ${\cal Y}^n$. 
Since $\varphi^{(n)}({\cal K}^{(n)}) \subseteq {\cal S}_{\Gamma}^{(n)}$,
the output $X^n=\varphi^{(n)}(K_n)$ of the encoder function 
$\varphi^{(n)}$
must satisfy 
$$
X^n\in {\cal S}^{(n)}_{\Gamma}
\mbox{ or equivalent to } \Gamma \geq \frac{1}{n}c(X^n) 
$$
with probability 1. Then we have 
\begin{align}
 &p_{X^nY^n}\left\{ R \leq \frac{1}{n}\log
 \frac{W^n(Y^n|X^n)}{Q_{Y^n}(Y^n)}+\eta
 \right\}
\notag\\
&=p_{X^nY^n}\hugel 
  R \leq \frac{1}{n} \log \frac{W^n(Y^n|X^n)}{Q_{Y^n}(Y^n)}+\eta,
\notag\\
& \qquad \qquad\quad \Gamma  \geq \frac{1}{n} c(X^n) \huger.
\label{eqn:azsadPPPa}  
\end{align}
From (\ref{eqn:azsadPPP}) and (\ref{eqn:azsadPPPa}), we have 
(\ref{eqn:azsad}) in Lemma \ref{lm:Ohzzz}.
\hfill\IEEEQED }

\irrComRevA{Let ${\cal P}^n({\cal Y})$ be a set of all distributions 
$Q^n=\{Q_t$ $\}_{t=1}^n$ on ${\cal Y}^n$ having the form
$$
Q^n(y^n)=\prod_{t=1}^nQ_t(y_t)\:\mbox{ for }y^n
\in {\cal Y}^n. 
$$}
From Lemma \ref{lm:Ohzzz}, we have the following lemma. 
\begin{lm}\label{lm:OhzzzB}
For any $\eta>0$ and for 
any $(\varphi^{(n)},$ $\psi^{(n)})$ satisfying 
\irPkAu{ 
$\varphi^{(n)}({\cal K}^{(n)}) 
\subseteq {\cal S}_{\Gamma}^{(n)}$ 
and } 
$(1/n)\log |{\cal K}_n| \geq R, $
\begin{align}
& {\rm P}_{\rm c}^{(n)}
(\varphi^{(n)},\psi^{(n)}{}|W)
 \leq p_{X^nY^n}
\hugel
\nonumber\\
& 
R\leq \frac{1}{n}
\sum_{t=1}^n
\log \frac{W(Y_t|X_t)}{Q_{t}(Y_t)}+\eta, 
\left. 
\Gamma \geq \frac{1}{n}\sum_{t=1}^nc(X_t)
\right\}
\nonumber\\
&+{\rm e}^{-n\eta}.
\label{eqn:azsadPPPz}
\end{align}
\irPkAu{
In (\ref{eqn:azsadPPPz}), we can choose any probability 
distribution $\{ Q_{t} \}_{t=1}^n$ on 
${\cal P}^n({\cal Y})$.}
\end{lm}

{\it Proof:} \irOAuc{ In (\ref{eqn:azsad}) in Lemma \ref{lm:Ohzzz}, 
we choose $Q_{Y^n}$ having the form 
\beq
Q_{Y^n}(Y^n)=\prod_{t=1}^nQ_{t}(Y_t).
\label{eqn:asDD}
\eeq
By definition it is obvious that  
\beq
\frac{c(X^n)}{n}=\frac{1}{n}\sum_{t=1}^nc(X_t).
\label{eqn:asDDz}
\eeq
From (\ref{eqn:asDD}), (\ref{eqn:asDDz}), 
and the bound (\ref{eqn:azsad}) 
in Lemma \ref{lm:Ohzzz}, we have (\ref{eqn:azsadPPPz}) 
in Lemma \ref{lm:OhzzzB}.
\hfill\IEEEQED}
\newcommand{\OmiT}{
\begin{align}
& {\rm P}_{\rm c}^{(n)}
(\varphi^{(n)},\psi^{(n)}{}|W)
\leq p_{X^nY^n}\hugel
\nonumber\\
& 
R\leq \frac{1}{n}
\sum_{t=1}^n
\log \frac{W(Y_t|X_t)}{Q_t(Y_t)}+\eta,
\Gamma\geq \frac{1}{n}\sum_{t=1}^n c(X_t)
\huger
\nonumber\\
&\qquad  +{\rm e}^{-n\eta},
\nonumber
\end{align}
completing the proof. 
}
\newcommand{\Apdb}{
}

We use the following lemma, which is well known as 
\irOlgRevB{
the Cram\`er's bound \cite{BucklewBook90} 
in the large deviation principle.
}


\begin{lm}
\label{lm:Ohzzzb}
For any real valued random variable $Z$ and any $\theta \geq 0$, 
we have
$$
\Pr\{Z \geq a \}\leq 
\exp
\left[
-\left(
\theta a -\log {\rm E}[\exp(\theta Z)]
\right) 
\right].
$$
\end{lm}

Here we define a quantity which serves as an exponential
upper bound of ${\rm P}_{\rm c}^{(n)}(\varphi^{(n)},$ 
$\psi^{(n)}{}|W)$. 
Let ${\cal P}^{(n)}(W)$ be a 
set of all probability distributions 
${p}_{X^nY^n}$ on 
$\empty {\cal X}^n$
$\times {\cal Y}^n$
having the form:
\begin{align*}
& {p}_{X^nY^n}(x^n,y^n)
=\prod_{t=1}^n 
{p}_{X_t|X^{t-1}}(x_t|x^{t-1})W(y_t|x_t).
\end{align*}
For simplicity of notation we use the notation $p^{(n)}$ 
for $p_{X^nY^n}$ $\in {\cal P}^{(n)}$
$(W)$. 
For $p^{(n)}$ $\in {\cal P}^{(n)}(W)$ 
and $Q^n=\{Q_{t}\}_{t=1}^n$ $\in {\cal P}^n(\cal Y)$, we define 
\begin{align*}
& \irr{\Omega^{(\mu,\lambda)}(p^{(n)},Q^{n})}
\\
& \defeq \log
{\rm E}_{p^{(n)}}
\left[
\prod_{t=1}^n \frac{W^{\lambda}(Y_t|X_t)
{\rm e}^{-\mu \lambda c(X_t)}}{Q_{t}^{\lambda}(Y_t)}
\right].
\end{align*}
By Lemmas \ref{lm:OhzzzB} and \ref{lm:Ohzzzb}, 
we have the following proposition. 
\begin{pro}
\label{pro:Ohzzp}
For any $\lambda\irOlgRevBb{, \mu \geq 0}$, any 
\irbRevA{$Q^n=\{Q \}_{ \irbRevAb{t=1} }^n$} 
\irbRevA{$\in {\cal P}^n({\cal Y})$}, 
and any $(\varphi^{(n)},\psi^{(n)})$  
satisfying 
\irPkAu{ 
$\varphi^{(n)}({\cal K}^{(n)}) \subseteq {\cal S}_{\Gamma}^{(n)}$ 
and } 
$(1/n)\log |{\cal K}_n| \geq R,$ 
\irbRevA{ there exists $p^{(n)}\in {\cal P}^{(n)}(W)$ 
such that}
\begin{align*}	
& {\rm P}_{\rm c}^{(n)}(\varphi^{(n)},\psi^{(n)}{}|W)
\\
&\leq 2\exp
\left\{
-n\frac{\ds 
\lambda(R -\mu\Gamma)-\frac{1}{n}
\irr{\Omega^{(\mu,\lambda)}(p^{(n)},Q^{n})}
}
{1+\lambda}
\right\}.
\end{align*}
\end{pro}

{\it Proof:} Under the condition
$(1/n)\log |{\cal K}_n| \geq R,$
we have the following chain of inequalities: 
\begin{align}
& 
{\rm P}_{\rm c}^{(n)}(\varphi^{(n)},\psi{}|W)
\MLeq{a} p_{X^nY^n} \hugel
\nonumber\\
& \quad R \leq \frac{1}{n}
\sum_{t=1}^n
\log \frac{W(Y_t|X_t)}{Q_t(Y_t)}+\eta, 
\Gamma \geq \frac{1}{n}\sum_{t=1}^nc(X_t)
\huger 
\nonumber\\ 
&\quad +{\rm e}^{-n\eta} 
\nonumber\\ 
&\leq p_{X^nY^n}\Hugel
(R-\mu\Gamma)-\eta 
\leq \frac{1}{n}
\sum_{t=1}^n
\log \left[\frac{W(Y_t|X_t)}{Q_t(Y_t)}\right]
\nonumber\\
& \qquad\qquad
\left. 
-\frac{ \mu }{n}\sum_{t=1}^n c(X_t)\right\}+{\rm e}^{-n\eta} 
\nonumber\\
&\MLeq{b} 
\exp\Bigl[n\Bigl\{-\lambda (R-\mu\Gamma)+\lambda \eta 
\left.\left.
+\frac{1}{n}
\irr{\Omega^{(\mu,\lambda)}(p^{(n)},Q^{n})}
\right\}\right]
\nonumber\\
&\quad +{\rm e}^{-n\eta}.
\label{eqn:aaabv}
\end{align}
Step (a) follows from Lemma \ref{lm:OhzzzB}.
Step (b) follows from Lemma \ref{lm:Ohzzzb}.
\irOlgRevBb{
When $n\lambda(R -\mu\Gamma)
\leq \Omega^{(\mu,\lambda)}(p^{(n)},Q^{n})$, the bound 
we wish to prove is obvious. In the following argument 
we assume that 
$n \lambda (R -\mu\Gamma)> \Omega^{(\mu,\lambda)}(p^{(n)},Q^{n})$.}
We choose $\eta$ so that 
\beqa
-\eta&=& -\lambda (R-\mu\Gamma) + \lambda\eta 
        +\frac{1}{n}\irr{\Omega^{(\mu,\lambda)}(p^{(n)},Q^{n})}.
\label{eqn:aaappp}
\eeqa
Solving (\ref{eqn:aaappp}) with respect to $\eta$, we have 
\beqno
\eta=
\frac{\ds 
\lambda (R-\mu\Gamma) -
\frac{1}{n}\irr{\Omega^{(\mu,\lambda)}(p^{(n)},Q^{n})}
}
{1+\lambda}.
\eeqno
For this choice of $\eta$ and (\ref{eqn:aaabv}), we have
\begin{align*}
& {\rm P}_{\rm c}^{(n)}
(\varphi^{(n)},\psi^{(n)}{}|W)
\leq 2{\rm e}^{-n\eta}
\\
&=2\exp
\left\{ 
-n\frac{\ds
\lambda (R-\mu\Gamma) 
-\frac{1}{n}\irr{\Omega^{(\mu,\lambda)}(p^{(n)},Q^{n})}}
{1+\lambda}
\right\},
\end{align*}
completing the proof. 
\hfill \IEEEQED

Set 
\begin{align}
& \overline{\Omega}^{(\mu,\lambda)}(W)
\notag\\
&\defeq  
\sup_{n\geq 1}
\max_{\scs {p}^{(n)} 
     \atop{\scs 
           \in {\cal P}^{(n)}(W)
          }
     }
\min_{\scs Q^n \in {\cal P}^n({\cal Y})}
\frac{1}{n}
\irr{\Omega^{(\mu,\lambda)}(p^{(n)},Q^{n})}.
\label{eqn:aaDD}
\end{align}
By the above definition of $\overline{\Omega}^{(\mu,\lambda)}(W)$ 
and Proposition \ref{pro:Ohzzp}, we have 
\begin{align}
G^{(n)}(R,\Gamma|W)
\geq&  
\frac{
\lambda (R-\mu\Gamma)-\overline{\Omega}^{(\mu,\lambda)}(W)
}{1+\lambda}-\frac{1}{n}\log 2.
\label{eqn:aaaxc}
\end{align}
Then from (\ref{eqn:aaaxc}), we obtain the following corollary. 
\begin{co} 
\label{co:corOne}
For any $\mu, \lambda \geq 0$, we have 
$$
G^*(R,\Gamma|W)\geq 
\frac{\lambda (R-\mu\Gamma) -\overline{\Omega}^{(\mu,\lambda)}(W)
} {1+\lambda}.
$$
\end{co}

We shall call $\overline{\Omega}^{(\mu,\lambda)}(W)$ 
the communication potential. The above corollary implies that 
the analysis of $\overline{\Omega}^{(\irr{\mu,}\lambda)}($$W)$ 
leads to an establishment of a strong converse theorem for 
the DMC. In the following argument we drive an explicit upper 
bound of $\overline{\Omega}^{(\irr{\mu,}\lambda)}(W)$. 
The following two propositions are mathematical cores 
to prove \irrTypo{Theorems \ref{Th:main0} and \ref{Th:main}.}
\begin{pro}\label{pro:mainpro0}
\irOlgRevB{ The NH method yields that} 
for any $\mu, \lambda \geq 0$, we have 
$
\overline{\Omega}^{(\mu,\lambda)}(W)
\leq \tilde{\Omega}^{(\mu,\lambda)}(W).
$
\end{pro}

\begin{pro}\label{pro:mainpro} \ \irOlgRevB{The improved 
recursive method yields that}
for any $\mu, \lambda \geq 0$, we have 
$\overline{\Omega}^{(\mu,\lambda)}(W)\leq $ $\Omega^{(\mu,\lambda)}(W).$
\end{pro}

\irbRevA{
As we mentioned previously, we have   
${\Omega}^{(\mu,\lambda)}(W)=\tilde{{\Omega}}^{(\mu,\lambda)}(W)$,
which implies that if one of the above two propositions holds, 
the other holds automatically. To compare the improved recursive 
method with the NH method and demonstrate that the former 
is quite different from the latter, we separately 
prove Propositions \ref{pro:mainpro0} and \ref{pro:mainpro}.
}
\irPkAu{
Furthermore, to compare the improved recursive method with 
the {\prev} recursive method, we prove the following 
proposition.
\begin{pro}\label{pro:mainproz}
The recursive method yields that for 
any $\mu, \lambda\geq 0$, we have 
$
\overline{\Omega}^{(\mu,\lambda)}(W)
\leq \hat{\Omega}^{(\mu,\lambda)}(W).
$
\end{pro}
}

We first prove Proposition \ref{pro:mainpro0} and 
\irPkAu{give a remark on the NH method. 
Using Proposition \ref{pro:mainpro0} and 
Corollary \ref{co:corOne}, we prove Theorem \ref{Th:main0}. 
We next prove Propositions \ref{pro:mainpro} and \ref{pro:mainproz}.
We further give a remark on the recursive and the improved 
recursive methods. Using Proposition \ref{pro:mainpro} 
and Corollary \ref{co:corOne}, we prove Theorem \ref{Th:main}. 
Similarly, using Proposition \ref{pro:mainproz} 
and Corollary \ref{co:corOne}, we prove 
Proposition \ref{pro:oldrecursive}. In the end of this section
we discuss the difference between the NH method and 
the improved recursive method.}

{\it Proof of Proposition \ref{pro:mainpro0}:} 
We first choose $\irrAE{\tilde{Q}} \in {\cal P}({\cal Y})$ arbitrary. 
Using $\irrAE{\tilde{Q}}  \in {\cal P}({\cal Y})$, we choose 
$Q^n \in {\cal P}^n({\cal Y})$ so that
$$
Q^n(y^n)=\irrAE{\tilde{Q}^n(y^n)}=\prod_{t=1}^n\irrAE{\tilde{Q}}(y_t)
$$
For this choice of $Q^n\irrAE{=\tilde{Q}^n(y^n)}$, 
we have the following chain 
of inequalities:
\begin{align}
& \exp\left[\Omega^{(\mu,\lambda)}(p^{(n)},\irrAE{\tilde{Q}}^n)\right]
\notag\\
&=\sum_{x^n \in {\cal X}^n }p_{X^n}(x^n)
\prod_{t=1}^n
\left(
\sum_{y_t \in {\cal Y}}
\frac{W^{1+\lambda}(y_t|x_t){\rm e}^{-\irrTypo{\mu\lambda} c(x_t)}}
     {\irrAE{\tilde{Q}}^{\lambda}(y_t)}
\right)
\notag\\
&\leq 
\left(
\max_{x\in {\cal X}}
\sum_{y \in {\cal Y}}
\frac{W^{1+\lambda}(y|x){\rm e}^{-\irrTypo{\mu \lambda} c(x)}}
     {\irrAE{\tilde{Q}}^{\lambda}(y)}
\right)^n
\label{eqn:AssPx}\\
&=
\left(
\max_{q_X\in {\cal P}({\cal X})}
\sum_{x \in {\cal X}}q_X(x)
\sum_{y \in {\cal Y}}
\frac{W^{1+\lambda}(y|x){\rm e}^{-\irrTypo{\mu\lambda} c(x)}}
     { \irrAE{\tilde{Q}}^{\lambda}(y)}
\right)^n
\notag\\
&=\exp\left[n
\left\{ 
\max_{q_X\in {\cal P}({\cal X})}
\Omega^{(\mu,\lambda)}(q_X, \irrAE{\tilde{Q}} |W)\right\}
\right].
\label{eqn:Zasddp}
\end{align}
From (\ref{eqn:Zasddp}), we have
\begin{align}
& \min_{{{Q}^n}\in {\cal P}^{n}({\cal Y})}
\frac{1}{n}\Omega^{(\mu,\lambda)}(p^{(n)},Q^n)
\leq \frac{1}{n}\Omega^{(\mu,\lambda)}
(p^{(n)},\irrAE{\tilde{Q}}^n)
\notag\\
&\leq \max_{q_X\in {\cal P}({\cal X})}
\Omega^{(\mu,\lambda)}(q_X,\irrAE{\tilde{Q}}|W).
\label{eqn:Zsdd000}
\end{align}
Since we have (\ref{eqn:Zsdd000}) 
for any $p^{(n)}\in {\cal P}^{(n)}(W)$, we have 
\beq
\overline{\Omega}^{(\mu,\lambda)}(W)
\leq \max_{q_X\in {\cal P}({\cal X})}
\Omega^{(\mu,\lambda)}(q_X,\irrAE{\tilde{Q}}|W).
\label{eqn:Zsdd00z}
\eeq
Since we have (\ref{eqn:Zsdd00z}) for any 
$\irrAE{\tilde{Q}} \in {\cal P}({\cal Y})$, we have 
the bound $\overline{\Omega}^{(\mu,\lambda)}(W)$ 
$\leq \tilde{\Omega}^{(\mu,\lambda)}(W)$ for 
any $\mu,\lambda\geq 0$.
\hfill\IEEEQED

\irPkAu{
\begin{rem}
In (\ref{eqn:aaDD}), the quantity 
$\overline{\Omega}^{(\mu,\lambda)}(W)$ 
has a form of $\max$-$\min$ optimization on 
$\Omega^{(\mu,\lambda)}(p^{(n)},Q^{n})$. On the other hand,
the single letterized upper bound $\tilde{\Omega}^{(\mu,\lambda)}(W)$ 
of $\overline{\Omega}^{(\mu,\lambda)}(W)$ has a form of 
$\min$-$\max$ \irPkAubc{optimization} on ${\Omega}^{(\mu,\lambda)}(q_X,Q|W)$.
This suggests that we have an exchange of the maximization and 
the minimization in the evaluation of upper bounds 
of $\overline{\Omega}^{(\mu,\lambda)}(W)$.
In fact, such an exchange 
appears in the right member of (\ref{eqn:AssPx}).     
This exchange implies the relaxation of evaluation 
on  upper bounds of $\overline{\Omega}^{(\mu,\lambda)}(W)$.
\end{rem}
}

{\it Proof of Theorem \ref{Th:main0}:}
From Corollary \ref{co:corOne} and Proposition \ref{pro:mainpro0}, we 
have $G^*(R,\Gamma|W)$ $\geq$$\tilde{G}^{(\mu,\lambda)}(R,\Gamma|W)$ for any 
$\mu, \lambda\geq 0$. Hence we have the bound 
$G^*(R,\Gamma|W)$$\geq$$\tilde{G}(R,\Gamma|W)$. 
\hfill\IEEEQED 

We next prove \irrTypo{Proposition}s \ref{pro:mainpro} and 
\ref{pro:mainproz}. Some preliminaries are 
necessary for the proof 
of this proposition. For each $t=1,2,\cdots,n$, 
define the function of 
$(x_t,y_t)$
$\in {\cal X}$
$\times {\cal Y}$
by 
\beqno
f_{{}Q_t}^{(\mu, \lambda)}
(x_t,y_t)
&\defeq& 
\frac{W^{\lambda}(y_t|x_t){\rm e}^{-\mu \lambda c(x_t)}}{Q_t^{\lambda}(y_t)}. 
\eeqno
For each $t=1,2,\cdots,n$, we define the probability distribution
\beqno
{p}_{X^tY^t;Q^t}^{(\mu,\lambda)}
&\defeq& 
\left\{
{p}_{X^tY^t;Q^t}^{(\mu,\lambda)}(x^t,y^t)
\right\}_{(x^t,y^t)
\in  \empty {\cal X}^t \times {\cal Y}^t} 
\eeqno
by 
\begin{align*}
& 
{p}_{X^t Y^t\irr{;Q^t}}^{(\mu,\lambda)}(x^t,y^t) 
\defeq 
C_t^{-1} p_{X^tY^t}(x^t,y^t)
\prod_{i=1}^t f_{{}Q_i}^{(\mu,\lambda)}(x_i,y_i)
\\
&=
C_t^{-1} p_{X^t}(x^t)
\prod_{i=1}^t \{W(y_i|x_i)
f_{{}Q_i}^{(\mu,\lambda)}(x_i,y_i)\},
\end{align*}
where
\beqno
C_t&\defeq&
{\rm E}_{p_{X^tY^t}}\left[
\prod_{i=1}^t f_{{}Q_i}^{(\mu,\lambda)}(X_i,Y
_i)
\right]
\eeqno
are constants for normalization. For each $t=1,2,\cdots,n$, set 
\beq 
\Phi_{t,Q^t}^{(\mu,\lambda)}\defeq C_tC_{t-1}^{-1},
\label{eqn:defa}
\eeq
where we define $C_0=1$. Then we have the following lemma.
\begin{lm}\label{lm:keylm}
\beqa 
& &\Omega^{(\mu,\lambda)}(p^{(n)}, Q^{n})
=\sum_{t=1}^n \log \Phi_{t,Q^{t}}^{(\mu,\lambda)}.
\label{eqn:defazzz}  
\eeqa
\end{lm}

{\it Proof}: From (\ref{eqn:defa}) we have
\beq
\log \Phi_{t,Q^t}^{(\mu,\lambda)}
=\log C_t - \log C_{t-1}. 
\label{eqn:aaap}
\eeq
Furthermore, by definition we have 
\beq
\irr{\Omega^{(\mu,\lambda)}(
p^{(n)},Q^{n})}
= \log C_n, C_0=1. 
\label{eqn:aaapq}
\eeq
From (\ref{eqn:aaap}) and (\ref{eqn:aaapq}), 
(\ref{eqn:defazzz}) is obvious. \hfill \IEEEQED

The following lemma is useful for the computation of 
$\Phi_{t,Q^t}^{(\mu,\lambda)}$ for $t=1,2,\cdots,n$.
\begin{lm}\label{lm:aaa}
For each $t=1,2, \cdots,n$, and for any 
$($ $x^t, y^t)\in $ 
$\empty {\cal X}^t$
$\times {\cal Y}^t$
we have
\begin{align}
& p_{X^t Y^t;Q^t}^{(\mu,\lambda)}(x^t,y^t)
\nonumber\\
&=(\Phi_{t,Q^t}^{(\mu,\lambda)})^{-1}
p_{X^{t-1} Y^{t-1};Q^{t-1} }^{(\mu,\lambda)}
(x^{t-1},y^{t-1})
\nonumber\\
& \quad \times 
p_{X_t|X^{t-1}}(x_t|x^{t-1})W(y_t|x_t)
f_{{}Q_t}^{(\mu,\lambda)}(x_t,y_t).
\label{eqn:satt}
\end{align}
Furthermore, we have
\beqa
& &\Phi_{t,Q^t}^{(\mu,\lambda)}
=\sum_{x^t,y^t} 
p_{{}X^{t-1}Y^{t-1};Q^{t-1}}^{(\mu,\lambda)}
(x^{t-1},y^{t-1})
\nonumber\\
& &\quad \times 
p_{X_t|X^{t-1}}(x_t|x^{t-1})W(y_t|x_t)
f_{Q_t}^{(\mu,\lambda)}(x_t,y_t).
\label{eqn:sattb}
\eeqa
\end{lm}

{\it Proof: } By the definition of 
${p}_{{}X^tY^{t};Q^t}^{(\mu,\lambda)}$$($$x^t,y^t)$, 
$t=1,2,$ $\cdots,n$, we have 
\begin{align}
&p_{X^tY^{t}; Q^t}^{(\mu,\lambda)}(x^t,y^{t})
\nonumber\\
&=C_t^{-1}
p_{X^t}(x^t)
\prod_{i=1}^t \{W(y_i|x_i)
f_{{}Q_i}^{(\mu,\lambda)}(x_i,y_i)\}.
\label{eqn:azaq}
\end{align} 
Then we have the following chain of equalities:
\begin{align} 
& p_{{}X^tY^t;Q^t}^{(\mu,\lambda)}(x^t,y^t)
\nonumber\\
&
\MEq{a}
C_t^{-1} 
\irrTypo{p_{X^t}(x^t)}
\prod_{i=1}^t \{ W(y_i|x_i)
f_{{}Q_i}^{(\mu,\irrTypo{\lambda})}
(x_i,y_i)\}
\nonumber\\
&=C_t^{-1}
p_{X^{t-1}}(x^{t-1})
\prod_{i=1}^{t-1} \{W(y_i|x_i)
f_{{}Q_i}^{(\mu,\lambda)}
(x_i,y_i)\}
\nonumber\\
&\quad\times 
p_{X_t|X^{t-1}}(x_t|x^{t-1})
W(y_t|x_t)
f_{{}Q_t}^{(\mu,\lambda)}
(x_t,y_t)
\nonumber\\
&\MEq{b}
C_t^{-1}C_{t-1}
p_{{}X^{t-1}Y^{t-1};Q^{t-1}}^{(\mu,\lambda)}
(x^{t-1},y^{t-1})
\nonumber\\
&\quad \times 
p_{X_t|X^{t-1}}(x_t|x^{t-1})W(y_t|x_t)
f_{{}Q_t}^{(\mu,\lambda)}(x_t,y_t)
\nonumber\\
&=(\Phi_{t,Q^t}^{(\mu,\lambda)})^{-1}
p_{{}X^{t-1}Y^{t-1};Q^{t-1}}^{(\mu,\lambda)}
(x^{t-1},y^{t-1})
\nonumber\\
&\quad \times 
p_{X_t|X^{t-1}}(x_t|x^{t-1})
W(y_t|x_t)
f_{{}Q_t}^{(\mu,\lambda)}
(x_t,y_t).
\label{eqn:daaaq}
\end{align}
Steps (a) and (b) follow from (\ref{eqn:azaq}). 
From (\ref{eqn:daaaq}), we have 
\begin{align} 
& \Phi_{t,Q^t}^{(\mu,\lambda)}
 p_{X^tY^t;Q^t}^{(\mu,\lambda)}(x^t,y^t)
\label{eqn:daxx}\\
&=p_{X^{t-1}Y^{t-1};Q^{t-1}}^{(\mu,\lambda)}
(x^{t-1},y^{t-1})
\nonumber\\
& \quad\times 
p_{X_t|X^{t-1}}(x_t|x^{t-1})
W(y_t|x_t)
f_{{}Q_t}^{(\mu,\lambda)}
(x_t,y_t).
\label{eqn:daaxx}
\end{align}
Taking summations of (\ref{eqn:daxx}) and (\ref{eqn:daaxx}) 
with respect to $x^t,$ $y^t$, we obtain 
\begin{align*}
\Phi_{t,Q^t}^{(\mu,\lambda)}
=&\sum_{x^t,y^t} 
p_{X^{t-1}Y^{t-1};Q^{t-1}}^{(\mu,\lambda)}(x^{t-1},y^{t-1})
\\
&\times p_{X_t|X^{t-1}}(x_t|x^{t-1})W(y_t|x_t)
  f_{{}Q_t}^{(\mu,\lambda)}(x_t,y_t),
\end{align*}
completing the proof.
\hfill \IEEEQED

We set 
\begin{align*}
p_{X_t;Q^{t-1}}^{(\mu,\lambda)}(x_t)
&=\sum_{x^{t-1},y^{t-1}}
p_{X^{t-1}Y^{t-1};Q^{t-1}}^{(\mu,\lambda)}(x^{t-1},y^{t-1})
\\
&\quad \times p_{X_t|{}X^{t-1}}(x_t|x^{t-1}).
\end{align*}
Then by (\ref{eqn:sattb}) in Lemma \ref{lm:aaa} and the definition 
of
$f_{Q_t}^{(\mu,\lambda)}$
$(x_t,y_t)$, we have 
\begin{align}
\Phi_{t,Q^{t}}^{(\mu,\lambda)}
&=\sum_{x_t,y_t}
p_{X_t;Q^{t-1}}^{(\mu,\lambda)}(x_t)
W(y_t|x_t)
\nonumber\\
& \quad \times 
\frac{W^{\lambda}(y_t|x_t){\rm e}^{-\mu \lambda c(x_t)} }
{Q_t^{\lambda}(y_t)}. 
\label{eqn:aasor}
\end{align}
In the following we prove \irrTypo{Proposition} \ref{pro:mainpro}.  

{\it Proof of \irOlgRevBb{Proposition} \ref{pro:mainpro}:} 
In (\ref{eqn:aasor}), \irbRevA{we denote the probability 
distribution 
$$
\left\{ p_{X_t;Q^{t-1}}^{(\mu,\lambda)}(x_t)
\right\}_{x_t\in {\cal X}}
$$
by $q_{X_t}$.}
Note that $q_{X_t}$ is a function of $Q^{t-1}$. We define a joint distribution 
$q_t=$ $q_{X_tY_t}$ on ${\cal X}\times {\cal Y}$ by 
$$
q_t(x_t,y_t)=q_{X_tY_t}(x_t,y_t)=q_{X_t}(x_t)W(y_t|x_t).
$$
Then, \irbRevA{from (\ref{eqn:aasor})}, we have
\begin{align}
\Phi_{t,Q^{t}}^{(\mu,\lambda)}
={\rm E}_{q_t}\left[\frac{W^{\lambda}(Y_t|X_t){\rm e}^{-\mu\lambda c(X_t)} }
{Q_t^{\lambda}(Y_t)}\right].
\label{eqn:AsWw}
\end{align}
 
We define $Q^n=\{Q_t\}_{t=1}^n$ recursively. 
For each $t=1,2,\cdots, n$,
we choose $Q_t$ so that it minimizes 
$\Phi_{t,Q^{t}}^{(\mu,\lambda)}$. 
Let $Q_{{\rm opt},t}$ be one of the \irPkAu{minimizers} on 
the above optimization problem. 
We set $Q_{{\rm opt}}^t\defeq \{Q_{{\rm opt},i}\}_{i=1}^t$.
Note that 
$Q_{{\rm opt},t}$ can be determined recursively depending 
on the $t-1$ previous \irPkAu{minimizers} $Q_{\rm opt}^{t-1}$.
Then, we have the following:
\begin{align}
& 
\log \Phi_{t,Q_{\rm opt}^{t}}^{(\mu,\lambda)}
=\log {\rm E}_{q_t}\left[\frac{W^{\lambda}
(Y_t|X_t){\rm e}^{-\mu\lambda c(X_t)}}
{Q_{{\rm opt},t}^{\lambda}(Y_t)}\right]
\notag\\
&= \min_{ Q \in {\cal P}({\cal Y})}
\irr{\Omega^{(\mu,\lambda)}(q_{X_t},Q|W)}
\notag\\
&\leq  \max_{q_{X_t}}\min_{ Q \in {\cal P}({\cal Y})}
\irr{\Omega^{(\mu,\lambda)}(q_{X_t},Q|W)}
=\Omega^{(\mu,\lambda)}(W).
\label{eqn:Zsdd}
\end{align}
Hence we have the following:
\begin{align}
& \min_{Q^n \in {\cal P}^n({\cal Y})}
\frac{1}{n}
\irr{\Omega^{(\mu,\lambda)}(p^{(n)},Q^{n})}
\leq \frac{1}{n}
\irr{\Omega^{(\mu,\lambda)}(p^{(n)},Q_{\rm opt}^{n})}
\notag\\
&\MEq{a}
\frac{1}{n}\sum_{t=1}^n \log \Phi_{t,Q_{\rm opt}^{t}}^{(\mu,\lambda)}
\MLeq{b}\Omega^{(\mu,\lambda)}(W).
\label{eqn:Zsss}
\end{align}
Step (a) follows from (\ref{eqn:defazzz}). 
Step (b) follows from (\ref{eqn:Zsdd}).
Since (\ref{eqn:Zsss}) holds for any $n\geq 1$ and for any
$p^{(n)}\in {\cal P}^{(n)}(W)$, we have the bound 
$\overline{\Omega}^{(\mu,\lambda)}(W)
\leq \Omega^{(\mu,\lambda)}(W)$ for any $\mu, \lambda \geq 0$.
\newcommand{\ZassPa}{
\beqno
& &\overline{\Omega}^{(\mu,\lambda)}(W)
\\
&=&\sup_{n \geq 1} \max_{p^{(n)}\in {\cal P}^{(n)}(W)}
\min_{Q^n\in {\cal P}^n({\cal Y})}
\frac{1}{n}\irr{\Omega^{(\mu,\lambda)}(p^{(n)},Q^{n})}
\nonumber\\
&\leq & \Omega^{(\mu,\lambda)}(W),
\eeqno
completing the proof.
}
\hfill\IEEEQED

{\it Proof of Proposition \ref{pro:mainproz}:} By the {\prev} 
recursive method we show that $\hat{\Omega}^{(\mu,\lambda)}(W)$ 
serves as an explicit upper bound of 
$\overline{\Omega}^{(\mu,\lambda)}(W)$. 
We start our argument from (\ref{eqn:AsWw}).
We set $q^t\defeq \{q_i\}_{i=1}^t$.
For each $t=1,2,\cdots,n$,
we choose $Q_t$ so that \irOlgRevBb{$Q_t=q_{Y_t}=q_{X_t}W$.}
Note that $q_{Y_t}$ can be determined 
recursively depending on the $t-1$ previous $q^{t-1}$.
Then, we have the following:
\begin{align}
& 
\log \Phi_{t, \{ q_{Y_i} \}_{i=1}^{t} }^{(\mu,\lambda)}
=\log {\rm E}_{q_t}
\left[\frac{W^{\lambda}(Y_t|X_t){\rm e}^{-\mu\lambda c(X_t)}}
{q_{Y_t}^{\lambda}(Y_t)}\right]
\notag\\
&= \Omega^{(\mu,\lambda)}(q_{X_t},q_{X_t}W|W)
  =\hat{\Omega}^{(\mu,\lambda)}(q_{X_t}|W)
\notag\\
&\leq  \max_{q_{X_t}} \hat{\Omega}^{(\mu,\lambda)}(q_{X_t}|W)
=\hat{\Omega}^{(\mu,\lambda)}(W).
\label{eqn:ZsddpZZ}
\end{align}
Hence, we have the following:
\begin{align}
& \min_{Q^n \in {\cal P}^n({\cal Y})}
\frac{1}{n}
     \Omega^{(\mu,\lambda)}(p^{(n)},Q^{n})
\leq \frac{1}{n}
     \Omega^{(\mu,\lambda)}(p^{(n)},\{ q_{Y_t} \}_{t=1}^{n})
\notag\\
&\MEq{a}
\frac{1}{n}\sum_{t=1}^n \log 
\Phi_{t, \{ q_{Y_t} \}_{t=1}^{n}}^{(\mu,\lambda)}
\MLeq{b}\hat{\Omega}^{(\mu,\lambda)}(W).
\label{eqn:Zsssa}
\end{align}
Step (a) follows from (\ref{eqn:defazzz}). 
Step (b) follows from (\ref{eqn:ZsddpZZ}).
Since (\ref{eqn:Zsssa}) holds for any $n\geq 1$ and for any
$p^{(n)}\in {\cal P}^{(n)}(W)$, we have the bound 
$\overline{\Omega}^{(\mu,\lambda)}(W) \leq  
\hat{\Omega}^{(\mu,\lambda)}(W)$ for any $\mu, \lambda \geq 0$.
\newcommand{\ZassP}{
\beqno
& &\overline{\Omega}^{(\mu,\lambda)}(W)
\\
&=&\sup_{n \geq 1} \max_{p^{(n)}\in {\cal P}^{(n)}(W)}
\min_{Q^n\in {\cal P}^n({\cal Y})}
\frac{1}{n}\irr{\Omega^{(\mu,\lambda)}(p^{(n)},Q^{n})}
\Omega_{p^{(n)}||Q^{n}}^{(\mu,\lambda)}(X^nY^n)
\nonumber\\
&\leq & 
\hat{\Omega}^{(\mu,\lambda)}(W),
\eeqno
completing the proof. 
}
\hfill\IEEEQED

\irPkAu{
\begin{rem}
In the {\prev} recursive algorithm we choose $Q_t$
so that $Q_t=q_{X_t}W$. This choice of $Q_t$ 
is sufficient for the lower bound $\hat{G}(R,\Gamma|W)$
to serve as an exponent function of 
the strong converse theorem for the DMC with input cost. 
However, to obtain the tight lower bound, 
the choice of $Q_t, t=1,2,\cdots,n$ 
so that it minimizes $\Phi_{t,Q^{t}}^{(\mu,\lambda)}$
is necessary. Similar improvement of the recursive method 
seems to be possible for the multiterminal 
communication systems 
\cite{Oohama15a},
\cite{Oohama15b},
\cite{OhIsita16abc},
\cite{oohama:18},  
\cite{oohama:19}, 
where 
the recursive method is used to derive a lower 
bound of the optimal exponent function.  
\end{rem}
}

{\it Proof of Theorem \ref{Th:main}:}
From Corollary \ref{co:corOne} and Proposition \ref{pro:mainpro}, we 
have $G^*(R,\Gamma|W)$ $\geq$$G^{(\mu,\lambda)}(R,\Gamma|W)$ 
for any $\mu, \lambda\geq 0$. Hence we have the bound 
$G^*(R,\Gamma|W)$$\geq$$G(R,\Gamma|W)$. 
\hfill\IEEEQED  

{\it Proof of Proposition \ref{pro:oldrecursive}:}
From Corollary \ref{co:corOne} and Proposition \ref{pro:mainproz}, we 
have $G^*(R,\Gamma|W)$ $\geq$$\hat{G}^{(\mu,\lambda)}(R,\Gamma|W)$ for any 
$\mu, \lambda\geq 0$. Hence we have the bound 
$G^*(R,\Gamma|W)$$\geq$$\hat{G}(R,\Gamma|W)$. 
\hfill\IEEEQED  

\irPkAu{

In the remaining part of this section we discuss about
the difference between the NH method and the improved 
recursive method. As you can see from the proof of 
Proposition \ref{pro:mainpro0}, the proof by NH method is very simple. 
This is due to an exchange between maximization and the minimization. 
This exchange implies a relaxation of the original optimization problem.      
However, if the minimax \eXchange theorem is established, the evaluation 
does not become loose and tight results are obtained.

On the other hand, the improved recursive method preserves 
the max-min structure in the single characterization of the 
multil-letter lower bound. Therefore, the resulting single 
letterized lower bound has a form of max-min optimization.
For that reason, we do not need the minimax \eXchange theorem to 
guarantee the tightness. 

In the following we explain that the NH method can not be applied 
to a multi-terminal communication system by giving an example of 
\irPkAubc{the} degraded broadcast channel \irPkAubc{(the DBC)} 
investigated 
by the author \cite{Oohama15a}. 
In \cite{Oohama15a}, the author \irPkAubc{ derived} a 
\irPkAubc{ strong} converse exponent function which 
is positive for rates outside the capacity region 
of the DBC. To obtain this result the author used the recursive method.
In the DBC, we have a proposition similar to Proposition \ref{pro:Ohzzp}. 
In this proposition, a lower bound of the optimal exponent 
function on the correct probability of decoding is characterized 
with a quantity like $\overline{\Omega}^{(\mu,\lambda)}(W)$. 
We denote this quantity by $\overline{\Omega}_{\rm DBC}$. 
Similar to $\overline{\Omega}^{(\mu,\lambda)}(W)$, 
$\overline{\Omega}_{\rm DBC}$ has a form of $\max$-$\min$ 
optimization on the multi-letter expression of 
an objective function like $(1/n)\Omega^{(\mu,\lambda)}(p^{(n)}, Q^n)$.
If we use the NH method to obtain a single 
letterized upper bound of $\overline{\Omega}_{\rm DBC}$, 
we can obtain the result similar to 
Proposition \ref{pro:mainpro0}. 
Concretely, we have the following bound: 
$$
\overline{\Omega}_{\rm DBC} \leq \tilde{\Omega}_{\rm DBC}, 
$$
where the function $\tilde{\Omega}_{\rm DBC}$ has a form of $\min$-$\max$ 
optimization on some objective function. In the case of the DBC, however,
this objective function is not convex with respect to 
variables related to the minimization or the maximization. 
This implies that the minimax \eXchange theorem 
may not hold in the case of the DBC. Hence, when the NH method 
is applied to the DBC, we can not eliminate the possibility 
that it is truly a loose evaluation. This is a main reason 
why the NH method can not be applicable to the proofs of 
strong converse theorems for multi-terminal 
communication systems.
}
\newcommand{\meMo}{

}

\section{Equivalence of Two Exponent Functions} 
\label{sec:ThreeExp}

\LemmaForProposition 

\irPkOh{
In the remaining part of this section we prove 
Property \ref{pr:prTwo} part c). }
\irOlgComRevB{We first observe 
that for each $\Gamma>\Gamma_0$, 
we have the following parametric expression of 
$G(R,\Gamma \Vl W)$: 
\begin{align}
G(R,\Gamma \Vl W)
&\MEq{a}\sup_{\scs \mu\geq 0,
\atop{\scs \rho\in [0,1]}} 
\left[\rho(R-\mu\Gamma)-J^{(\mu,\rho)}(W)\right]
\notag\\
&\MEq{b}\max_{\scs \mu\geq 0,
\atop{\scs \rho\in [0,1]}} 
\left[\rho(R-\mu\Gamma)-J^{(\mu,\rho)}(W)\right].
\label{eqn:prmtExG}
\end{align}
Step (a) follows from Proposition \ref{pro:pro1z}.
Step (b) follows Property \ref{pr:ApdLmTwo} part d). 
We next provide another expression of 
$\hat{G}(R,\Gamma|W)$, which is analogous 
to the parametric expression (\ref{eqn:prmtExG}) 
of $G(R,\Gamma \Vl W)$. To this end we set 
for $\rho\in [0,1)$ and $\mu \geq 0$, 
define 
\begin{align}
& \hat{J}^{(\mu,\rho)}(q_X|W)
 \defeq \overline{\rho}
 \Omega^{\left(\mu,\frac{\rho}{\overline{\rho}}\right)}
 (q_X,q_XW\Vl W)
\notag\\
& =\log
\left[\sum_{(x,y) \in {\cal X}\times {\cal Y}} 
q_X(x)
\left\{\frac{W(y|x){\rm e}^{-\mu\rho c(x)}}{(q_XW)^{\rho}(y)}
\right\}^{\frac{1}{\overline{\rho}}} 
\right]^{\overline{\rho}}.
\notag
\end{align}
When $\rho=1$, we define
\begin{align*}
 \hat{J}^{(\mu,1)}(q_X|W)\defeq& \lim_{\rho \uparrow 1 }
  \clrTypoRevB{\hat{J}}^{(\mu,\rho)}(q_X|W)
\\
=&\log 
\max_{\scs (x,y)\in {\cal X}\times {\cal Y}:
                    \atop{\scs \irrComRevA{q_X(x)W(y|x)>0}} }
\frac{W(y|x){\rm e}^{-\irrTypo{\mu}c(x)}}{(q_XW)(y)}. 
\end{align*}
Furthermore, set 
\begin{align*}
\hat{J}^{(\mu,\rho)}(W) & \defeq 
\max_{q_X\in {\cal P}({\cal X})} \irb{\hat{J}}^{(\mu,\rho)}(q_X|W),
\\
\hat{G}^{(\mu,\rho)}(R,\Gamma\Vl W) & \defeq
\rho(R-\mu\Gamma) -\hat{J}^{(\mu,\rho)}(W). 
\end{align*}
We have the following property.
\begin{pr}\label{pr:ApdLmThr}
$\quad$
\begin{itemize}

 \item[a)] For each fixed $\mu\geq 0$, 
 $\clrTypoRevB{\hat{J}}^{(\mu,\rho)}(W)$ is left continuous  
 at $\rho=1$.

 \item[b)] For each fixed $\mu\geq 0$, 
 $\clrTypoRevB{\hat{J}}^{(\mu,\rho)}(W)$ is a convex function of 
 $\rho \in [0,1)$.

 \item[c)] For each fixed $\rho\in [0,1]$, 
  $\clrTypoRevB{\hat{J}}^{(\mu,\rho)}(W)$ 
  is a convex function of $\mu \geq 0$.

 \item[d)] For $R\geq 0$ and $\Gamma > \Gamma_0$,  
the supremum in the definition of $\hat{G}(R,\Gamma\Vl W)$
can be replaced with the maximum. That is,  
\begin{align*}
& \sup_{\scs \mu \geq 0, \atop{\scs \irPkOh{\rho} 
  \in [0,1]}}
\hat{G}^{(\mu,  \irPkOh{\rho})}(R,\Gamma \Vl W)
= \max_{\scs \mu \geq 0, \atop{\scs \irPkOh{\rho}\in [0,1]}}
\hat{G}^{(\mu,\irPkOh{\rho})}(R,\Gamma \Vl W).
\end{align*}
\end{itemize}
\end{pr}

\clrMjRvOh{
We can prove this property in manners quite 
parallel with those of the proof of the Properties 
\ref{pr:ApdLmTwo}. In the proof of the part b) 
we also use Property \ref{pr:prTwo} part b) 
and Lemma \ref{lm:lmSddQ}.
Details of the proofs are given in Appendix \ref{sub:ApdLmTwo}.
}
From (\ref{eqn:hatGBda}), for each $\Gamma>\Gamma_0$, 
we have the following 
parametric expression of 
$\hat{G}(R,\Gamma|W)$: 
\begin{align}
&\hat{G}(R,\Gamma|W)=
\sup_{\scs \mu\geq 0,
\atop{\scs \lambda \geq 0}}
\frac{\ds \lambda(R-\mu \Gamma)-\hat{\Omega}^{(\mu,\lambda)}(W)}
{1+\lambda} 
\notag\\
&=\sup_{\scs \mu \geq 0, 
\atop{\scs \rho=\frac{\lambda}{1+\lambda}\in [0,1)}}
\left[ \rho(R-\mu\Gamma) -\hat{J}^{(\mu,\rho)}(W)
\right] 
\notag\\
&\MEq{a}\sup_{\scs \mu \geq 0, 
\atop{\scs \rho\in [0,1]}}
\left[ \rho(R-\mu\Gamma) -\hat{J}^{(\mu,\rho)}(W)
\right] 
\notag\\
&\MEq{b}\max_{\scs \mu \geq 0, \atop{\scs \rho \in [0,1]}}
\left[ \rho(R-\mu\Gamma) -\hat{J}^{(\mu,\rho)}(W)
\right]. 
\label{eqn:prmtExhatG}
\end{align}
Step (a) follows from Property \ref{pr:ApdLmThr} part a).
Step (b) follows from Property \ref{pr:ApdLmThr} part d). 
The parametric expression (\ref{eqn:prmtExG}) of 
$G(R,\Gamma|W)$ and the expression  (\ref{eqn:prmtExhatG})
of $\hat{G}(R,\Gamma|W)$ are useful for 
proving Property \ref{pr:prTwo} part c). 
}

\irPkOh{
We proceed to the next step for the proof. We consider
the case where $W$ is the symmetric erasure channel
given by (\ref{eqn:Erasure}). For 
$q\in {\cal P}({\cal X})$, set ${\rm Supp}(q) 
\defeq \{x \in {\cal X}: q(x)>0\}$. 
In the following argument, for simplicity 
of notation, we set 
${\cal X}_+\defeq {\rm Supp}(q_X)$. 
We compute ${J}^{(\mu,\rho)}(q_X|W)$ 
and $\hat{J}^{(\mu,\rho)}(q_X|W)$   
for the symmetric erasure channel given by (\ref{eqn:Erasure}). 
In the following argument we fix any nonnegative $\mu$. 
For each $\rho \in [0,1)$, define the three quantities 
$\Upsilon_i=\Upsilon_i(\rho),i=0,1,2$ by  
\begin{align}
\Upsilon_0(\rho) &\defeq \sum_{x\in {\cal X}_{+}} 
q_{X}(x)
  \left[\ts \frac{{\rm e}^{-{\mu c(x)}}}{q_X(x)}
 \right]^{\frac{\rho}{\overline{\rho}}}
\\
\Upsilon_1(\rho) & \defeq \sum_{x\in {\cal X}_{+}} 
q_{X}^{\overline{\rho}}(x) {\rm e}^{-{\mu\rho c(x)}},
\\
\Upsilon_2(\rho) & \defeq \sum_{x \in {\cal X}_{\scs +} }
q_X(x){\rm e}^{-\mu \frac{\rho}{\overline{\rho}} c(x)}.
\end{align}
Then we have
\begin{align}
&
\hat{J}^{(\mu,\rho)}(q_X|W)
={\overline{\rho}} \log \left[
 \overline{\theta}\Upsilon_0(\rho)
 +\theta\Upsilon_2(\rho) \right],
\label{eqn:hatJcomp}\\
&{J}^{(\mu,\rho)}(q_X|W)
= \log\left[\overline{\theta}\Upsilon_1(\rho)
+\theta\Upsilon_2
^{\overline{\rho}}(\rho)\right].
\label{eqn:Jcomp}
\end{align}
For each $\rho\in [0,1)$, set
\begin{align*}
 \kappa(\rho) \defeq & 
\left[\Upsilon_1(\rho)\right]^{-1}
\Upsilon_2^{\overline{\rho}}(\rho),      
\\
 \zeta(\rho)\defeq &
 \left[\overline{\theta}\Upsilon_1(\rho)+\theta
 \Upsilon_2^{\overline{\rho}}(\rho)
\right]^{-1}
 \left[
\overline{\theta}
 \Upsilon_1^{\frac{1}{\overline{\rho}}(\rho)}
  +\theta\Upsilon_2(\rho)
 \right]^{\overline{\rho}}
\notag\\
=&\left[\overline{\theta}+\theta\kappa(\rho)\right]^{-1}
\left[\overline{\theta}
+\theta
  \kappa^{\frac{1}{\overline{\rho}}}(\rho)\right]
 ^{\overline{\rho}}.
\end{align*}
We have the following property on 
$\Upsilon_i(\rho),i=0,1,2$, 
$\kappa(\rho)$, and $\zeta(\rho)$.   
\begin{pr}
\label{pr:PrOnUpsilonsEtc}
We assume that $|{\cal X}_{+}|\geq 2$. Under this assumption 
we have the following:  
\begin{itemize}
\item[a)] For each $\rho \in (0,1)$, we have the following 
\begin{align}
 \Upsilon_0(\rho)\geq 
 \Upsilon_1^{\frac{1}{\overline{\rho}}}(\rho),\:\; 
 \Upsilon_1 > \Upsilon_2^{\overline{\rho}}.
\label{eqn:UpBds}
\end{align}
\item[b)] 
 For each $\rho\in (0,1)$, we have the following:
\begin{align}
0\leq \kappa(\rho)<1,\: \zeta(\rho) >1. 
\label{eqn:UpBdsTwo}
\end{align}
Let 
\begin{align*}
& q_{X,\min}\defeq 
\min_{x \in{\cal X}_{+}}q_X(x),\: 
\Gamma_{\rm min}\defeq \min_{x \in{\cal X}_{+}}c(x),\
\\
&x^{\ast}\defeq \argmin_{x \in {\cal X}_{+}}c(x).
\end{align*}
Then we have the following:
\begin{align}
& [q_{X,\min}{\rm e}^{\mu \Gamma_{\rm min}}]^{\overline{\rho}}
 \sum_{x\in {\cal X_+}}{\rm e}^{-\mu c(x)}
\notag\\
& \leq {\Upsilon_1(\rho)}
\leq [{\rm e}^{\mu \Gamma_{\max}}]^{\overline{\rho}}
\sum_{x\in {\cal X_+}}{\rm e}^{-\mu c(x)},
\label{eqn:UpsilonOneBd}\\
&q_X^{\overline{\rho}}(x^{\ast}){\rm e}^{-\mu \Gamma_{\rm min}}
\leq \Upsilon_2^{\overline{\rho}} 
\leq \left[{\rm e}^{\mu\Gamma_{\rm min}}\right]^{\overline{\rho}}
 {\rm e}^{-\mu \Gamma_{\rm min}}.
\label{eqn:UpsilonTwoBd}
\end{align}
From (\ref{eqn:UpsilonOneBd}) and (\ref{eqn:UpsilonTwoBd}), 
we have
\begin{align}
&\left[\frac{q_X(x^{\ast})}
    {{\rm e}^{\mu\Gamma_{\max}}}
\right]^{\overline{\rho}}
\frac{{\rm e}^{-\mu \Gamma_{\rm min}}}
{\ds \sum_{x\in {\cal X_+}}{\rm e}^{-\mu c(x)}}
\notag\\
&\leq \kappa(\rho)
\leq 
q^{-\overline{\rho}}_{X,\min}\cdot 
     \frac{{\rm e}^{-\mu \Gamma_{\rm min}}}
{\ds \sum_{x\in {\cal X_+}}{\rm e}^{-\mu c(x)}}.
\label{eqn:FracUpsilonBd}
\end{align}
\item[c)] We set
$$
\kappa(1)\defeq \lim_{\rho\uparrow 1}\kappa(\rho),\:
\zeta(1) \defeq \lim_{\rho\uparrow 1}\zeta(\rho).
$$
Then we have the followings:
\begin{align}
 \kappa(1)&=\frac{{\rm e}^{-\mu \Gamma_{\rm min}}}
{\ds \sum_{x\in {\cal X_+}}{\rm e}^{-\mu c(x)}}<1,
\label{eqn:LimKappa}\\
\zeta(1)&=\frac{1}{\overline{\theta}+\theta \kappa(1)}
\label{eqn:LimZeta}\\
&
=\frac{\ds \sum_{x\in {\cal X_+}}{\rm e}^{-\mu c(x)}}
  {\ds \overline{\theta}\sum_{x\in {\cal X_+}}{\rm e}^{-\mu c(x)} 
  +\theta {\rm e}^{-\mu\Gamma_{\rm min}} }>1.
\label{eqn:LimZetab}
\end{align} 
\end{itemize}
\end{pr}
}

\irPkOh{
Proof of this property is given in 
Appendix \ref{sub:ApdPrfPrUpEtc}.
}
\newcommand{\ApdPrfPrUpEtc}{
\subsection{
Proof of Property \ref{pr:PrOnUpsilonsEtc}
}\label{sub:ApdPrfPrUpEtc}
In this appendix we prove 
Property \ref{pr:PrOnUpsilonsEtc}.

{\it Proof of Property \ref{pr:PrOnUpsilonsEtc}:}
We first prove the two bounds of (\ref{eqn:UpBds}) 
in the part a). We first prove the first 
inequality in (\ref{eqn:UpBds}). 
On a lower bound of $\Upsilon_0$, we have the following:
\begin{align}
& \Upsilon_0(\rho) = \sum_{x\in {\cal X}_{+}} 
q_{X}(x) {\rm e}^{-\mu \frac{\rho}{\overline{\rho}} c(x)}
\cdot 
[q_X(x)]^{-\frac{\rho}{\overline{\rho}}}
\notag\\
&\MGeq{a}\left[
\sum_{x\in {\cal X}_{+}} 
q_{X}^{\overline{\rho}}(x) {\rm e}^{-\mu {\rho} c(x)}
\right]^{\frac{1}{\overline{\rho}}}
\left[
\sum_{x\in {\cal X}_{+}} q_X(x)
\right]^{-\frac{\rho}{\overline{\rho}}}
=\Upsilon_1^{\frac{1}{\overline{\rho}}}(\rho).
\notag
\end{align}
In (a), we have used the reverse H\"older inequality
$$
\sum_i a_ib_i\geq 
\left( \sum_{i} a_i^{\frac{1}{\alpha}}\right)^{\alpha}
\left(\sum_{i} b_i^{\frac{1}{\beta}}\right)^{\beta}
$$
which holds for \irrTypoRevA{nonnegative} $a_i,b_i$ 
and for $\alpha+\beta=1$ such that either 
$\alpha >1$ or $\beta >1$.
In our case we have applied the inequality to 
$$
\left.
\ba{rcl}
i &\to & x,\:(\alpha,\beta) \to 
\left(\frac{1}{\overline{\rho}}, 
-{\ts \frac{\rho}{\overline{\rho}}}\right),
\vspace*{1mm}\\
a_i&\to & 
 q_X(x){\rm e}^{-\mu \frac{\rho}{\overline{\rho}}c(x)}, 
  b_i \to \left[q_X(x)\right]^{-\frac{\rho}{\overline{\rho}}}. 
\ea
\right\}
$$
We next prove the second inequality 
in (\ref{eqn:UpBds}). For each $\rho\in (0,1)$, 
we have the following:
\begin{align*}
&\Upsilon_2^{\overline{\rho}}(\rho)
 =\left[\sum_{x \in {\cal X}_{\scs +} }
q_X(x){\rm e}^{-\mu \frac{\rho}{\overline{\rho}} c(x)}
\right]^{\overline{\rho}}
\\
&\ML{a}\sum_{x \in {\cal X}_{\scs +} }
q_X^{\overline{\rho}}(x){\rm e}^{-\mu {\rho} c(x)}
=\Upsilon_1(\rho). 
\end{align*}
Step (a) follows from $|{\cal X}_{+}|\geq 2$ 
and the strict inequality
$$
\left[\sum_{i\in {\cal X}_{+}} a_i\right]^{\overline{\rho}}
<\sum_{i\in {\cal X}_{+}} a_i^{\overline{\rho}} 
$$
we have for $a_i>0,i\in {\cal X}_{+}$ and $\rho\in (0,1)$. 

We next prove the part b). The bound $0<\kappa(\rho)< 1$ is 
obvious from the second bound in \irb{(\ref{eqn:UpBds})}.
From $0<\kappa(\rho)< 1$ and the fact 
that for each $\rho\in (0,1)$, $a^{\overline{\rho}}$ 
is strictly concave with respect to $a\geq 0$, we have 
$$
 \overline{\theta}+\theta \kappa(\rho)
< \left[
\overline{\theta}
  +\theta\kappa^{\frac{1}{\overline{\rho}}}(\rho)
 \right]^{\overline{\rho}},
$$
from which we have the bound $\zeta(\rho)>1$ for $\rho\in (0,1)$.
We prove the two bounds in (\ref{eqn:UpsilonOneBd}). 
On lower bounds of $\Upsilon_1$, we have the 
following chain of inequalities:
\begin{align*}
&\Upsilon_1(\rho) 
 = \sum_{x\in {\cal X}_{+}} 
q_{X}^{\overline{\rho}}(x) {\rm e}^{-{\mu\rho c(x)}}   
\\
& =
\sum_{x\in {\cal X}_{+}} 
[q_{X}(x) {\rm e}^{\mu c(x)}]^{\overline{\rho}}
{\rm e}^{-{\mu c(x)}} 
\\
&\geq [q_{X,\min }{\rm e}^{\mu\Gamma_{\rm min}}]^{\overline{\rho}}
\sum_{x\in {\cal X}_{+}} {\rm e}^{-{\mu c(x)}}.
\end{align*}
On an upper bound of $\Upsilon_1$, we have the 
following chain of inequalities:
\begin{align*}
&\Upsilon_1(\rho) 
=\sum_{x\in {\cal X}_{+}} 
[q_{X}(x) {\rm e}^{\mu c(x)}]^{\overline{\rho}}
{\rm e}^{-{\mu c(x)}} 
\\
&\leq [ {\rm e}^{\mu \Gamma_{\max}} ]^{\overline{\rho}}
\sum_{x\in {\cal X}_{+}} {\rm e}^{-\mu c(x)}.
\end{align*}
Thus the two bound in (\ref{eqn:UpsilonOneBd}) are proved.
We prove the two bounds in (\ref{eqn:UpsilonTwoBd}).
On lower bounds of $\Upsilon_2^{\overline{\rho}}$, 
we have the following chain of inequalities:
\begin{align*}
&\Upsilon_2^{\overline{\rho}}(\rho) = 
\left[\sum_{x \in {\cal X}_{\scs +} }
q_X(x){\rm e}^{-\mu \frac{\rho}{\overline{\rho}} c(x)}
\right]^{\overline{\rho}} 
\\
& \geq \left[ q_X(x^{\ast}) {\rm e}^{-\mu 
 \frac{\rho}{\overline{\rho}}\Gamma_{\rm min}}
  \right]^{\overline{\rho}}  
 \irb{\geq} q_X^{\overline{\rho}}(x^{\ast})
  {\rm e}^{-\mu\Gamma_{\rm min}}.
\end{align*}
On upper bounds of $\Upsilon_2^{\overline{\rho}}$, 
we have the following chain of inequalities:
\begin{align*}
&\Upsilon_2^{\overline{\rho}}(\rho) = 
\left[\sum_{x \in {\cal X}_{\scs +} }
q_X(x){\rm e}^{-\mu \frac{\rho}{\overline{\rho}} c(x)}
\right]^{\overline{\rho}}  
\\
& \leq {\rm e}^{-\mu\rho\Gamma_{\rm min}}
  = \left[{\rm e}^{\mu 
  \Gamma_{\rm min}}\right]^{\overline{\rho}}
    {\rm e}^{-\mu\Gamma_{\rm min}}.
\end{align*}
Thus the two bound in (\ref{eqn:UpsilonTwoBd}) are proved.
We finally prove the part c). 
We observe upper and lower bounds 
of $\kappa(\rho)$ in (\ref{eqn:FracUpsilonBd}) for the case where
$\rho \in (0,1)$ tends to one. 
Since we have
$$
\lim_{\rho\uparrow 1}
\left[\frac{q_X(x^{\ast})}
    {{\rm e}^{\mu\Gamma_{\max}}}
\right]^{\overline{\rho}}
=\lim_{\rho\uparrow 1}q^{-\overline{\rho}}_{X,\min}=1
$$
in (\ref{eqn:FracUpsilonBd}), 
we have the equality (\ref{eqn:LimKappa}).
We prove \clrTypoOh{the} equality (\ref{eqn:LimZeta}). 
We first observe that from (\ref{eqn:FracUpsilonBd}), 
we have 
\begin{align}
0<\kappa^{\frac{1}{\overline{\rho}}}(\rho)
& \MLeq{a} q_{X,\min}^{-1}
  \kappa^{\frac{1}{\overline{\rho}}}(1)
  \ML{b}q_{X,\min}^{-1}.
\label{eqn:UpKappa}
\end{align}
Step (a) follows from (\ref{eqn:FracUpsilonBd}) and (\ref{eqn:LimKappa}). 
Step (b) follows form $0< \kappa(1)<1$.
From (\ref{eqn:UpKappa}), we have 
\begin{align}
\overline{\theta}^{\overline{\rho}}\leq 
\left[\irb{\overline{\theta}}+\theta\kappa^{\frac{1}
{\overline{\rho}}}(\rho)\right]^{\overline{\rho}}
\leq  \left[\overline{\theta}
+\theta q_{X,\min}^{-1}\right]^{\overline{\rho}}. 
\label{eqn:ZetaBdOne}
\end{align}
From (\ref{eqn:ZetaBdOne}), we have 
\begin{align}
\frac{\overline{\theta}^{\overline{\rho}}}
{\overline{\theta}+\theta \kappa(\rho)}
\leq \zeta(\rho)\leq 
\frac{\left[\overline{\theta} +\theta q_{X,\min}^{-1}\right]^{\overline{\rho}}}
{\overline{\theta}+\theta \kappa(\rho)}.
\label{eqn:LimZetaTwoz}
\end{align}
Since we have
$$
\lim_{\rho\uparrow 1} 
\frac{\overline{\theta}^{\overline{\rho}}}
     {\overline{\theta}+\theta \kappa(\rho)}
=\lim_{\rho\uparrow 1}
\frac{\left[\overline{\theta} +\theta q_{X,\min}^{-1}\right]^{\overline{\rho}}}
{\overline{\theta}+\theta \kappa(\rho)}
=\frac{1}{\overline{\theta}+\theta \kappa(1)}
$$
in (\ref{eqn:LimZetaTwoz}), 
we have the equality (\ref{eqn:LimZeta}). 
\hfill\IEEEQED
}

As a corollary of Lemma \ref{lm:lmSddQ}, we have 
the following. 
\begin{co}
\label{co:Subopt} 
\irPkOh{
We consider the case where $W$ is 
the symmetric erasure channel given 
by (\ref{eqn:Erasure})}.
\irPkOh{Then we have the following three results:}
\begin{itemize}
\item[a)]\irOlgComRevB{
$J^{(\mu,\rho)}(W) 
=\max_{q_X\in {\cal P}({\cal X})}
J^{(\mu,\rho)}(q_X| W)$ is attained by some 
$q_X \in {\cal P}({\cal X})$ with 
$|{\rm Supp}(q_X)| \geq 2$. }
\item[b)]\irPkOh{
For any 
$\mu \geq 0$, any $\rho\in (0,1]$, and 
any $q_X$ $\in {\cal P}({\cal X})$ with 
$|{\rm Supp}(q_X)|\geq 2$, 
\begin{align}
&\hat{J}^{(\mu,\rho)}(q_X|W)
>J^{(\mu,\rho)}(q_X|W).
\label{eqn:Aszpp}
\end{align}}
\item[c)]\irPkOh{
For any $\mu \geq 0$ and any 
$\rho\in (0,1]$, $\hat{J}^{(\mu,\rho)}(W)$ 
is strictly larger than $J^{(\mu,\rho)}(W)$.
}
\end{itemize}
\end{co}  

{\it Proof:} \irPkOh{
The part c) can easily be obtained by the parts a) and b).
We only prove the parts a) and b).}

\irOlgComRevB{We first prove 
the part a). }
\irOlgComRevB{
We consider the case 
where $\rho\in (0,1)$. The case $\rho=1$ will 
separately be discussed later. 
We start from the case of 
$|{\cal X}_{+}|=1$. 
Let ${\rm Supp}(q_X) =\{a\},a\in {\cal X}$. 
Then we have the following:
\begin{align}
& \exp\left[ J^{(\mu,\rho)}(q_X|W) \right]
=\sum_{y\in {\cal Y}}
\left[W^{\frac{1}{\overline{\rho}}}
(y|a){\rm e}^{-\mu\frac{\rho}{\overline{\rho}}c(\clrTypoRevB{a})} 
\right]^{\overline{\rho}}
\notag\\
& = \exp\left[-\mu\rho c(a)\right],
\notag
\end{align}
from which we have  
\begin{align}
& \max_{q_X:|{\cal X}_{+}|=1} 
J^{(\mu,\rho)}(q_X| W)
=-\min_{a\in {\cal X}}{{\mu\rho}c(a)}
\notag\\
&  
=-{\mu\rho}\Gamma_0.
\end{align}
We next consider the case where 
$|{\cal X}_{+}|=2$. 
Let ${\cal X}_{+}=\{1,2\}$. Set $c_i\defeq c(i),i=1,2.$ 
Without loss of generality we may assume 
that $c_1=\Gamma_0$. We set $q_X(1)=\overline{q},q_X(2)={q}$.
Computing $\exp[J^{(\mu,\rho)}(q_X|W)]$, we 
have the following: 
\begin{align}
& \exp\left[
  J^{(\mu,\rho)}(q_X|W)\right]
=\theta\left[
\overline{q}{\rm e}^{-{\mu \frac{\rho}{\overline{\rho}}c_1} }  
+        {q}{\rm e}^{-{\mu \frac{\rho}{\overline{\rho}}c_2} }  
  \right]^{\overline{\rho}}
\notag\\  
& \quad 
    +\overline{\theta} 
\left[\overline{q}^{\overline{\rho} }
    {\rm e}^{-{\mu\rho c_1}}
               +q^{\overline{\rho}}
    {\rm e}^{-{\mu \rho c_2}}
    \right].
\end{align}
For $q\in [0,1]$, set
\begin{align*}
 \Xi(q)\defeq & 
\theta\left[
\overline{q}{\rm e}^{-{\mu \frac{\rho}{\overline{\rho}}c_1} }  
+        {q}{\rm e}^{-{\mu \frac{\rho}{\overline{\rho}}c_2} }  
  \right]^{\overline{\rho}}
\notag\\  
& 
    +\overline{\theta} 
\left[\overline{q}^{\overline{\rho} }
    {\rm e}^{-{\mu\rho c_1}}
               +q^{\overline{\rho}}
    {\rm e}^{-{\mu \rho c_2}}
    \right].
\end{align*}
From its formula it is obvious that $\Xi(q)$ 
is strictly concave with respect to $q\in [0,1]$.
Furthermore, we have the following:
\begin{align}
& \Xi(0)={\rm e}^{-\mu\rho c_1},\:
  \Xi(1)={\rm e}^{-\mu\rho c_2},
\notag\\
&\lim_{q\downarrow 0} \Xi^{\prime}(q)=+\infty,\:
\lim_{q\uparrow 1} \Xi^{\prime}(q)=-\infty.
\notag
\end{align}
Those imply that 
\begin{align}
&\exp\left[\max_{q_X:|{\cal X}_{+}|=1} 
J^{(\mu,\rho)}(q_X| W)\right]
\notag\\
&=\max_{q_X:|{\cal X}_{+}|=1} 
\exp\left[J^{(\mu,\rho)}(q_X| W)\right]
\notag\\
& ={\rm e}^{-{\mu\rho}{}\Gamma_0}
< \max_{q_X:|{\cal X}_{+}|=2} 
\exp\left[J^{(\mu,\rho)}(q_X|W)\right]
\notag\\
&=\exp\left[
\max_{q_X:|{\cal X}_{+}|=2} 
J^{(\mu,\rho)}(q_X|W)\right].
\label{eqn:JoneJtwoa}
\end{align}
From (\ref{eqn:JoneJtwoa}), we have that
for the choice of ${\cal X}_{+}$ such that 
${\cal X}_{+}=\{1,2\}$ with $c(1)=\Gamma_0$,
$$
\max_{q_X:|{\cal X}_{+}|=1} 
J^{(\mu,\rho)}(q_X| W)<
\max_{q_X:|{\cal X}_{+}|=2} 
J^{(\mu,\rho)}(q_X|W), 
$$
implying that we have the part a) 
for $\rho \in (0,1)$.
We next consider the case of $\rho=1$. 
When $|{\cal X}_{+}|=1$, we have 
\begin{align}
&\max_{q_X:|{\cal X}_{+}|=1} 
J^{(\mu,1)}(q_X| W)
=-\min_{a\in {\cal X}}{{\mu}c(a)}
=-{\mu}\Gamma_0.
\end{align}
On the other hand when $|{\cal X}_{+}|=2$, we have 
\begin{align}
& \exp\left[J^{(\mu,1)}(q_X|W)\right]
=\clrTypoRevB{\theta}{\rm e}^{-\mu \Gamma_0}
+\clrTypoRevB{\overline{\theta}}
\left[{\rm e}^{-\mu \Gamma_0}+{\rm e}^{-\mu c_2}\right] 
\notag\\
& ={\rm e}^{-\mu \Gamma_0}+
\clrTypoRevB{\overline{\theta}}{\rm e}^{-\mu c_2}
> {\rm e}^{-\mu \Gamma_0}.
\label{eqn:CompJThr}
\end{align}
From (\ref{eqn:CompJThr}), we have that
for some  
$q_X \in {\cal P}({\cal X})$ with ${\cal X}_{+}=\{1,2\}$ 
and $c(1)=\Gamma_0$,
$$
J^{(\mu,1)}(q_X|W)> 
-{\mu}\Gamma_0=\max_{q_X:|{\cal X}_{+}|=1} 
J^{(\mu,1)}(q_X| W),
$$
implying that we have the part a) also for $\rho=1$.
}

\irPkOh{

\clrTypoRevB{We next prove the part b).
For each $\rho\in (0,1)$,
we have the following chain of inequalies:}
\begin{align}
&\hat{J}^{(\mu,\rho)}(q_X|W)-{J}^{(\mu,\rho)}(q_X|W)
\notag\\
&\MEq{a}\log
\left[
\overline{\theta}\Upsilon_0 +\theta \Upsilon_2
 \right]^{\overline{\rho}}
 -\log \left[ \overline{\theta}\Upsilon_1 
   +\theta\Upsilon_2^{\overline{\rho}}
  \right]
 \notag\\
&\MGeq{b}\log
\left[
\overline{\theta}\Upsilon_1^{\frac{1}{\overline{\rho}}} 
+\theta \Upsilon_2
 \right]^{\overline{\rho}}
 -\log\left[ \overline{\theta}\Upsilon_1
 +\theta\Upsilon_2^{\overline{\rho}} 
  \right]
\notag\\
& =\log \zeta(\rho) \MG{c}0.
\label{eqn:LbofJb}   
\end{align}
Step (a) follows from (\ref{eqn:hatJcomp}) 
and (\ref{eqn:Jcomp}). 
Step (b) follows from the first inequality of 
(\ref{eqn:UpBds}) in Property \ref{pr:PrOnUpsilonsEtc} 
part a). Step (c) follows from the second 
inequality of (\ref{eqn:UpBdsTwo}) in Property 
\ref{pr:PrOnUpsilonsEtc} part b). 
We now consider the case of $\rho=1$. By letting 
$\rho\in (0,1)$ arbitrary close to one in (\ref{eqn:LbofJb}), 
we have the following bound: 
\begin{align}
&\hat{J}^{(\mu,1)}(q_X|W)-{J}^{(\mu,1)}(q_X|W)
\notag\\
&= \lim_{\rho\uparrow 1}\hat{J}^{(\mu,\rho)}(q_X|W)
  -\lim_{\rho\uparrow 1}{J}^{(\mu,\rho)}(q_X|W)
\notag\\  
&=\lim_{\rho\uparrow 1}\left[
 \hat{J}^{(\mu,\rho)}(q_X|W)-{J}^{(\mu,\rho)}(q_X|W)
 \right]
\notag\\ 
&\MGeq{a}\lim_{\rho\uparrow 1}\zeta(\rho)
=\log \zeta(1) \MG{b}0.
\label{eqn:LbofJc}   
\end{align} 
Step (a) follows from (\ref{eqn:LbofJb}). 
Step (b) follows from the equality 
of (\ref{eqn:LimZetab}) in Property \ref{pr:PrOnUpsilonsEtc} 
part c). Thus we have (\ref{eqn:Aszpp}) for any 
$\mu \geq 0$, any $\rho\in (0,1]$, 
and any $q_X \in {\cal P}({\cal X})$ 
with $|{\rm Supp}(q_X)|\geq 2$. 
\hfill\IEEEQED
}

\irOlgComRevB{
{\it Proof of Property \ref{pr:prTwo} part c):}
For each $\Gamma > \Gamma_0$, we use the parametric 
expression (\ref{eqn:prmtExG}) of $G(R,\Gamma\Vl W)$ 
and the expression (\ref{eqn:prmtExhatG}) 
of $\hat{G}(R,\Gamma\Vl W)$.   
Let $(\mu^{*},\rho^{*})$ be the pair attaining 
the maximum of $\hat{G}^{(\mu,\rho)}(R,\Gamma|W)$
for $\mu\geq 0$, $\rho\in \clrTypoRevB{[0,1]}$ 
in the definition of $\clrTypoOh{\hat{G}}(R,\Gamma|W)$. 
Note that by Property \ref{pr:prTwo} part a), under 
the assumption that $R>C(\Gamma\Vl W)$, we have 
\begin{align}
\hat{G}(R,\Gamma \Vl W)
=\hat{G}^{(\mu^{\ast},\rho^{\ast})}(R,\Gamma \Vl W)>0.
\label{eqn:PrLmTwOa}
\end{align}
Since we have that for any $\mu \geq 0$,
$
\hat{G}^{(\mu,0)}(R,\Gamma|W)=0,
$
the pair $(\mu^{\ast},\rho^{\ast})$ satisfying 
(\ref{eqn:PrLmTwOa}) must be 
$\mu^{\ast}\geq 0$ and $\rho^{\ast} \in (0,1]$.
Then on $\hat{G}(R,\Gamma|W)$, we have the following 
chain of inequalities:
\begin{align*}
&\hat{G}(R,\Gamma\Vl W)
 =\hat{G}^{(\mu^{\ast},\rho^{\ast})}(R,\Gamma \Vl W)
\\
&=\rho^{\ast}(R-\mu^{\ast}\Gamma)-
\hat{J}^{(\mu^{\ast},\rho^{\ast})}(W)
\\
&\ML{a}\rho^{\ast}(R-\mu^{\ast}\Gamma)-
  {J}^{(\mu^{\ast},\rho^{\ast})}(W)
\leq {G}(R,\Gamma\Vl W).
\end{align*}
Step (a) follows from $\rho^{\ast}\in (0,1]$ 
and Corollary \ref{co:Subopt} part c).
\hfill\IEEEQED
}


\section{\irrTypoRevA{Extension} to General Memoryless Channels}
\label{sec:GeneralCase}
\noindent

In this section we consider a stationary general memoryless 
channel \irrTypo{(GMC)}, where ${\cal X}$ and ${\cal Y}$ are real lines. 
The GMC is specified with a noisy channel $W$.
We assume that for each \irOlgTypoRevB{$x\in {\cal X}$}, 
$W$ has a \irrTypoRevA{conditional} 
density function $W({\rm d}y|x)$. Except for 
Theorem \ref{Th:DK} 
and Proposition \ref{pro:pro1b}, the results 
we have presented so far also hold for this general case.
Let $q_X$ be a probability measure on ${\cal X}$ having  
the density $q_X({\rm d}x)$. 
Let $Q$ be a probability measure on ${\cal Y}$ having  
the density $Q({\rm d}y)$. 
In the case of GMC, the definitions of $\Omega^{(\mu,\lambda)}(q_X,Q|W)$ 
and $\Omega^{(\mu,\lambda)}(W)$ are
\begin{align*}
& \Omega^{(\mu,\lambda)}(q_X,Q|W)
\\
&  \defeq \log
\left[\int \int {\rm d}x {\rm d}y
q_{X}(x)
\frac{ W^{1+\lambda}(y|x){\rm e}^{-\mu \lambda c(x)}}
{Q^\lambda(y)}\right],
\\
& \Omega^{(\mu,\lambda)}(W)
\defeq 
\max_{q_X}
\min_{Q}
\Omega^{(\mu,\lambda)}(q_X,Q|W).
\end{align*}
For GMC $W$, we define the exponent functions 
$G_{\rmOH}^{(\mu,\lambda)}($ 
$R,\Gamma|W)$ and $G_{\rmOH}(R,\Gamma|W)$
in a manner similar to the definitions 
of those exponent functions in the case of DMC. 
The following theorem is a generalization 
of Theorem \ref{Th:main} to the case of GMC.
\begin{Th}\label{th:General}
For any GMC $W$, we have 
\beqa
& &G^*(R,\Gamma|W) 
\geq G_{\rmOH}(R,\Gamma|W).
\label{eqn:GmainIeq000z}
\eeqa
\end{Th}

We next describe a lemma which is a generalization 
of Lemma \ref{lm:lmSddQ} to the case of GMC. For 
$\rho \in [0,1)$, define   
\begin{align*}
& J^{(\mu, \rho)}(q_X|W)
\\
&  \defeq \log
\int {\rm d}y
\left[\int {\rm d}x
q_X(x)\left\{W(y|x){\rm e}^{-\mu \clrTypoOh{\rho} c(x)}\right\}
^{\frac{1}{\clrTypoOh{\overline{\rho}}}} 
\right]^{\clrTypoOh{\overline{\rho}}}.
\end{align*}
Then we have the following lemma.
\begin{lm}\label{lm:GlmSddQ} For any probability 
\irrTypoRevA{density} function 
$q_X=\irrTypo{q_X({\rm d}x)}$ on ${\cal X}$, we have 
$$
\min_{Q}\Omega^{(\mu,\lambda)}(q_X,Q|W)
=(1+\lambda)J^{(\mu,\frac{\lambda}{1+\lambda})}(q_X|W).
$$
The probability density function $Q$ attaining 
$(1+\lambda)$ $J^{(\mu,\frac{\lambda}{1+\lambda})}($$q_X|W)$ 
is given by 
$$
Q(y)=\kappa \left[ 
\int {\rm d} x q_X(x) 
W^{1+\lambda}(y|x){\rm e}^{-\mu \lambda c(x)}\right]^{\frac{1}{1+\lambda}},
$$
where $\kappa$ is a constant for normalization, having the form
\begin{align}
\kappa^{-1}
&=\int {\rm d} y  
\left[\int {\rm d} x q_X(x) 
W^{1+\lambda}(y|x){\rm e}^{-\mu \lambda c(x)} 
\right]^{\frac{1}{1+\lambda}}
\notag\\
&=\exp\left[ {J}^{(\mu, \frac{\lambda}{1+\lambda})}(q_X|W) \right].
\label{eqn:GAzzXc}
\end{align}
%
\end{lm}

For GMC $W$, we define the exponent functions 
$G_{\rm AR}^{(\mu,\lambda)}($$R,\Gamma|W)$ and $G_{\rm AR}(R,\Gamma|W)$
in a manner similar to the definitions 
of those exponent functions in the case of DMC. 
From Lemma \ref{lm:GlmSddQ}, 
we have the following proposition, which 
is a generalization of Proposition \ref{pro:pro1z} to the case of GMC. 
\begin{pro}\label{pro:Gpro1z} For any GMC $W$ and 
for any $\mu,\lambda\geq 0$, we have 
the following:
\begin{align}
G_{\rmOH}^{(\mu,\lambda)}(R,\Gamma|W)
&=G_{\rm AR}^{(\mu,\frac{\lambda}{1+\lambda})}(R,\Gamma|W).
\label{eqn:GmainEq00a}
\end{align}
In particular, we have 
\begin{align}
& G_{\rmOH}(R,\Gamma|W)=G_{\rm AR}(R,\Gamma|W).
\label{eqn:GmainEq0xx}
\end{align}
\end{pro}

From Theorem \ref{th:General} and Proposition \ref{pro:Gpro1z}, 
we immediately obtain the following result. 
\begin{Th}\label{th:GenaralB} For any GMC $W$, we have 
\beqa
& &G^*(R,\Gamma|W) 
\geq G_{\rmOH}(R,\Gamma|W)
    =G_{\rm AR}(R,\Gamma|W).
\label{eqn:GmainIeq000}
\eeqa
\end{Th}

Theorem \ref{Th:DK} is related to the upper bound of $G^*(R,\Gamma|W)$. 
Proof of this theorem depends heavily on a finiteness of ${\cal X}$. 
We have no result on the upper bound of $G^{*}(R,\Gamma|W)$ and the 
tightness of the bound $G_{\rmOH}(R,\Gamma|W)$. In the case of GMC, 
$G_{\rmOH}(R,\Gamma|W)$ and $G_{\rm AR}(R,\Gamma|W)$ are not computable 
since those are variational problems. On the other hand, 
$G_{\rmOH}(R,\Gamma|W)$ has a min-max expression. In 
\cite{Oohama17}, the author succeeded in obtaining an 
explicit form of $G(R,\Gamma|W)$ for additive white Gaussian noise 
channels \irrAE{(AWGNs)} by utilizing the min-max property of
$G_{\rmOH}(R,\Gamma|W)$.

%

\section*{\empty}

\appendix

\ApdaAAAA
\ApdaDirectDK

\ApdONE
\irOlgComRevB{\PrApdLmTwo}
\irPkOh{\ApdPrfPrUpEtc}

\bibliographystyle{IEEEtran}
\bibliography{SingleStConvIeiceMain}

\end{document}